\documentclass{aa}  

\usepackage{graphicx}
\usepackage{txfonts}
\usepackage{amsmath}
\usepackage{enumerate}
\usepackage{natbib}
\newcommand{\Teff} {$T_\mathrm{eff}$}

\begin{document} 

   \title{Discovery of binarity, spectroscopic frequency analysis, and mode identification of the $\delta$\,Sct star 4\,CVn \thanks{This paper includes data taken at The McDonald Observatory of The University of Texas at Austin.}\fnmsep\thanks{The software package FAMIAS, developed in the framework of the FP6 European Coordination Action HELAS (http://www.helas-eu.org/), has been used in this research.}}

   \author{V.\,S. Schmid\inst{1}\fnmsep\thanks{Aspirant PhD Fellow, Fonds voor Wetenschappelijk Onderzoek -- Vlaanderen (FWO), Belgium}
          \and N. Theme\ss l\inst{2,3}
          \and M. Breger\inst{3,4}
          \and P. Degroote\inst{1}\fnmsep\thanks{Postdoctoral fellow of the Fonds voor Wetenschappelijk Onderzoek -- Vlaanderen (FWO), Belgium}
          \and C. Aerts\inst{1,5}
          \and P.\,G. Beck\inst{1,6}
          \and A. Tkachenko\inst{1}\fnmsep$^{\star\star\star\star}$
          \and \\T. Van Reeth\inst{1}
          \and S. Bloemen\inst{5}
          \and J. Debosscher\inst{1}
          \and B.\,G. Castanheira\inst{4}
          \and B.\,E. McArthur\inst{4}
          \and P.\,I. P\'apics\inst{1}
          \and V. Fritz\inst{3}
          \and \\R.\,E. Falcon\inst{4}}

          \titlerunning{Discovery of binarity, spectroscopic frequency analysis, and mode identification of the $\delta$\,Sct star 4\,CVn}
          \authorrunning{V.\,S. Schmid et al.}
          
          \institute{
          Institute of Astronomy, KU Leuven, Celestijnenlaan 200D, B -- 3001 Leuven, Belgium
          \and 
          Max Planck Institut f\"ur Sonnensystemforschung, Justus-von-Liebig-Weg 3, 37077 G\"ottingen, Germany
          \and
          Department of Astrophysics, University of Vienna, T\"{u}rkenschanzstra\ss e 17, A -- 1180 Vienna, Austria
          \and 
          Department of Astronomy, University of Texas at Austin 2515 Speedway, Stop C1400
          Austin, Texas 78712-1205
          \and
          Department of Astrophysics/IMAPP, Radboud University Nijmegen, P.O. Box 9010,
          6500 GL Nijmegen, The Netherlands
          \and
          Laboratoire AIM, CEA/DSM -- CNRS -- Universit\'e Denis Diderot -- IRUF/SAp, 91191 Gif-sur-Yvette Cedex, France
          }

   \date{Received 18 February 2014 / Accepted 10 July 2014}

  \abstract{More than 40 years of ground-based photometric observations of the $\delta$\,Sct star 4\,CVn revealed 18 independent oscillation frequencies, including radial as well as non-radial p-modes of low spherical degree $\ell\leq2$. From 2008 to 2011, more than 2000 spectra were obtained at the 2.1-m Otto-Struve telescope at the McDonald Observatory. We present the analysis of the line-profile variations, based on the Fourier-parameter fit method, detected in the absorption lines of 4\,CVn, which carry clear signatures of the pulsations. From a non-sinusoidal, periodic variation of the radial velocities, we discovered that 4\,CVn is an eccentric binary system, with an orbital period $P_{orb}=124.44\pm0.03~$d and an eccentricity $e=0.311\pm0.003$. We firmly detect 20 oscillation frequencies, 9 of which are previously unseen in photometric data, and attempt mode identification for the two dominant modes, $f_1=7.3764~\mathrm{d}^{-1}$ and $f_2=5.8496~\mathrm{d}^{-1}$, and determine the prograde or retrograde nature of 7 of the modes. The projected rotational velocity of the star, $v_{eq}\sin i\simeq106.7~\mathrm{km\,s}^{-1}$, translates to a rotation rate of $v_{eq}/v_{crit}\geq33\%$. This relatively high rotation rate hampers unique mode identification, since higher-order effects of rotation are not included in the current methodology. We conclude that, in order to achieve unambiguous mode identification for 4\,CVn, a complete description of rotation and the use of blended lines have to be included in mode-identification techniques.}
   
   \keywords{techniques: spectroscopic -- stars: variables: delta Scuti -- stars: individual: 4\,CVn -- stars: fundamental parameters -- binaries: spectroscopic}

   \maketitle
%

\section{Introduction}

In the Hertzsprung-Russell diagram the group of $\delta$\,Sct pulsators is located in the classical instability strip. They are on or slightly above the main sequence moving toward the giant branch after the hydrogen in their cores is depleted. This evolutionary stage makes them interesting from an astrophysical point of view and important objects for asteroseismic studies, since their interior structures undergo large changes. Comprehensive reviews on $\delta$\,Sct stars and on asteroseismology are given by \citet{Breger2000a}, and \citet{Aerts2010}, respectively.

New advances in the field of A- and F-type stars were achieved by space-based telescopes, such as MOST \citep{Walker2003}, CoRoT \citep{Auvergne2009}, and most recently \textit{Kepler} \citep{Borucki2010} that provide high-precision time series photometry. For example, see \citet{Antoci2011}, who reported the first detections of solar-like oscillations in a $\delta$\,Sct star, and \citet{Murphy2013}, who presented a detailed asteroseismic study of an SX Phe star in a binary and also derived masses for both components. \citet{Grigahcene2010} and \citet[and references therein]{Uytterhoeven2011} found that $\gamma$\,Dor/$\delta$\,Sct hybrids are a much more common phenomenon than expected from ground-based observations. Nevertheless, extensive ground-based observing campaigns are still necessary. They can reveal phenomena which are hidden in white-light photometric data of space missions that are comparatively short-lived, such as variations occurring on very long time scales of years and decades, as well as differences between different passbands. High-precision spectroscopic observations allow us to study the features of stellar oscillations in line-profile variations (LPVs) and give important information on the abundances of stellar photospheres. In this paper we present the results of long-term spectroscopic observations of 4\,CVn, one of the best studied $\delta$\,Sct stars.

The star 4\,CVn \citep[$V=6.04$ mag, \Teff~$=6800\pm150~\mathrm{K}$, $\log g=3.34\pm0.20$, $\log L/\mathrm{L_\odot}=1.550\pm0.070$, $v_{eq}\sin i\geq120~\mathrm{km\,s}^{-1}$;][]{Lenz2010, Castanheira2008} has been observed photometrically for more than 40 years. The analysis of the extensive data set by \citet{Breger1999, Breger2000, Breger2008} led to the discovery of a complex pulsation pattern consisting of $18$ independent pulsation modes (Table~\ref{T_photo}). The main range of pulsation frequencies lies between 4 and 10 d$^{-1}$. Additional combination frequencies have been detected in lower and higher frequency ranges. Almost all modes show variations in amplitude and frequency that occur over several years and decades. Amplitude variations have been observed in many $\delta$~Sct stars. With sufficient frequency resolution, modes with varying amplitude and phase can be resolved as two close frequencies, which produce a beating pattern \citep{BregerBischof2002,BregerPamyatnykh2006}. Correlations between the amplitude and phase shifts can confirm a beating of two close, intrinsic modes as opposed to true amplitude variability of a single mode. For 4\,CVn the frequencies $6.1170~\mathrm{d}^{-1}$ and $6.1077~\mathrm{d}^{-1}$ are confirmed to be intrinsic modes producing amplitude variations and phase shifts through beating \citep{Breger2010}. However, for frequencies with amplitude and phase variability on time bases longer than one year, the coverage of the beating cycle is incomplete. It can thus not be ruled out that they are single modes with true amplitude variability.

\begin{table}
\caption{Photometric frequencies and amplitudes, and preliminary mode identification of 4\,CVn.}
\label{T_photo}	
\centering 
\begin{tabular}{l c c | c c} 
\hline\hline 
 \multicolumn{2}{c}{Frequency} & Amplitude & $\ell$ & $m$ \\	
 & & $y$ filter & & \\
 & d$^{-1}$ & mmag & &  \\
 \hline
$\nu_1$ & 8.595 & 15.3 & 1 & 1 \\
$\nu_2$ & 7.375 & 11.6 & 1 & -1 \\
$\nu_3$ & 5.048 & 10.7 & 1 & -1 \\
$\nu_4$ & 6.117 & 9.2 & & \\
$\nu_5$ & 5.851 & 10.1 & 2 & 1 \\
$\nu_6$ & 5.532 & 6.4 & 2 & \\
$\nu_7$ & 6.190 & 5.7 & & \\
$\nu_8$ & 6.976 & 5.0 & 0 & 0 \\
$\nu_9$ & 4.749 & 3.2 & & \\
$\nu_{10}$ & 7.552 & 3.3 & 1 & \\
$\nu_{11}$ & 6.750 & 0.9 & & \\
$\nu_{12}$ & 6.440 & 1.6 & & \\
$\nu_{13}$ & 5.986 & 0.8 & & \\
$\nu_{14}$ & 7.896 & 0.8 & & \\
$\nu_{15}$ & 5.134 & 0.8 & & \\
$\nu_{16}$ & 5.314 & 0.8 & & \\
$\nu_{17}$ & 6.404 & 0.4 & & \\
$\nu_{18}$ & 6.680 & 0.2 & & \\
\hline
\end{tabular}
\tablefoot{Frequencies and amplitudes taken from \citet{Breger1999}. The $\ell$ values are based on multicolor photometry by \citet{Lenz2010} and the $m$ values are the preliminary results of the spectroscopic mode identification of the 2008 data set by \citet{Castanheira2008}, ignoring the binarity of the pulsator.}
\end{table}

Mode identification can be achieved by interpretation of spectroscopic LPVs. The expansion and contraction of the star as well as displacements of stellar-surface elements, caused by the oscillations, lead to periodic Doppler shifts in the profiles of absorption lines. By comparing the observed properties of LPVs to theoretical predictions, the harmonic degree $\ell$ and the azimuthal order $m$ of the mode can be derived. Moreover, fitting with synthetic line profiles allows the determination of other stellar parameters and line-profile parameters, such as the projected rotational velocity $v_{eq}\sin i$, and the inclination $i$ (the angle between the pulsation axis and the line-of-sight; we assume that the pulsation axis coincides with the rotation axis). Unlike brightness or radial velocity, which are values integrated over the stellar surface, LPVs are less prone to partial cancellation effects and therefore allow for the detection of high-degree modes \citep[for details see][Chapter~6]{Aerts2010}.

Spectroscopic data has been gathered at the McDonald Observatory in Texas, USA from January 2008 until June 2011 (see Sect.~\ref{S_observations}). Preliminary results of the frequency analysis and mode identification of the season of 2008 were published by \citet{Castanheira2008}. They assigned $\ell$ and $m$ values to five different modes previously detected in photometry, summarized in Table~\ref{T_photo}. No high-degree modes were found. Furthermore, they measured $v_{eq}\sin i \geq120~\mathrm{km\,s}^{-1}$ for 4\,CVn.

A subsequent analysis by \citet{Breger2010} revealed correlations between the frequency variations and the azimuthal orders of the modes. While the prograde modes of 4\,CVn show an increasing frequency followed by a decline after the year 1991, its retrograde modes have a decrease in frequency with a rising value after 1991. Radial modes, on the other hand, show no or very little frequency variations. \citet{Breger2010} interpreted this contrary behavior of prograde and retrograde modes as a change in rotational splitting.

\begin{figure}
\includegraphics[width=88mm]{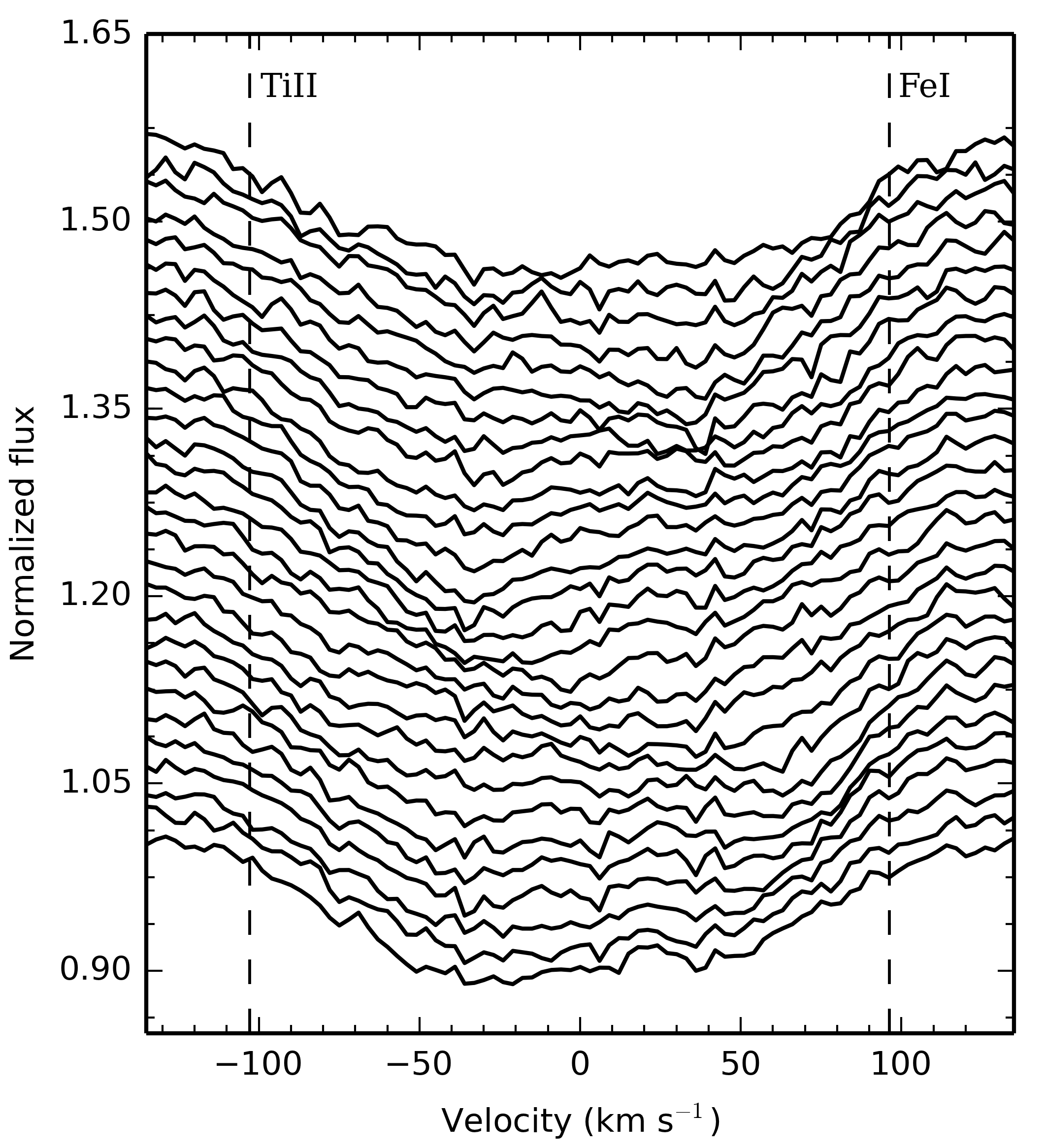}
\caption{Line-profile variations in the \ion{Fe}{II} line at $4508.288\mathrm{~\AA}$. The flux is normalized to 1 and is displayed with an offset. Shown are 29 spectra from the night of 31$^\mathrm{st}$ of March, 2010. The vertical, dashed lines mark the metal lines \ion{Ti}{II} at $4506.743~\mathrm{\AA}$ ($-102.81~\mathrm{km\,s}^{-1}$) and \ion{Fe}{II} at $4509.735~\mathrm{\AA}$ ($96.29~\mathrm{km\,s}^{-1}$), respectively.}
\label{F_lpv}
\end{figure}

For this paper we analyzed the complete data set of spectroscopic observations. First of all, we discovered the star to be a spectroscopic binary. After orbital subtraction, we performed mode identification by applying the Fourier-parameter fit method \citep{Zima2006} to the LPVs present in metal absorption lines in the 4\,CVn spectrum (see Fig.~\ref{F_lpv}). Additionally, we present an abundance analysis and confirm the solar-like metallicity of the star.

\section{Data and observations}
\label{S_observations}

The Sandiford \'Echelle Spectrograph \citep{McCarthy1993} is an instrument mounted on the 2.1-m Otto-Struve Cassegrain reflector telescope at the McDonald Observatory, Texas, USA. It has a resolving power of $R=60\,000$ and is therefore ideally suited to observe the relatively bright star 4\,CVn ($V=6.04\mathrm{~mag}$) with a high signal-to-noise ratio (S/N) and a high spectral resolution. Depending on the atmospheric conditions we chose an integration time of 600 to 900 s, leading to a mean S/N of $\sim200$ at a temporal resolution of $4\%$ and $10\%$ for the two main modes present in the photometric data, $\nu_1=8.594~\mathrm{d}^{-1}$ and $\nu_2=5.048~\mathrm{d}^{-1}$, respectively (see Table~\ref{T_photo}). The instrument provides a continuous wavelength coverage in the observed range of $4200\mathrm{~\AA}$ to $4700\mathrm{~\AA}$. In total, 2129 spectra were obtained throughout 86 nights in the period of January 2008 until June 2011 (see Table~\ref{T_obs}) leading to a time span of $\Delta T=1248\mathrm{~d}$ and thus to a frequency resolution of $1/\Delta T=0.0008\mathrm{~d}^{-1}$. For precise wavelength calibration, comparison frames of a Th-Ar lamp were obtained after each exposure.

\begin{table}
\caption{Observation log of obtained spectroscopy of 4\,CVn}
\label{T_obs}	
\centering 
\begin{tabular}{c c c c c} 
\hline\hline 
Season & Start date & End date & nights & N \\	
\hline 
2008 & 2008-Jan-19 & 2008-May-20 & 33 & 874 \\
2010 & 2010-Feb-18 & 2010-May-26 & 33 & 767 \\
2011 & 2010-Dec-14 & 2011-Jun-21 & 20 & 488 \\
\hline
Total & 2008-Jan-19 & 2011-Jun-21 & 86 & 2129 \\
\hline
\end{tabular} 
\tablefoot{Nights denotes the number of nights observed in each specific year; N is the number of scientific spectra taken of 4\,CVn. Not all spectra were used in the analysis due to cosmic-ray hits or very low S/N.}
\end{table}

Standard IRAF routines for the reduction of \'echelle spectra were used to carry out the data reduction and wavelength calibration. Subsequently, the spectra were normalized in a similar manner as described by \citet{Papics2012}, i.e., by fitting a cubic spline to manually chosen continuum points. They were then cut around a suitable line for further analysis. This line, an absorption feature of \ion{Fe}{II} at $4508.288\mathrm{~\AA}$, was carefully chosen from the available spectral range, as explained below.

\subsection{Line selection}
\label{S_lineselection}

Certain criteria have to be fulfilled for the selection of an absorption line for the analysis of the LPVs, such as a sufficient depth and sharpness. For sharp lines, the pulsational broadening is dominant over other line-broadening mechanisms across the whole profile. It is also important that the line is unblended and does not overlap with any other atomic line. Metal lines are therefore better suited for mode identification from LPVs than hydrogen or helium lines \citep[for more details see][Chapter~6]{Aerts2010}. Different spectral lines originate from different layers in the atmosphere and thus probe different pulsation modes. Therefore, the signatures of the oscillations might be smeared out in the line profiles of blended lines. Moreover, the displacement of the line-forming regions by the pulsations leads to slight variations of the local temperature, which in turn lead to equivalent-width ($EW$) variations. This will affect the outcome of the mode identification and has to be taken into account.

\citet{Castanheira2008} found a projected rotational velocity of $v_{eq}\sin i\geq120\mathrm{~km\,s}^{-1}$ for 4\,CVn, which means that the star is a moderately fast rotator. This aggravates the situation of dense blue spectra of F stars with only very small continuum regions, as each absorption line is rotationally broadened. To select suitable lines for our study, we calculated an atmospheric model with effective temperature \Teff~$=6800$~K, surface gravity $\log g=3.32$, and solar metallicity \citep{BregerPamy2002,Lenz2010}, using the \textsc{LLmodel} code \citep{Shulyak2004}. With the \textsc{SynthV} \citep{Tsymbal1996} code we computed a synthetic spectrum and checked each line separately. The \ion{Fe}{II} line around $4508.288\mathrm{~\AA}$ is the only line which is unblended except for two contributions in the line wings of \ion{Ti}{II} at $4506.743\mathrm{~\AA}$ ($27.2\%$ of the line depth of \ion{Fe}{II}) and \ion{Fe}{I} at $4509.735\mathrm{~\AA}$ ($39.8\%$). The position of the two blending features are marked by vertical, dashed lines in Fig.~\ref{F_lpv}.

Additionally, \citet{Castanheira2008} used another \ion{Fe}{II} line around $4549.474\mathrm{~\AA}$ for mode identification of the 2008 data set of 4\,CVn in their analysis. We refrained from using this line, since it turned out to be heavily blended by at least five other absorption features, of which \ion{Fe}{I} at $4549.466\mathrm{~\AA}$ and \ion{Ti}{II} at $4549.617\mathrm{~\AA}$ give the strongest contribution to this line blend.

\section{Binarity and radial velocity variations}\label{S_binary}

After the spectra were corrected for the barycentric velocity shift, a dominant non-sinusoidal variation was visible in the first moment of the line, which represents the radial velocity \citep[for a definition, see][]{Aerts1992}. This periodicity can be explained by a binary component and the movement of the primary around the center of mass of the system. No spectral features of the secondary could be detected in the data. This is further discussed in Sect.~\ref{S_specana}.

\begin{figure}
\includegraphics[width=88mm]{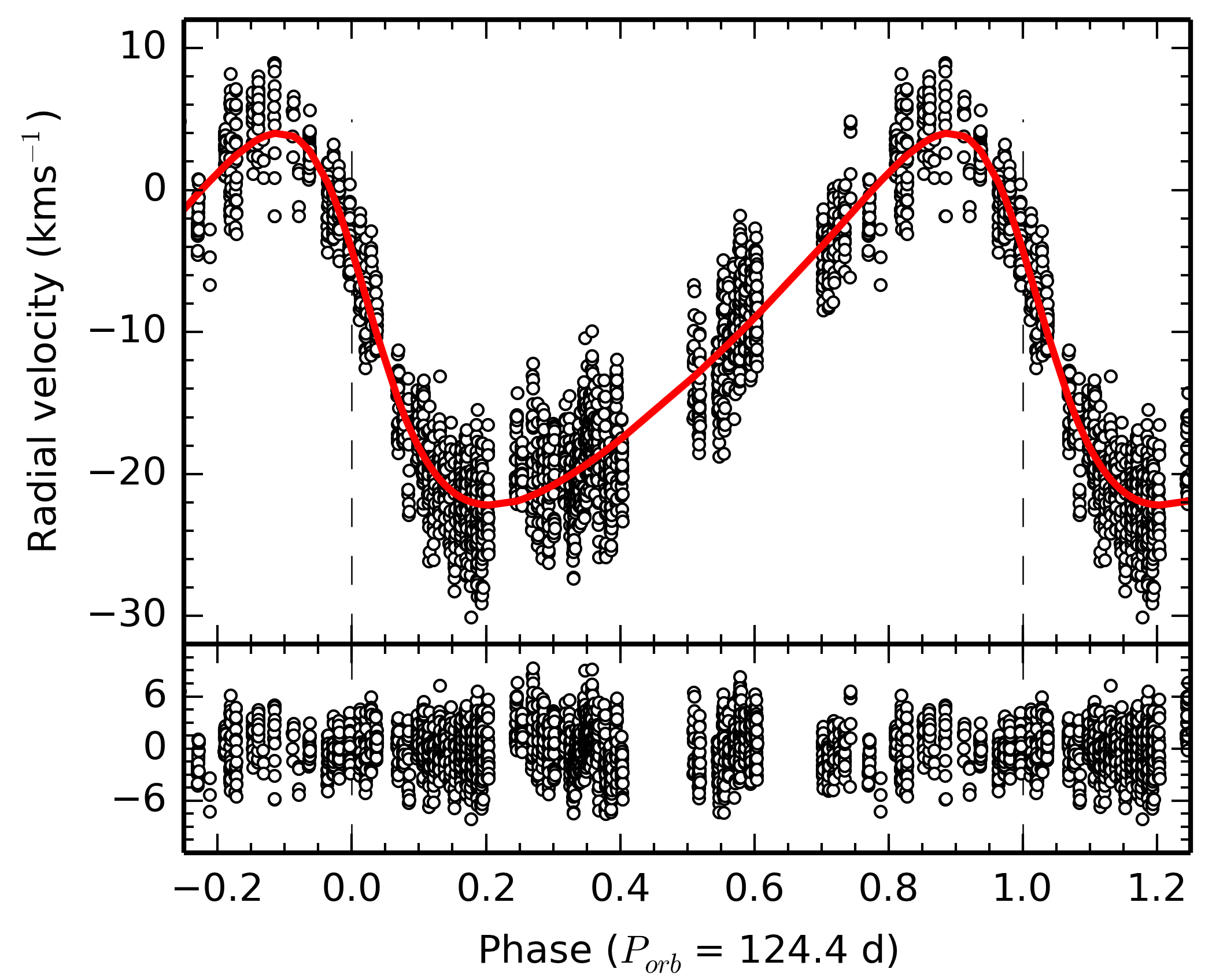}
\caption{Upper Panel: The observed radial velocities (open circles) fitted with the orbital model (solid line) with the parameters displayed in Table~\ref{T_rvorb} and phased with the orbital period $P_{orb}=124.4~$d. Lower Panel: Residuals after subtracting the binary model with an RMS of $2.8~\mathrm{km\,s}^{-1}$. The pulsation signal (see Table~\ref{T_spectro}) is visible on top of the variation due to binarity and remains present in the residuals. The vertical, dashed lines mark the beginning and the end of one phase.}
\label{F_binary}
\end{figure}

Since the first moment of the line depends somewhat on the wavelength range within which it is calculated, we created five different data sets with different dispersion ranges. We then fitted a Keplerian orbit to each data set. The best solution and errors of the parameters were estimated with a Markov Chain Monte Carlo (MCMC) method implemented in the package \texttt{emcee} \citep{Foreman-Mackey2013}. To account for the scatter due to pulsations, we assumed an uncertainty of $\sigma_\mathrm{RV}=2.8~\mathrm{km\,s}^{-1}$ (the RMS value of the residuals after subtraction of the binary model, see Table~\ref{T_rvorb}) for each radial velocity measurement. To account for the dependence of the radial velocities on the dispersion range, we fitted all five data sets combined. All five models are consistent within the uncertainties of each solution. Figure~\ref{F_binary} shows the orbital model fitted to the radial velocities obtained from the dispersion range $-128.0~\mathrm{km\,s}^{-1}$ to $106.0~\mathrm{km\,s}^{-1}$ and the residuals after subtraction of the mean orbit. The results are shown in Table~\ref{T_rvorb}.

Due to the binary motion of the pulsating star we expect the frequencies to show a periodic shift at the base of the orbital period \citep[for an explanation of the Doppler shift of oscillation frequencies, see][]{Shibahashi2012}. For the observed photometric frequencies a shift of $\sim0.0002~\mathrm{d}^{-1}$ to $\sim0.0004~\mathrm{d}^{-1}$ can be estimated. Since the analyzed data set has a frequency resolution of $0.0008~\mathrm{d}^{-1}$, we cannot resolve the expected frequency shifts due to the binary motion. We also did not detect a variation based on the orbital period of the times of maxima in an $(O-C)$ diagram of the first moment.

From the mean binary orbit, a correction value was calculated for each spectrum to shift the data to the reference frame of the primary. The frequency analysis and mode identification of the pulsations was subsequently performed on the data corrected for the binary motion.

\begin{table}
\caption{Orbital Parameters of 4\,CVn.}
\label{T_rvorb}
\centering
\begin{tabular}{l | r@{$\pm$}l }
\hline\hline
$P_{orb}$ (days) & 124.44 & 0.03 \\
$T$ (JD) & 2\,454\,605 & 10.3 \\
$e$  & 0.311 & 0.003 \\
$\omega$ (deg) & 70.2 & 0.7 \\
$K_1$ ($\mathrm{km\,s}^{-1}$) & 13.24 & 0.05 \\
$\gamma$ ($\mathrm{km\,s}^{-1}$) & -10.44 & 0.03 \\
\hline
RMS ($\mathrm{km\,s}^{-1}$) & \multicolumn{2}{c}{2.8} \\
\hline
\end{tabular}
\end{table}

\section{Revision of fundamental parameters}
\label{S_specana}

Since the data gathered at McDonald Observatory are limited in wavelength range, we obtained additional observations with the HERMES spectrograph \citep{Raskin2011}, a high-resolution spectrograph ($R = 85\,000$) mounted at the 1.2-m Mercator telescope at La Palma, Spain. We obtained three spectra equally spread over the night of the 3$^\mathrm{rd}$ of February, 2013. The reduction was performed by the HERMES pipeline and normalization was done in the same way as for the McDonald spectra, i.e., by fitting a cubic spline to the continuum region \citep{Papics2012}. We extracted a wavelength range of $4500\mathrm{~\AA}$ to $6000\mathrm{~\AA}$ from the data.

For the analysis of the average spectrum of 4\,CVn, we used the GSSP program package \citep{Tkachenko2012, Lehmann2011}. The code relies on a comparison between observed and synthetic spectra computed in a grid of $T_\mathrm{eff}$, $\log g$, micorturbulence $\xi$, metallicity [m/H], and $v_{eq} \sin i$, and finds the optimum values of these parameters from a minimum in $\chi^2$. Individual abundances of chemical elements are adjusted in the second step, by fixing the metallicity to the above determined value but keeping the other atmospheric parameters free. The errors of measurement (1-$\sigma$ confidence level) are calculated from the $\chi^2$ statistics using the projections of the hypersurface of the $\chi^2$ from all grid points of all parameters onto the parameter in question. For a more detailed discussion on the estimation of the uncertainties, see \citet{Tkachenko2013}. Strong LPVs detected in the spectrum of 4\,CVn, as well as possible contributions from the secondary component, are an additional source of uncertainties which are not taken into account in the spectral analysis.
Our analysis is based on the grid of atmosphere models computed using the most recent version of the \textsc{LLmodels} code \citep{Shulyak2004}. For the calculation of the synthetic spectra, we used the \textsc{SynthV} code \citep{Tsymbal1996}, and information on atomic lines has been extracted from the Vienna Atomic Line Database \citep[VALD,][]{Kupka2000}.

\begin{table}
\caption{Fundamental and atmospheric parameters of 4\,CVn.}
\label{T_fundpars}
\centering 
\tabcolsep=12pt
\begin{tabular}{l | r@{$\,\pm\,$}l} 
\hline\hline 
\Teff (K) & 6875 & 120\tablefootmark{a} \\
$\log g$ (dex) & 3.30 & 0.35\tablefootmark{a} \\
$[\mathrm{m/H}]$ (dex) & -0.05 & 0.15\tablefootmark{a} \\
$v_{eq} \sin i$ (km\,s$^{-1}$) & 109 & 3\tablefootmark{a} \\
$\xi$ (km\,s$^{-1}$) & 4.00 & 0.45\tablefootmark{a} \\
ST & \multicolumn{2}{c}{F2 III-IV\tablefootmark{a}} \\
\hline
$\pi$ (mas) & 10.51 & 0.40 \\
$\log L/\mathrm{L_\odot}$ & 1.47 & 0.05\tablefootmark{b} \\
Radius (R$_\odot$) & \multicolumn{2}{c}{3.7--4.1\tablefootmark{b}} \\
Mass (M$_\odot$) & \multicolumn{2}{c}{1.0--2.0\tablefootmark{b}} \\
\hline
\end{tabular}
\tablefoot{\tablefoottext{a}{Parameters derived via spectral analysis.}\\
\tablefoottext{b}{Parameters based on combined photometric and spectroscopic observations.}}
\end{table}

\begin{table}
\caption{Abundances of individual chemical elements of 4\,CVn.}
\label{T_abund}
\centering
\begin{tabular}{c c c | c c c}
\hline\hline
Element & Value & Sun & Element & Value & Sun \\
 & dex & dex & & dex & dex \\
 \hline
 Fe & -4.69(15) & -4.59 & Cr & -6.47(20) & -6.40 \\
 Mg & -4.41(25) & -4.51 & Mn & -6.65(40) & -6.65 \\
 Ti & -7.24(30) & -7.14 & Si & -4.41(40) & -4.53 \\
 Ni & -5.82(20) & -5.73 & C & -3.64(40) & -3.65 \\
 Ca & -5.84(25) & -5.73 & Ba & -9.52(50) & -9.87 \\
  & & & Sc & -8.57(50) & -8.99 \\
\hline
\end{tabular}
\tablefoot{Error bars are given in parenthesis in terms of last digits. Solar values are those derived by \citet{Grevesse2007}.}
\end{table}

\begin{figure*}
\includegraphics[width=180mm]{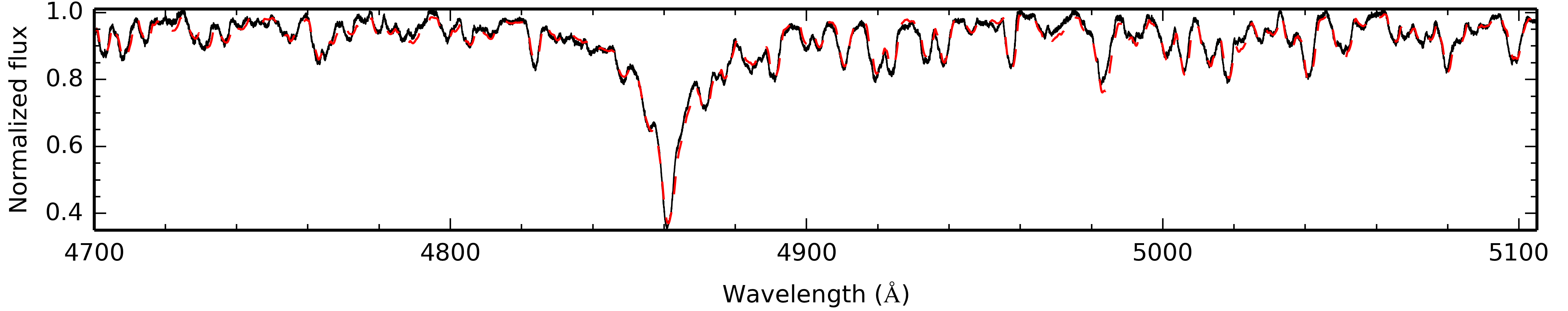}
\caption{Fit of the observed spectrum (solid) with a synthetic spectrum (dashed) computed using our optimized parameters (see Table~\ref{T_fundpars}).}
\label{F_spectralfit}
\end{figure*}

Table~\ref{T_fundpars} lists the atmospheric parameters of 4\,CVn we derived. The individual abundances of some chemical elements are given in Table~\ref{T_abund}. The spectral type and the luminosity class have been computed using an interpolation in the tables published by Schmidt-Kaler (1982). Figure~\ref{F_spectralfit} compares the observations with the best fit synthetic spectrum in a $300~\mathrm{\AA}$ wide wavelength range including the H$_\beta$ spectral line.

Moreover, we used the Hipparcos parallax of 10.51 mas \citep{VanLeeuwen2007} in combination with 2MASS and Geneva photometry to compute the luminosity and the radius of the star. We used a grid-based method, which also includes interstellar reddening as a free parameter, as described by \citet{Degroote2011}. By using the radius and the spectroscopic $\log g$ we are able to estimate the mass. We find an extinction $E(B-V)=0.02$~mag. Despite this value being close to zero, adding extinction to the calculations is important for the consistency of our estimated parameters, mass and radius. The results of the photometric analysis are summarised in Table~\ref{T_fundpars}.

\begin{figure}
\includegraphics[width=88mm]{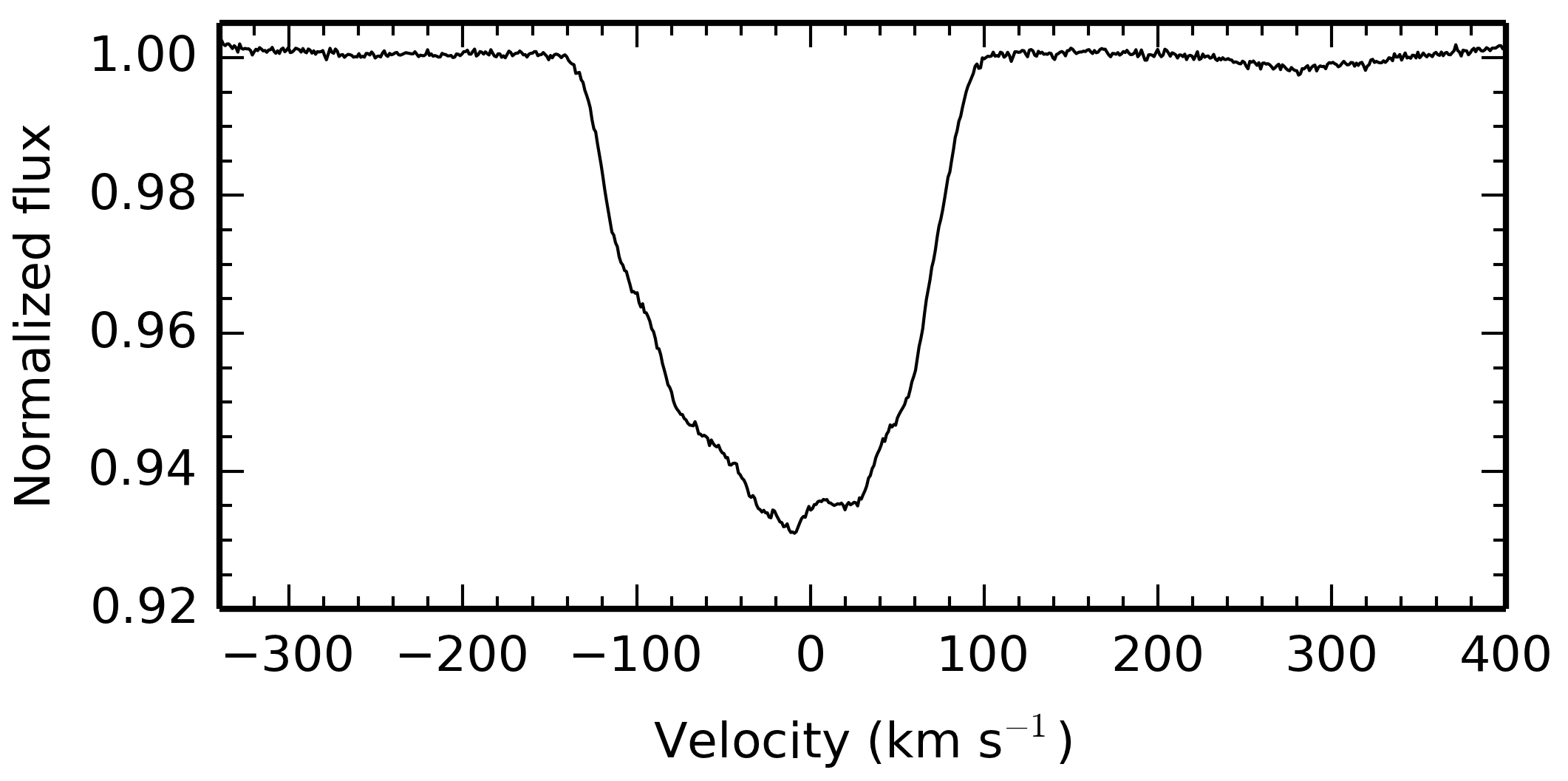}
\caption{Average line profile of 4\,CVn computed by means of the LSD technique.}
\label{F_LSD}
\end{figure}

We also used the above analysed observed spectrum to compute a mean profile with a high S/N by means of the Least-Squares Deconvolution technique (LSD) \citep{Donati1997, Kochukhov2010}. Figure~\ref{F_LSD} shows the LSD profile of 4\,CVn, the line mask was pre-computed based on the parameters listed in Table~\ref{T_fundpars}. Besides clear ``bumps'' in the center of the profile (dip) of the star, there is also an indication of another broad spectral line at $RV\sim280~\mathrm{km\,s}^{-1}$. Whereas the former is connected to the intrinsic variability of the star in terms of (high-degree) non-radial pulsations, the latter might be a signature of a (very) faint stellar companion. 

Given the measured semi-amplitude of the primary $K_1=13.24~\mathrm{km\,s}^{-1}$ and the measured systemic velocity $\gamma=-10.44~\mathrm{km\,s}^{-1}$, a radial velocity of $\sim~280~\mathrm{km\,s}^{-1}$ of the secondary at orbital phase $\phi=0.64$ would lead to a mass ratio of $q\leq0.05$. The line strength of the spectral feature around $RV\sim280~\mathrm{km\,s}^{-1}$ in the LSD profile is $\sim0.2\%$. If we assume a minimum contribution of $\sim0.2\%$ of the secondary to the system luminosity $\log L/\mathrm{L_\odot}=1.47$, we estimate $L_2\sim0.06~\mathrm{L_\odot}$, which corresponds to an early M-type main-sequence star of $\sim0.6~\mathrm{M_\odot}$ \citep[Appendix G]{CarrollOstlie2006}. A mass ratio of $q\leq0.05$ would then give a primary mass of $\sim12~\mathrm{M_\odot}$, which is too massive given the observed spectral type F2III-IV.

The additional spectral contribution is so faint that it could also be an effect of continuum normalization. As the spectral lines of 4\,CVn are broadened by fast rotation ($v_{eq}\sin i = 109~\mathrm{km\,s}^{-1}$), a continuum is not present everywhere in the wavelength range. Thus, inaccurate continuum normalization is a more likely explanation for the spectral feature at $RV\sim280~\mathrm{km\,s}^{-1}$ than a secondary signal.

\section{Frequency analysis}\label{S_freqana}

Before starting the analysis of the data we inspected each spectrum by eye. We rejected 93 observations, due to weak cosmic-ray hits within the line or poor S/N, leaving 2036 (797 in 2008, 758 in 2010, and 481 in 2011) good spectra. Since the periodogram of 4\,CVn is characterized by amplitude variations on short and long timescales, we split the whole data set in three parts, studying each season separately. Otherwise the variability of the oscillation pattern would lead to disturbing side peaks in the Fourier spectra and to uncertainties between real frequency peaks and aliasing. 

To search for significant periodicities in the LPVs we employed the Fourier-parameter fit method (FPF), which was developed by \citet{Zima2006} as an advancement of the pixel-by-pixel method \citep{Mantegazza2000}. It is implemented in the software package FAMIAS \citep{Zima2008}. For each bin in the dispersion range of the line a Fourier transform is performed and a Lomb-Scargle periodogram \citep{Lomb1976,Scargle1982} is calculated. An optimisation of the mode parameters is done by applying a least-squares fitting algorithm to the original spectra, again, for each bin separately and thereby computing zero point, amplitude, and phase across the line profile. Subsequently, these parameters are used for mode identification (see Sect.~\ref{S_modeid}). A Monte-Carlo perturbation approach has been used to calculate the errors on the amplitudes, assuming a Gaussian distribution of the noise of the amplitude profiles. The S/N of each frequency is calculated from the mean periodogram, prewhitened with the significant frequency peaks. Thereby, the noise ($\sigma_{\mathrm{res}}$) is calculated as the mean amplitude in a range of $10~\mathrm{d}^{-1}$ (covering most of the frequency spectrum) around the frequency peak in the periodogram. We have adopted a significance criterion of $A>4\sigma_{\mathrm{res}}$ \citep{Breger1993}. The results of our analysis are summarized in Table~\ref{T_spectro}.

\begin{table}
\caption{Spectroscopic frequencies and amplitudes of 4\,CVn.}
\label{T_spectro}	
\centering
\begin{tabular}{l r r r r r} 
\hline\hline 
 \multicolumn{2}{c}{Frequency} & \multicolumn{1}{c}{$A_\mathrm{whole~set}$} & \multicolumn{1}{c}{$A_{2008}$} & \multicolumn{1}{c}{$A_{2010}$} & \multicolumn{1}{c}{$A_{2011}$} \\	
 & d$^{-1}$ & km\,s$^{-1}$ & km\,s$^{-1}$ & km\,s$^{-1}$ & km\,s$^{-1}$  \\
 & $\pm0.0008$ & $\pm0.009$ & $\pm0.015$ & $\pm0.014$ & $\pm0.016$ \\
\hline 
$f_1$ & 7.3764 & 0.953 & 0.95 & 0.98 & 1.00 \\
$f_2$ & 5.8496 & 0.869 & 0.87 & 0.90 & 0.88 \\
$f_3$ & 5.0481 & 0.550 & 0.48 & 0.53 & 0.59 \\
$f_4$ & 8.5942 & 0.444 & 0.48 & 0.51 & 0.33 \\
$f_5$ & 5.5315 & 0.389 & 0.38 & 0.39 & 0.44 \\
$f_6$ & 8.6552 & 0.389 & 0.31 & 0.40 & 0.51 \\
$f_7$ & 8.1687\tablefootmark{a} & 0.379 & 0.33 & 0.41 & 0.48 \\
\hline
$f_8$ & 6.6801 & \multicolumn{4}{l}{whole set, 2010, 2011} \\
$f_9$ & 4.0743\tablefootmark{a} & \multicolumn{4}{l}{whole set, 2010, 2011\tablefootmark{c}} \\
$f_{10}$ & 6.1171 & \multicolumn{4}{l}{whole set, 2010} \\
$f_{11}$ & 6.975\tablefootmark{b} & \multicolumn{4}{l}{2008} \\
$f_{12}$ & 10.1702\tablefootmark{a} & \multicolumn{4}{l}{whole set, 2010, 2011} \\ 
$f_{13}$ & 6.1910 & \multicolumn{4}{l}{whole set, 2011} \\
$f_{14}$ & 12.4244\tablefootmark{a} & \multicolumn{4}{l}{whole set, 2010, 2011} \\
$f_{15}$ & 9.4113\tablefootmark{a} & \multicolumn{4}{l}{whole set, 2008, 2010, 2011} \\
$f_{16}$ & 9.7684\tablefootmark{a} & \multicolumn{4}{l}{whole set, 2008, 2010} \\
$f_{17}$ & 10.0372\tablefootmark{a} & \multicolumn{4}{l}{whole set, 2008} \\
$f_{18}$ & 6.4030 & \multicolumn{4}{l}{whole set} \\
$f_{19}$ & 13.388\tablefootmark{a} & \multicolumn{4}{l}{2008} \\
$f_{20}$ & 10.016\tablefootmark{a} & \multicolumn{4}{l}{2011} \\
\hline
\end{tabular}
\tablefoot{Amplitudes $A_\mathrm{whole~set}$ from the frequency analysis of the whole data set compared to the amplitudes of the different seasons $A_{2008}$, $A_{2010}$, and $A_{2011}$. The amplitudes have an average error of $0.009~\mathrm{km\,s}^{-1}$ for the whole data set, and of $0.015, 0.014, and 0.016~\mathrm{km\,s}^{-1}$ for the seasons 2008, 2010, and 2011, respectively. The frequency resolution $1/\Delta T$ is $0.0008~\mathrm{d}^{-1}$ for the whole data set and $0.009\text{, }0.01\text{, and }0.005~\mathrm{d}^{-1}$ for the seasons 2008, 2010, and 2011, respectively.\\
\tablefoottext{a}{Frequencies not detected in previous photometric studies.}\\
\tablefoottext{b}{Photometric radial mode \citep{Lenz2010}.}\\
\tablefoottext{c}{Alias peak of frequency detected.}} 
\end{table}

\subsection{The \ion{Fe}{II} line at $4508.288~\mathrm{\AA}$}

\begin{figure}
\includegraphics[width=88mm]{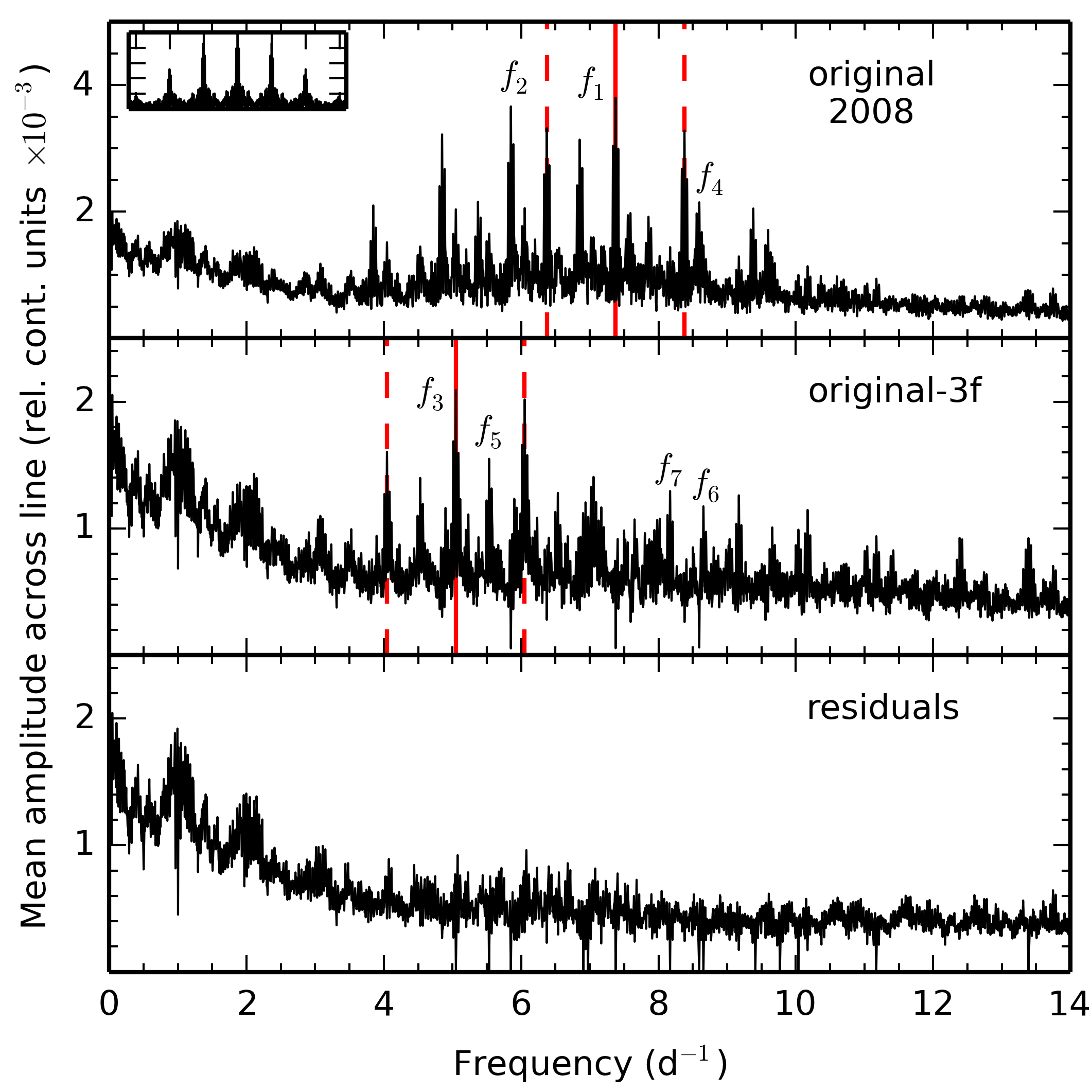}
\caption{Mean of the Lomb-Scargle periodograms per bin of the season of 2008. From top to bottom: mean Fourier spectra of the original data set, and after prewhitening of 3 and 14 frequencies, respectively. The red, solid lines mark the highest peak in each periodogram, while red, dashed lines mark the respective one-day aliases. Highest peaks are $f_1=7.3764~\mathrm{d}^{-1}$ (upper panel) and $f_3=5.0481~\mathrm{d}^{-1}$ (center panel). The inset in the upper panel displays the spectral window for the 2008 data set. }
\label{F_pergram2008}
\end{figure}

\begin{figure}
\includegraphics[width=88mm]{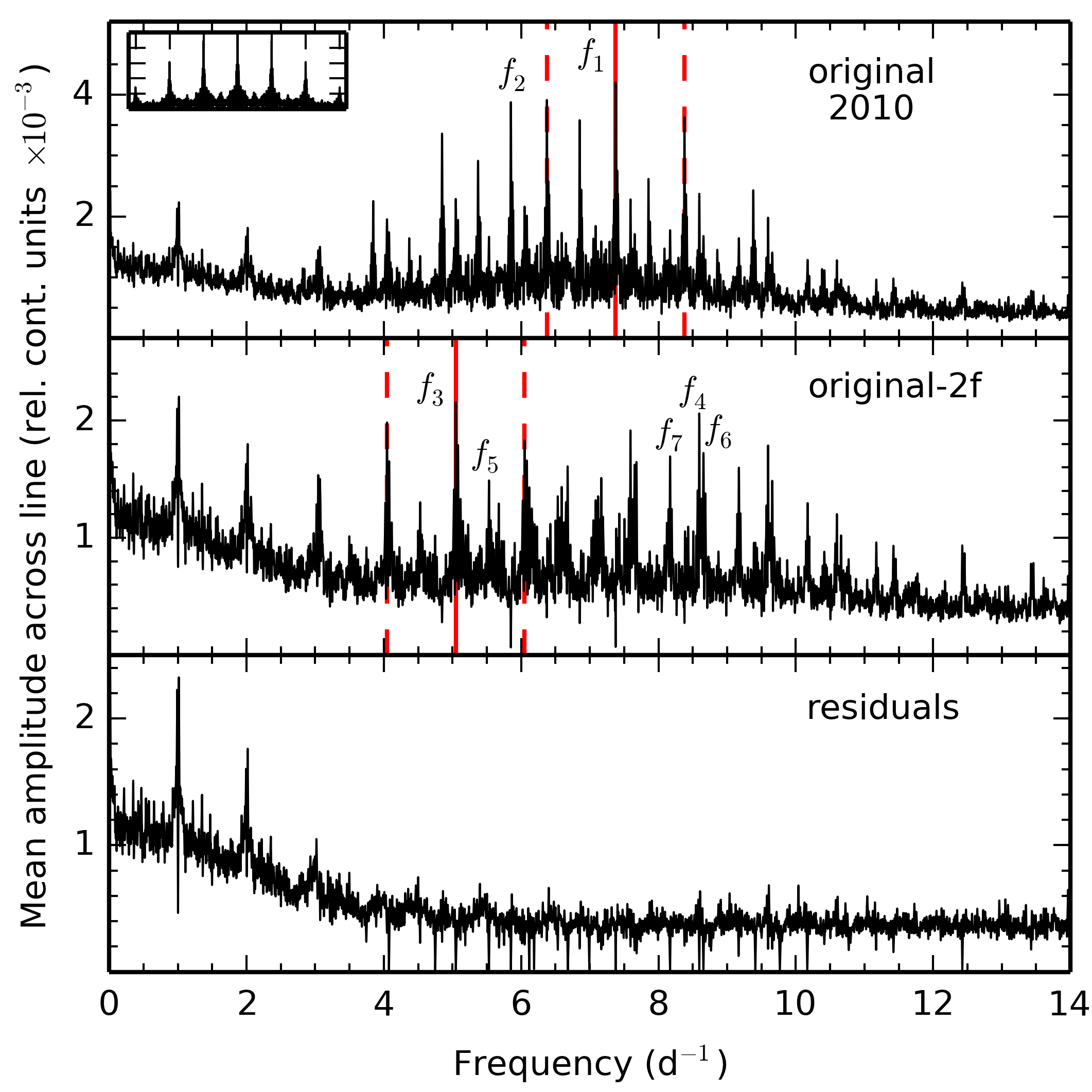}
\caption{Same as Fig.~\ref{F_pergram2008} for the season of 2010. From top to bottom: mean Fourier spectra of the original data set, and after prewhitening of 2 and 17 frequencies, respectively. The highest peaks marked by the red, solid lines are $f_1=7.3764~\mathrm{d}^{-1}$ (upper panel) and $f_3=5.0481~\mathrm{d}^{-1}$ (center panel).}
\label{F_pergram2010}
\end{figure}

\begin{figure}
\includegraphics[width=88mm]{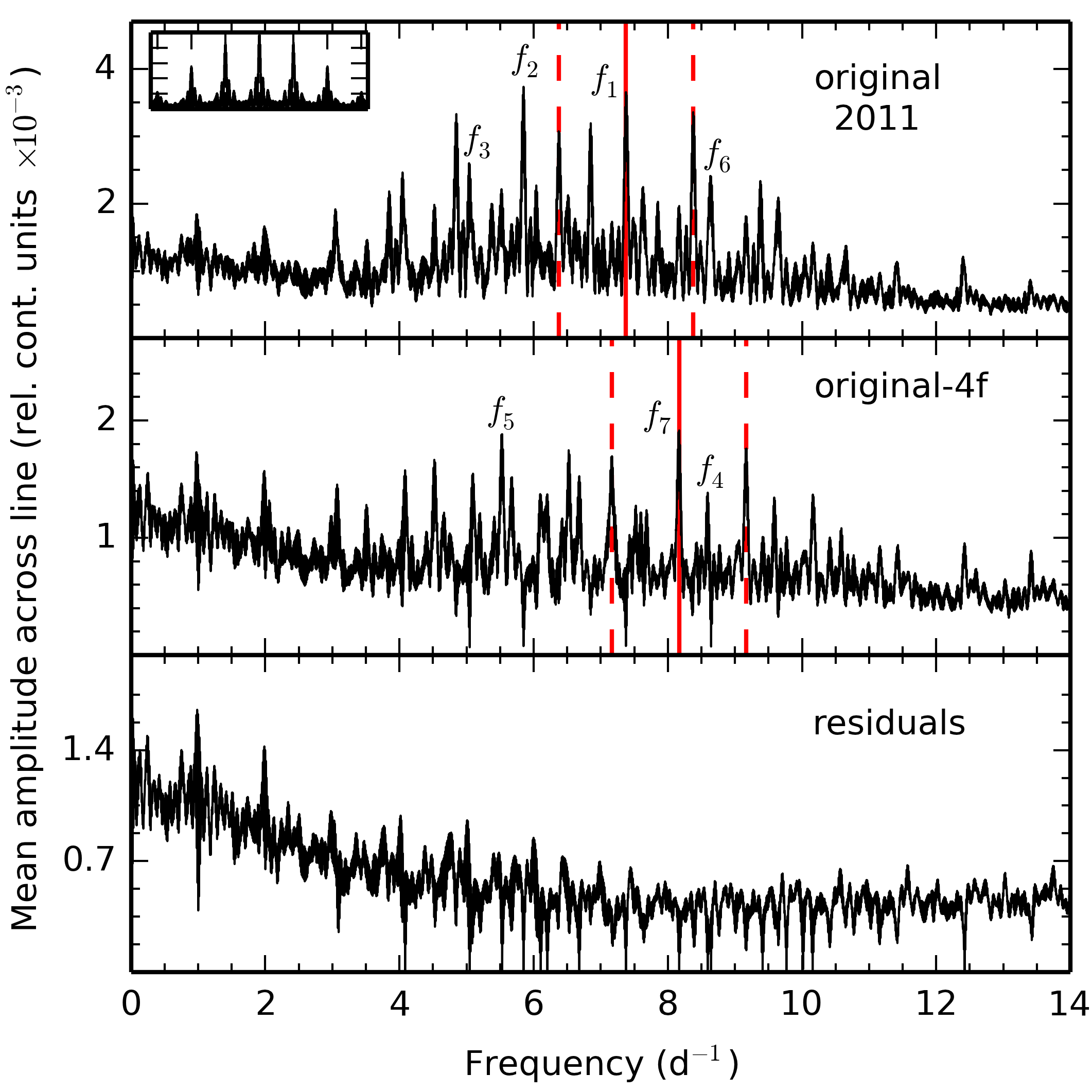}
\caption{Same as Figs.~\ref{F_pergram2008} and \ref{F_pergram2010} for the season of 2011. From top to bottom: Fourier spectra of the original data set, and after prewhitening of 4 and 16 frequencies, respectively. The highest peaks marked by the red, solid lines are $f_1=7.3764~\mathrm{d}^{-1}$ (upper panel) and $f_7=8.1687~\mathrm{d}^{-1}$ (center panel). }
\label{F_pergram2011}
\end{figure}

The increase in noise towards low frequencies in the periodograms (see Figs.~\ref{F_pergram2008}, \ref{F_pergram2010}, and \ref{F_pergram2011}) can be explained by instrumental effects or imperfect subtraction of the binary orbit. Given that the spectrograph is mounted at the telescope and not placed in a remote room where pressure and temperature could be kept constant to avoid spectral drift, some long-term variations are unavoidably present in the data. The high noise level can also be caused by instrumental noise and correlations of different noise sources. Residual variability on the basis of the binary orbit would cause frequency peaks around $0.008~\mathrm{d}^{-1}$. For those reasons, we discarded all low-frequency peaks below $2~\mathrm{d}^{-1}$ as being not trustworthy. Furthermore, the daily gaps due to the night-day rhythm led to significant one-day aliasing, which is visible in the spectral window of each season (see inset in the top panel of Figs.~\ref{F_pergram2008}, \ref{F_pergram2010}, and \ref{F_pergram2011} for season 2008, 2010, and 2011, respectively). To overcome these obstacles a prewhitening of each frequency was performed, before carrying out another Fourier transform.

Figures \ref{F_pergram2008}, \ref{F_pergram2010}, and \ref{F_pergram2011} display the mean of the periodograms that were computed for each bin and for the seasons 2008, 2010, and 2011, respectively. Different stages of prewhitening are shown and the 7 dominant frequencies are marked. We found 6 frequencies that are confirmed by photometric studies and are present in the periodograms of the whole data set and in all three seasons separately as well. These are $f_1=7.3764~\mathrm{d}^{-1}$, $f_2=5.8496~\mathrm{d}^{-1}$, $f_3=5.0481~\mathrm{d}^{-1}$, $f_4=8.5942~\mathrm{d}^{-1}$, $f_5=5.5315~\mathrm{d}^{-1}$, and $f_6=8.6552~\mathrm{d}^{-1}$ (Breger, Lenz \& Pamyatnykh, in prep.). Other photometrically significant frequencies could be detected in some seasons, while their amplitude was below the detection limit in other subsets. The frequency $f_{11}=6.975~\mathrm{d}^{-1}$, for example, could only be detected in 2008. Breger, Lenz \& Pamyatnykh (in prep.) confirm a rapid, steady decrease in amplitude for this frequency from 2008 to 2012. Also $f_8=6.6801~\mathrm{d}^{-1}$, $f_{10}=6.1171~\mathrm{d}^{-1}$, and $f_{13}=6.1910~\mathrm{d}^{-1}$ appear to be varying in amplitude, since they have only been detected in either the season of 2010 or 2011, or in both. However, the amplitude variability we detected for $f_8$, $f_{10}$, and $f_{13}$ does not agree with the amplitudes derived from the photometric data set. Since these frequencies have a low S/N, their amplitude could also be below the detection limit in other seasons.

Additionally, we found several frequencies, which were not detected in previous photometric studies by \citet{Breger1999, Breger2000, Breger2008}, among them $f_7=8.1688~\mathrm{d}^{-1}$, $f_9=4.0743~\mathrm{d}^{-1}$, $f_{15}=9.4113~\mathrm{d}^{-1}$, and $f_{16}=9.7684~\mathrm{d}^{-1}$. The frequency $f_7$ can be firmly detected with a comparatively high amplitude ($A>0.3~\mathrm{km\,s}^{-1}$) in all sets and subsets. Frequencies $f_9$, $f_{15}$, and $f_{16}$ have much lower amplitudes, and the amplitudes of $f_9$ and $f_{16}$ even lie below the detection limit of seasons 2008 and 2011, respectively. The consistency of the detection in the separate seasons and the sufficient S/N for these frequencies in the merged data of the three seasons led us to the conclusion that they correspond to real oscillation frequencies. The fact that these frequencies were not detected in photometry suggests that they are either high-degree modes with $\ell\ge2$ or are varying in amplitude and reached detectable amplitudes in the last five years.

\citet{Breger1999} could identify peaks above 10~d$^{-1}$ as combination frequencies of the main modes between 4 and 10~d$^{-1}$. The frequencies $f_{12}=10.1702~\mathrm{d}^{-1}$, $f_{14}=12.4244~\mathrm{d}^{-1}$, $f_{17}=10.0372~\mathrm{d}^{-1}$, $f_{19}=13.388~\mathrm{d}^{-1}$, and $f_{20}=10.016~\mathrm{d}^{-1}$ lie in this high frequency range, but could not be confirmed to be combinations of any of the detected frequency peaks with $4\leq f\leq10~\mathrm{d}^{-1}$. The amplitude and the phase profiles across the line of a mode offer an additional diagnostic to distinguish between noise peaks and real frequencies. If those profiles show a significant trend and not just random scatter, the frequency is likely to be a real pulsation mode. In Appendix~\ref{A_amp.phiProfiles} the amplitude and phase profiles across the line are displayed for the frequencies $f_8$ to $f_{20}$. While the amplitude profiles of $f_{12}$, $f_{14}$, $f_{17}$, $f_{19}$, and $f_{20}$ are rather noisy, the phase profiles of these peaks do resemble those of real mode frequencies, i.e., increasing or decreasing from the blue to the red edge of the line. They were thus added to Table~\ref{T_spectro}.

In order to test the influence of the line-cutting limits, we did a frequency analysis using different dispersion ranges for the \ion{Fe}{II} line 4508.288~$\mathrm{\AA}$, as explained in Sect.~\ref{S_binary}. All different values of frequency and amplitude for all five dispersion ranges are consistent within the 1$\sigma$ error bars.

\subsection{Analysis of LSD profiles}
\label{S_LSDmcd}

Due to the high rotation rate of the star, we only found one absorption line to be suitable for performing mode identification. In order to test the potential of using more information from the other spectral lines, we calculated LSD profiles for the McDonald spectra obtained in 2010 with the method by \citet{Kochukhov2010, Tkachenko2013b}. We made use of the line mask, which was computed for Sect.~\ref{S_specana}, to calculate a standard LSD profile from all the lines in the line mask, as well as an LSD profile of the Fe lines only, using the multi-profile technique, in order to take the blending of Fe-lines into account. Figures~\ref{F_LPVimage_single}, \ref{F_LPVimage_LSDnormal}, and \ref{F_LPVimage_LSDFe} in Appendix~\ref{A_LSDcomp} show the mean profile, the standard deviation, and a 2D color image of the LPVs, phase folded onto the frequency $f_1=7.3764~\mathrm{d}^{-1}$, for the \ion{Fe}{II} line at $4508.288~\mathrm{\AA}$, the standard LSD profiles, and the Fe-LSD profiles, respectively.

The shape and depth of an LSD profile depends on the line mask which is used for the computation. As it can be considered as an average of several lines (weak and strong lines of different elements), the depth will typically be less than that of a strong single line, such as \ion{Fe}{II} at 4508.288~$\mathrm{\AA}$. Also the overall shape of the LSD profile can differ from that of a single line. All these effects can be seen in the Figures in Appendix~\ref{A_LSDcomp}. Note that the amplitude scales in Fig.~\ref{F_LPVimage_single} (single \ion{Fe}{II} line) are different from the amplitude scales in Fig.~\ref{F_LPVimage_LSDnormal} (standard LSD profiles) and Fig.~\ref{F_LPVimage_LSDFe} (Fe-LSD profiles). The uneven structure in the continuum region in the lower panels of Figs.~\ref{F_LPVimage_LSDnormal} and \ref{F_LPVimage_LSDFe} is a consequence of the continuum normalization, which is complicated by the high $v_{eq}\sin i =109~\mathrm{km\,s} ^{-1}$ of 4\,CVn. We did not detect any feature in the LSD profiles that is moving according to the orbital period and could be connected to the companion of 4\,CVn.

\begin{figure}
\includegraphics[width=88mm]{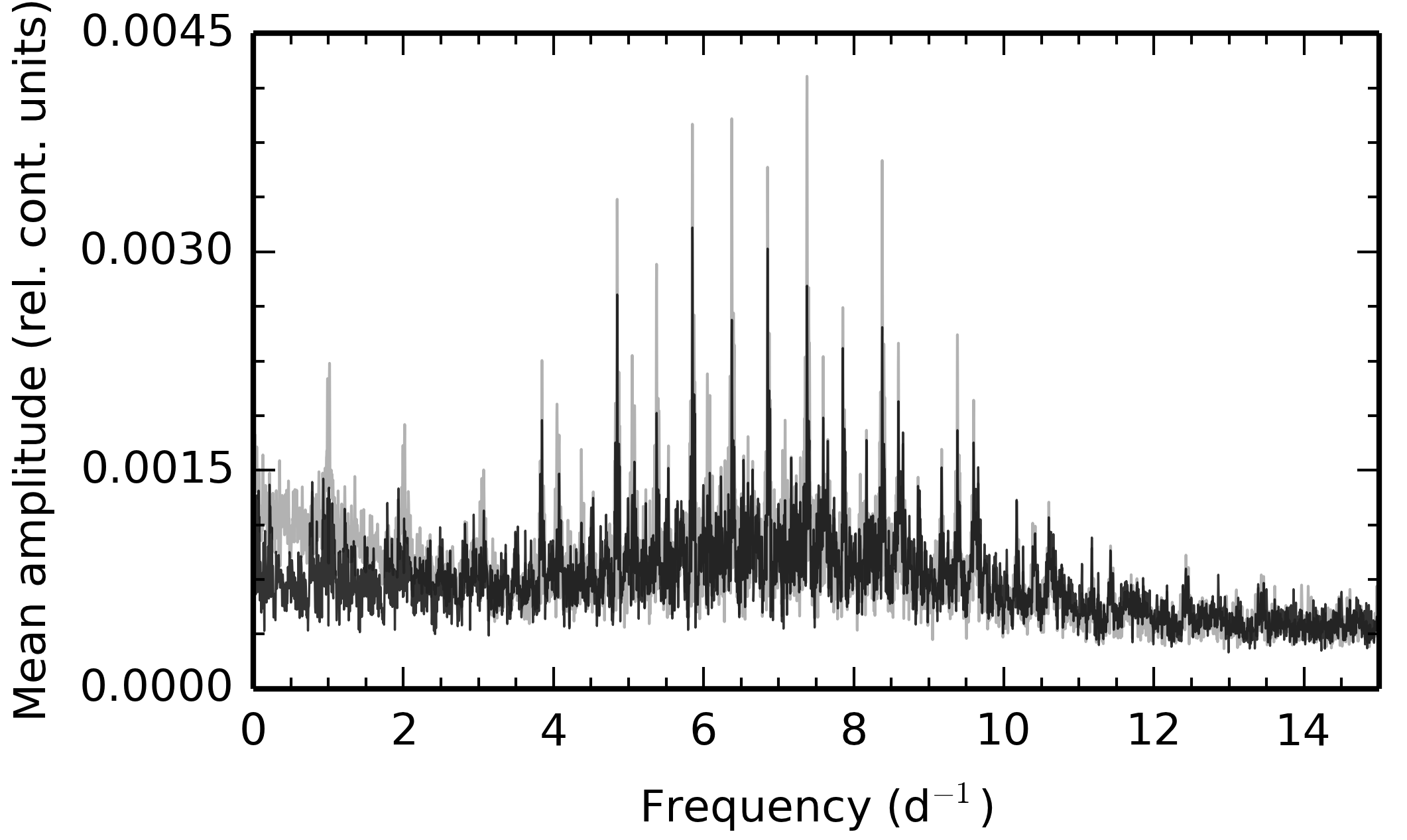}
\caption{Mean of the Lomb-Scargle periodogram per bin in the season of 2010 for the \ion{Fe}{II} line 4508.288~$\mathrm{\AA}$ (grey) and the standard LSD profiles for all lines in the line mask, rescaled to the depth of the \ion{Fe}{II} line (black).}
\label{F_fourierLSDsingle}
\end{figure}

As already explained in Sect.~\ref{S_lineselection}, different lines can be more or less sensitive to stellar pulsations. It is therefore possible that the pulsation signal is smeared out in an LSD profile. This can affect the phase, but also the amplitude of the pulsation modes. Figure~\ref{F_fourierLSDsingle} depicts the mean amplitude spectra for the \ion{Fe}{II} line 4508.288~$\mathrm{\AA}$ (grey) and for the standard LSD profiles (black). In order to compare the two periodograms, we had to rescale the LSD profiles to match the amplitude of the \ion{Fe}{II} line, as the pulsation amplitudes are expressed in units relative to the continuum of the line. The noise level in the periodogram of the rescaled LSD profiles is only lower in the low-frequency range and it contains less pulsation signal than the periodogram of the single line. This is reflected in the number of detected frequencies. While we could detect 14 significant ($A>4\sigma_\mathrm{res}$) frequencies in the LPVs of \ion{Fe}{II} (Table~\ref{T_spectro}), we could only find 8 of them in the LPVs of the LSD profile. The frequencies detected in the LSD profiles had a lower S/N, and the amplitudes dropped by almost $30\%$ for $f_1$ and by $\sim7\%$ for $f_2$.

The results are similar for the LSD profiles based only on the Fe lines. The noise level in the mean periodogram of the Fe-LSD profiles is slightly higher than in the mean periodogram of the standard LSD profiles, since less lines were used for the computation of the profiles. Consequently, the amplitudes and S/N of the detected frequencies are even lower than the values found from the analysis of the standard LSD profiles.

As the LSD profiles do not yield any improvement to the single line, we conducted the further analysis on the \ion{Fe}{II} line 4508.288~$\mathrm{\AA}$. It was already stressed by \citet[Chapter 4]{Aerts2010} that line-profile analysis works best on one carefully selected, isolated line, if available.

\section{Mode identification}
\label{S_modeid}

\subsection{Fourier-parameter fit method}

During the frequency analysis (see Sect. \ref{S_freqana}) the observed Fourier parameters zero point $Z_\lambda$, amplitude $A_\lambda$, and phase $\phi_\lambda$ across the line were obtained for each oscillation frequency. For mode identification the observed profiles are compared to theoretical profiles. Possible solutions can be compared by calculating a reduced $\chi^2$. The theoretical values are computed from synthetic line profiles, which result from an integration over a surface grid divided into $10\,000$ segments. The intrinsic profile of each surface element is assumed to be Gaussian, and the local variations of flux arise from variations of temperature and surface gravity. Temperature variations, which lead to a varying equivalent width $EW$, are neglected in a first approximation. Thus, we can also neglect the dependence of equivalent width on temperature and assume a constant $EW$. The method is described in detail by \citet[and references therein]{Zima2006,Zima2008}.

To obtain accurate mode identification, the FPF method requires the input of stellar parameters, such as mass, radius, metallicity, temperature, and surface gravity. Besides giving a solution for $(\ell,m)$ it also allows us to derive parameters of the line that are independent of pulsation, like, equivalent width $EW$ (only in first approximation), the width of the intrinsic Gaussian profile $\sigma$, and the velocity offset $dZ$, which is a measure of radial velocity. Additionally the inclination angle between pulsation axis and line-of-sight as well as the projected rotational velocity can be derived.

We recall that the method is most powerful to determine the $m$-values and not so much the $\ell$-values \citep{Zima2008}.

\subsection{Mode identification of $f_1=7.3764~\mathrm{d}^{-1}$ and $f_2=5.8496~\mathrm{d}^{-1}$}

\begin{figure}
\includegraphics[width=88mm]{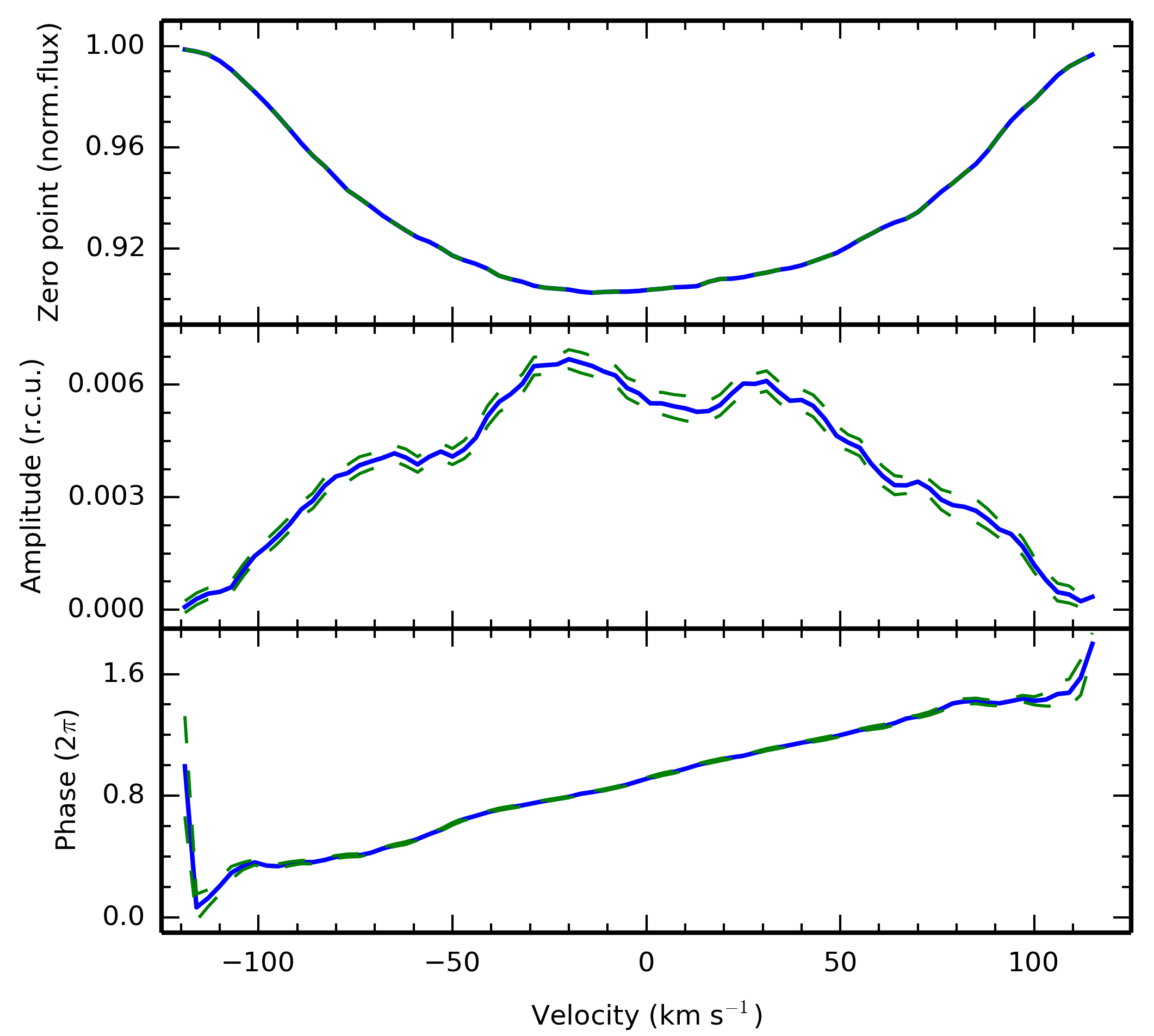}
\caption{Zero point (upper panel), amplitude (center panel), and phase (lower panel) across the line of the $f_1$, calculated with the phase-folded data set. Observations are shown as the blue, solid line, while the errors of the observations are the green, dashed lines.}
\label{F_phiF1zap}
\end{figure}

\begin{figure}
\includegraphics[width=88mm]{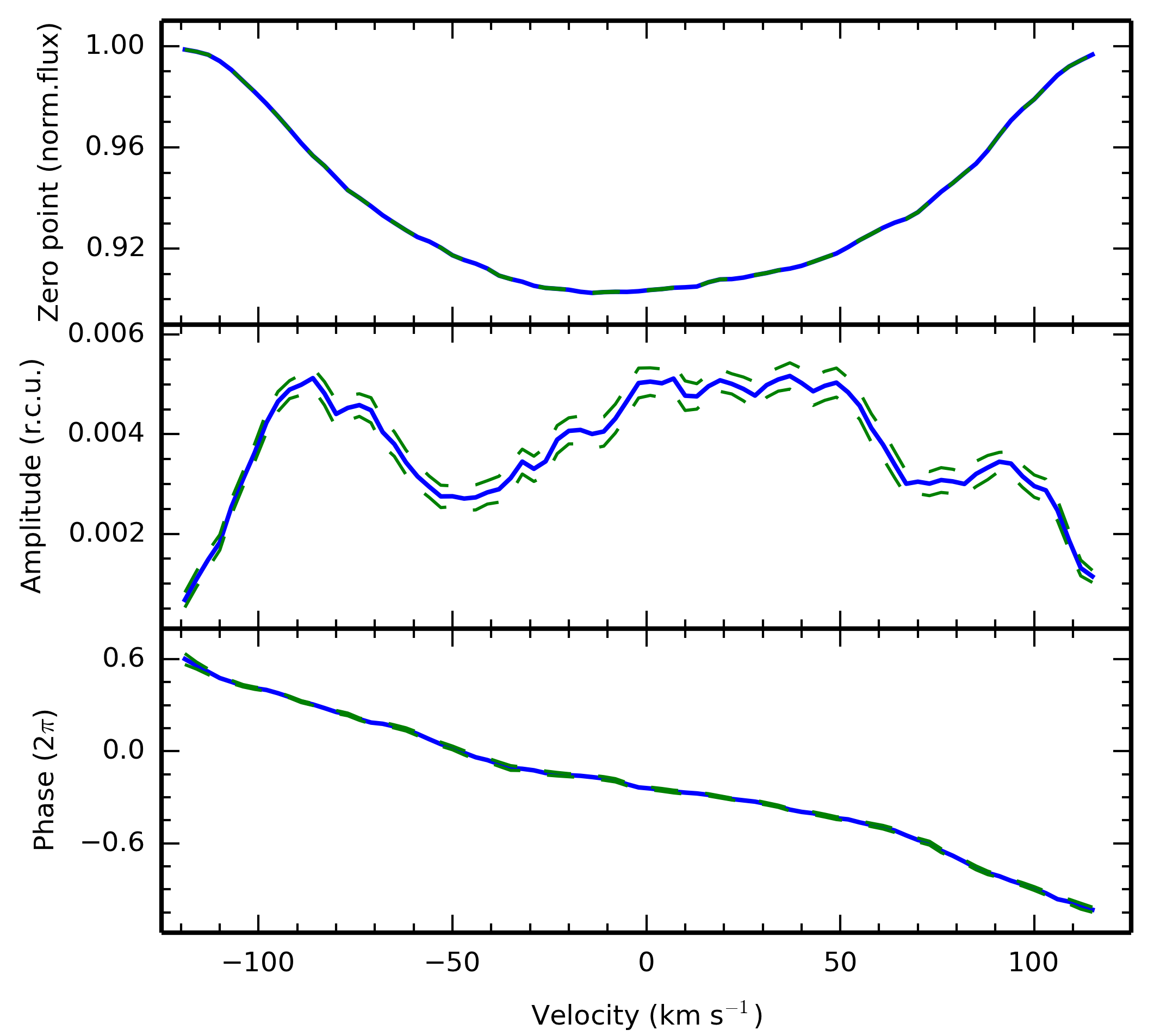}
\caption{Zero point (upper panel), amplitude (center panel), and phase (lower panel) across the line of the frequency $f_2$, calculated with the phase-folded data set. The colors and lines have the same meaning as in Fig.~\ref{F_phiF1zap}.}
\label{F_phiF2zap}
\end{figure}

Mode identification has been carried out for the two dominant frequencies $f_1=7.3764~\mathrm{d}^{-1}$ and $f_2=5.8496~\mathrm{d}^{-1}$. We folded the series of spectra on the phase of $f_1$ and $f_2$, respectively, and smoothed the LPVs in 50 phase bins. In this way we remove the signal of the several other p-modes which are excited with very low amplitudes and whose frequencies are hard to distinguish from noise and can therefore not be extracted from the data. Hence, we are able to treat the star as a mono-periodic pulsator during the mode identification. The zero point, amplitude, and phase profiles across the line obtained from the phase-folded data set are displayed in Figs.~\ref{F_phiF1zap} and \ref{F_phiF2zap} for $f_1$ and $f_2$, respectively.

We found the solution $(\ell_1,m_1)=(3,-2)$ for $f_1$. For $f_2$, mode identification was ambiguous; $(\ell_2,m_2)=(3,3)$ and $(\ell_2,m_2)=(2,2)$ are the best fits to $f_2$ depending on the stellar input parameters. However, other solutions cannot be excluded for both $f_1$ and $f_2$, since the $\chi^2$ differs by only a few percent. Furthermore, we receive a high $\chi^2$ ($\chi^2>28$) for all $\ell$ and $m$ combinations, which means that no solution provides a good fit to the observations. This will be discussed in Sect.~\ref{S_discussion}. All results are summarized in Appendix~\ref{A_MIresults}, in Tables~\ref{T_resultsMIset1}, \ref{T_resultsMIset2}, and \ref{T_resultsMIset3}. Figures~\ref{F_phiF1zapFit:l3m-2}, \ref{F_phiF1zapFit:l3m-3}, \ref{F_phiF1zapFit:l3m-1}, \ref{F_phiF1zapFit:l2m-2}, \ref{F_phiF1zapFit:l2m-1}, and \ref{F_phiF1zapFit:l1m-1} show the observed amplitude and phase profile of $f_1$ fitted with the modes $(3,-2)$, $(3,-3)$, $(3,-1)$, $(2,-2)$, $(2,-1)$, and $(1,-1)$, respectively. The fits to the observed amplitude and phase profile of $f_2$ with the modes $(3,3)$, $(3,2)$, $(3,1)$, $(2,2)$, $(2,1)$, and $(1,1)$ are shown in Figs. \ref{F_phiF2zapFit:l3m3}, \ref{F_phiF2zapFit:l3m2}, \ref{F_phiF2zapFit:l3m1}, \ref{F_phiF2zapFit:l2m2}, \ref{F_phiF2zapFit:l2m1}, and \ref{F_phiF2zapFit:l1m1}, respectively.

\subsubsection{Line-profile parameters}
Before fitting amplitude and phase across the line to identify the modes, the pulsation-independent line-profile parameters, projected rotational velocity $v_{eq}\sin i$, equivalent width $EW$, the width of the intrinsic Gaussian profile $\sigma$, and the velocity offset of the line center at 0 km\,s$^{-1}$ $dZ$ are constrained by fitting the zero-point profile with a non-pulsating model. However, these parameters varied when fitting the pulsation modes and were therefore left as free parameters during the mode identification. The ranges within which the fits were computed are summarized in Table~\ref{T_pulsind}.

\begin{table}
\caption{Line-profile parameters used for the mode identification of 4\,CVn.}
\label{T_pulsind}
\centering

\begin{tabular}{l | c}
\hline
\hline
Central wavelength ($\mathrm{\AA}$) & 4508.288 \\
$v_{eq}\sin i$ (km\,s$^{-1}$) & 106.0--107.3 \\
$EW$ (km\,s$^{-1}$) & 15.0--15.25 \\
$\sigma$ (km\,s$^{-1}$) & 8.0--9.5 \\
$dZ$ (km\,s$^{-1}$) & $-0.05$--0.5 \\
\hline
\end{tabular}
\tablefoot{The ranges of the line-profile parameters contain the best fits of the different sets of stellar parameters and the frequencies $f_1$ and $f_2$.}
\end{table}

\subsubsection{Three sets of stellar input parameters}
The photometric and spectroscopic analysis, presented in Sect.~\ref{S_specana}, suggests that the mass of 4\,CVn lies between 1 and 2 M$_\odot$, while the radius is between 3.7 and 4.1 R$_\odot$. The effective temperature of the star $T_\mathrm{eff}=6875\pm120~\mathrm{K}$, the surface gravity $\log g=3.3\pm0.35$ dex, and the metallicity is near solar.

\citet{BregerPamy2002} computed a model with the 18 photometrically observed frequencies of 4\,CVn, assuming $M=2.4~\mathrm{M_\odot}$, $\log L/\mathrm{L_\odot}=1.76$, \Teff~$=6800~K$, $\log g=3.32$, $V_\mathrm{rot}=82~\mathrm{km\,s}^{-1}$, and solar metallicity. The luminosity of this model is a factor of 2 higher than our deduced value $\log L/\mathrm{L}_\odot=1.47\pm0.05$.

Since the stellar parameters are not well constrained, we calculated the mode identification for three different sets of parameters, which are summarized in Table~\ref{T_stellar}. The values of set 1 and set 2 lie within the ranges of the results of our spectroscopic and photometric analysis, while the values for set 3 are based on the model by \citet{BregerPamy2002}. The inclination angle, $i$, was fitted within the range $23^\circ \leq i \leq 90^\circ$. We adopted a lower limit $i\sim23^\circ$, as it is the critical value, where the star rotates at the break-up velocity \citep[$v_{crit}=\sqrt{GM_\ast/R_e}$, where $R_e$ is the equatorial radius;][]{Townsend2004}. The critical velocity $v_{crit}$, which depends on the stellar mass and radius, is also given for each set in Table~\ref{T_stellar}.

It is not possible to distinguish between the three sets of stellar parameters. The best solutions of all three sets lie within $10\%$ of $\chi^2$ and are therefore equally possible. For $f_1$, the best fitting $(\ell_1,m_1)$ combination is always a $(3,-2)$ mode. The solution for $f_2$, however, shows a dependence on the stellar input parameters and differs for the three different sets (see Tables~\ref{T_resultsMIset1}, \ref{T_resultsMIset2}, \ref{T_resultsMIset3}).

\begin{table}
\caption{Stellar input parameters for the mode identification of 4\,CVn.}
\label{T_stellar}	
\centering 

\begin{tabular}{l | c | c | c}
\hline
\hline
 & Set 1\tablefootmark{a} & Set 2\tablefootmark{a} & Set 3\tablefootmark{b} \\
\hline
Mass ($\mathrm{M_\odot}$) & 2.0 & 1.5 & 2.4 \\
Radius ($\mathrm{R_\odot}$) & 3.72 & 3.75 & 5.6 \\
\Teff~($K$) & 7050 & 6950 & 6800 \\
$\log g$ (dex) & 3.6 & 3.45 & 3.32 \\
$[\mathrm{m/H}]$ (dex) & 0 & 0 & 0 \\
$v_{crit}~(\mathrm{km\,s}^{-1})$\tablefootmark{c} & 320.3 & 276.3 & 286.0 \\
\hline
Inclination (deg) & \multicolumn{3}{c}{23--90} \\
\hline
\end{tabular}
\tablefoot{\tablefoottext{a}{Values based on our spectroscopic and photometric analysis (see Sect. \ref{S_specana}).}\\
\tablefoottext{b}{Values based on \citet{BregerPamy2002}.}\\
\tablefoottext{c}{The critical velocity is given by $v_{crit}=\sqrt{GM_\ast/R_e}$.}}
\end{table}

\subsubsection{Inclination angle}
\begin{figure}
\includegraphics[width=88mm]{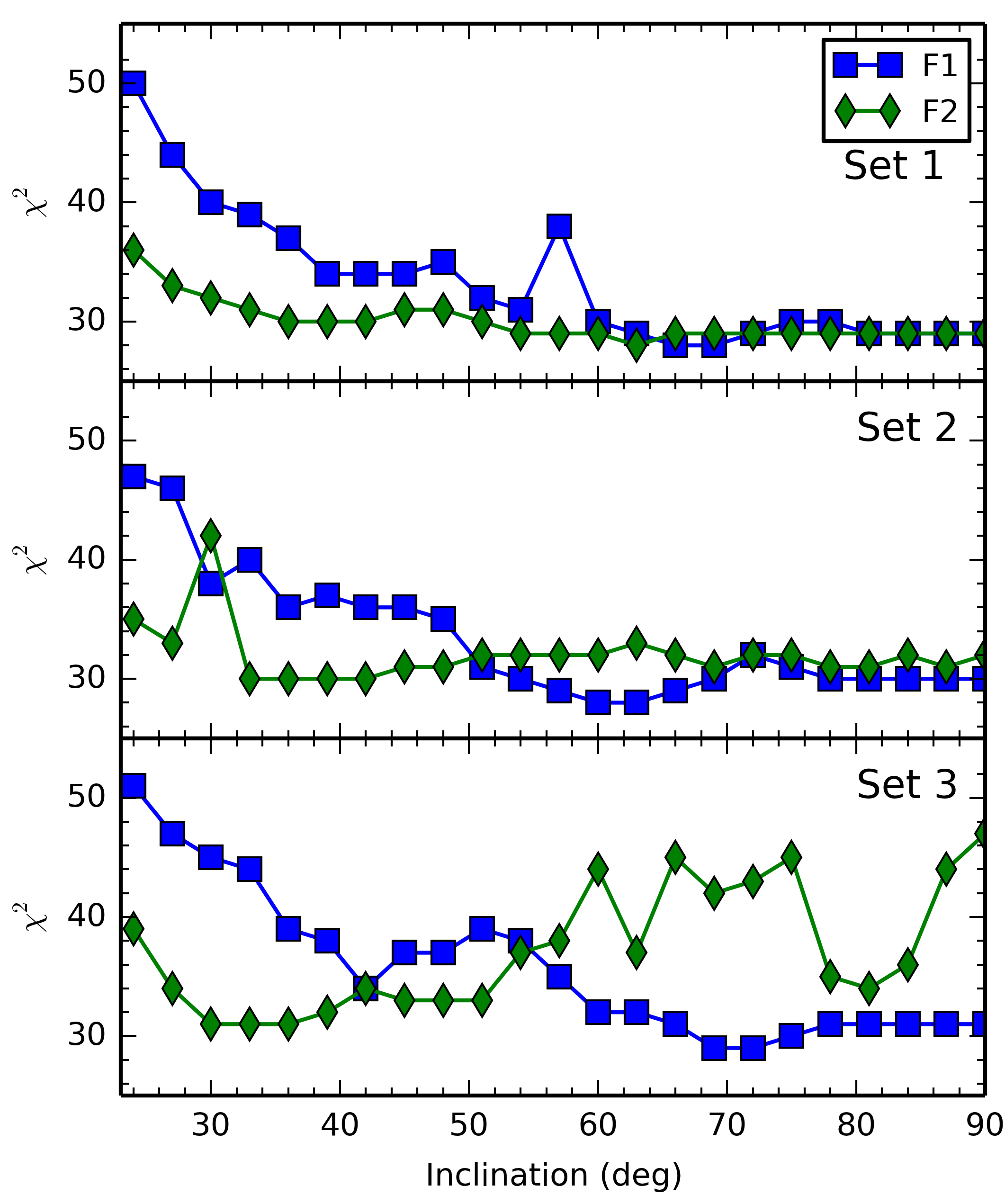}
\caption{The $\chi^2$ distribution of the inclination angle $i$ for $f_1$ (blue squares) and $f_2$ (green diamonds) and for set 1 (upper panel), set 2 (center panel), and set 3 (lower panel) of stellar parameters, respectively.}
\label{F_chi2inc}
\end{figure}

For our computations, we assume that the pulsation axis coincides with the rotation axis. The inclination angle $i$ is then the angle between this axis and the line-of-sight. It is thus not physically possible that $f_1$ and $f_2$ are observed at a different $i$. However, in Tables~\ref{T_resultsMIset1}, \ref{T_resultsMIset2}, and \ref{T_resultsMIset3} it can be seen that the value of $i$ varies between the solutions for the different sets of stellar parameters and the two frequencies. This is also illustrated in Fig.~\ref{F_chi2inc}. Displayed are, from top to bottom, the $\chi^2$ distribution of $i$ for set 1, set 2, and set 3 of stellar parameters, respectively. The minimum $\chi^2$ value is shown for each $3^\circ$ bin in inclination for $f_1$ (blue squares) and $f_2$ (green diamonds) in each panel. For set 3 the frequencies $f_1$ and $f_2$ show an opposite behavior. While $f_1$ favors higher inclination values around $\sim60^\circ$ to $\sim90^\circ$, mode identification for $f_2$ yields $i<50^\circ$. Set 1 and set 2 give a more consistent result for $i$, where $f_1$ and $f_2$ follow the same trend. On the other hand, there is also no clear minimum visible in $\chi^2$, neither for $f_1$ nor for $f_2$ in any of the three sets. The inclination is highly degenerate with the degree $\ell$ and the order $|m|$ of the oscillation mode, as it determines the visibility of a certain mode. Mode identification was ambiguous for $f_1$ and $f_2$ and certain effects hamper us from drawing clear conclusions, as will be discussed in Sect.~\ref{S_discussion}. Thus, we can also not find a consistent inclination for 4\,CVn.

\subsubsection{Temperature and equivalent-width variations}

So far we ignored variations of temperature in a first approximation and assumed a constant $EW$; thus, $d(EW)/d(T_\mathrm{eff})=0$.

To improve the fits of the amplitude and phase profiles we varied the parameter $d(EW)/d(T_\mathrm{eff})$ and other parameters related to the temperature dependence of the flux variation; at first within a very broad range, and gradually narrowing down the range. However, the parameters did not converge and the minimum in $\chi^2$ varied with each iteration, without improving the overall $\chi^2$ significantly. The final solution therefore had to be calculated with $d(EW)/d(T_\mathrm{eff})=0$, neglecting temperature variations.

\begin{table}
\caption{Identification of the azimuthal order for the seven dominant frequencies.}
\label{T_modeid}	
\centering 
\begin{tabular}{l c l} 
\hline\hline 
\multicolumn{2}{c}{Frequency} & \\
& $d^{-1}$ & \\
\hline
$f_1$ & 7.3764 & retrograde \\
$f_2$ & 5.8496 & prograde \\
$f_3$ & 5.0481 & retrograde or axisymmetric \\
$f_4$ & 5.5315 & prograde \\
$f_5$ & 8.5942 & prograde \\
$f_6$ & 8.6552 & prograde \\
$f_7$ & 8.1688 & high-degree, prograde \\
\hline
\end{tabular}
\end{table}

\subsection{Azimuthal order of the 7 most-dominant frequencies}

Mode identification has also been attempted for the frequencies $f_3$, $f_4$, $f_5$, $f_6$, and $f_7$. However, the amplitude of $f_3$ decreases by almost $40\%$ compared to $f_2$ and the amplitudes of the other frequencies are even lower. Therefore, phase folding of the spectra on the respective frequencies and smoothing the LPVs in phase bins did not remove the signal of the two most-dominant frequencies sufficiently. A consistent mode identification can thus not be achieved.

However, as the FPF method is sensitive to phase, the slope of the phase profile across the line allows us to distinguish between \textit{prograde} \citep[i.e., a wave traveling in the direction of rotation; positive $m$, following the definition by][]{Zima2008}, \textit{retrograde} (negative $m$), and \textit{axisymmetric} ($m=0$) modes. Our results are displayed in Table~\ref{T_modeid}.

We also attempted mode identification for $f_{12}=6.975~\mathrm{d}^{-1}$. Our results do not contradict the solution presented in \citet{Lenz2010}, who reported that it is a radial mode.

The very complex amplitude and phase profiles of $f_7$, i.e., containing many bumps, suggests that it is a high-degree mode, with $\ell \geq3$. For modes with a high $\ell$ value, partial cancellation would lead to very low amplitudes in photometric observations, where the brightness variation integrated over the stellar disk is measured. This mode was not detected in photometric data before, which supports our hypothesis of a high $\ell$ for $f_7$.

\section{Discussion}
\label{S_discussion}
\subsection{Mode visibility and rotational splitting}
The mode identification we attempted for the star 4\,CVn did not yield unambiguous results. The $\chi^2$ statistics give several possible solutions for $f_1$ and $f_2$, depending on the stellar input parameters. For frequency $f_1=7.3764~\mathrm{d}^{-1}$ the best solution yields $\ell_1=3$ and for frequency $f_2=5.8496~\mathrm{d}^{-1}$, $\ell_2=2$ or $\ell_2=3$. Theory predicts that modes with degree $\ell=3$ cancel out almost completely in photometric observations \citep[Chapter~6]{Aerts2010}. However, both $f_1$ and $f_2$ were detected with significant amplitudes in photometric data (see Table~\ref{T_photo}), which contradicts the results of our mode identification. The solutions $(\ell_1,m_1)=(2,-2)$ and $(\ell_2,m_2)=(2,2)$ would be in best agreement with the current theory and our results.

We, therefore, tested the hypothesis that $f_1$ and $f_2$ are the outer parts of a rotationally split quintuplet. Rotational splitting \citep{Ledoux1951} describes the shift of frequencies of modes of same degree $\ell$ and different order $|m|$ ($0\leq|m|\leq\ell$) due to the Coriolis force. It is dependent on $|m|$ and the rotational period of the star. With set 3 of stellar parameters and $i>80^\circ$ it would be possible that we are observing rotational splitting in 4\,CVn (see Table~\ref{T_resultsMIset3} and Fig.~\ref{F_chi2inc}).

\subsection{Rotation rate}
We conclude that 4\,CVn rotates at a significant fraction of its critical velocity. We measured a projected rotational velocity $v_{eq}\sin i\simeq106.7~\mathrm{km\,s}^{-1}$ for the star. Thus, its equatorial velocity $v_{eq}$ is at least $\sim33\%$ of its critical velocity $v_{crit}$. Since the inclination angle $i$ and the stellar mass and radius are poorly constrained, it is possible that it rotates at up to almost $\sim70\%$ of $v_{crit}$. Tables~\ref{T_resultsMIset1}, \ref{T_resultsMIset2}, and \ref{T_resultsMIset3} display the possible ranges of $i$ and $v_{eq}/v_{crit}$. 

\citet{Reese2013} showed that even a moderate rotation rate $v_{eq}/v_{crit}=0.3$ has large effects on the visibility of oscillation modes. While partial cancellation makes $\ell=3$ modes almost not detectable in photometric data for slow rotation ($v_{eq}/v_{crit}\leq0.2$), their amplitudes are as high as the amplitudes of modes with $\ell \leq2$ if $v_{eq}/v_{crit}=0.3$ \citep[see Fig.~7 of][]{Reese2013}. This would mean that our results of $\ell=3$ for $f_1$ and $f_2$ do not contradict the mode amplitudes observed in photometric light curves. However, fast rotation also alters the geometry of the modes \citep{Reese2009}. Since higher-order effects of rotation are not yet included in the methodology for mode identification, this could be an explanation for the poor fits we obtained. If this is the case, the amplitude and phase profiles across the line, as well as the amplitude ratios of multi-color photometry would have to be revisited, with a methodology that includes the effects of rotation on the pulsation modes.

\section{Summary}
In this work, we analyzed the line-profile variations of the $\delta$\,Sct star 4\,CVn based on the spectroscopic observations obtained between January 2008 and June 2011 at McDonald Observatory, Texas, USA. We discovered that the star is the primary component of an eccentric binary system with an orbital period $P_{orb}=124.44\pm0.03~$d and eccentricity $e=0.311\pm0.003$. No signal of the secondary could be detected in our data. The frequency analysis revealed 20 oscillation modes, 11 of which were already detected in photometric data sets by \citet{Breger1999,Breger2000,Breger2008}. 

By phase-folding the series of spectra onto the two dominant frequencies $f_1$ and $f_2$, respectively and smoothing the LPVs in phase bins, we removed the signal of other periodicities to prepare the data for mode identification. The best solution for frequency $f_1$ is an $(\ell_1,m_1)=(3,-2)$ mode, while the best solution for $f_2$ was either $(\ell_2,m_2)=(3,3)$ or $(\ell_2,m_2)=(2,2)$ depending on the stellar input parameters. For the calculations of the synthetic LPVs we used three different sets of stellar parameters:
\begin{enumerate}
\item $M=2.0~\mathrm{M_\odot}$, $R=3.72~\mathrm{R_\odot}$, $T_\mathrm{eff}=7050~$K, $\log g=3.6$, and [m/H]=0.
\item $M=1.5~\mathrm{M_\odot}$, $R=3.75~\mathrm{R_\odot}$, $T_\mathrm{eff}=6950~$K, $\log g=3.45$, and [m/H]=0.
\item $M=2.4~\mathrm{M_\odot}$, $R=5.6~\mathrm{R_\odot}$, $T_\mathrm{eff}=6800~$K, $\log g=3.32$, and [m/H]=0.
\end{enumerate}
Set 1 and 2 are based on the spectroscopic and photometric analysis we performed to revise the stellar parameters. Set 3 is based on a model calculated by \citet{BregerPamy2002}. No set of stellar parameters can be excluded, based on the $\chi^2$ statistics.

Relying on previous results by \citet{Castanheira2008}, \citet{Breger2010} claimed a relationship between the frequency variations of 4\,CVn and the azimuthal order $m$ of the modes, and especially the direction of the running wave (\textit{retrograde}, \textit{prograde} or \textit{axisymmetric}/\textit{radial} modes). The FPF method is most sensitive to the sign of $m$. Our results in Table~\ref{T_modeid} are in agreement with the directions of the waves used by \citet{Breger2010}.

The star is a fast rotator with a projected rotational velocity $v_{eq}\sin i\simeq106.7~\mathrm{km\,s}^{-1}$. Depending on the stellar input parameters and the inclination angle $i$ (which could not be constrained), the star is rotating at $>33\%$ of the critical velocity $v_{crit}$. The higher-order effects of rotation were not taken into account in our analysis, since they are not yet included in the methodology for mode identification.

A further limitation from the methodology is that unblended metal lines have to be used for mode identification. However, our observed spectral range (4500 to 4700~$\mathrm{\AA}$) is a very dense spectral region with many absorption lines and very little continuum for F stars, and almost all absorption lines are (heavily) blended. In an attempt to solve this problem, we calculated LSD profiles from the McDonald spectra, obtained in the season 2010. We found that the pulsation signal is smeared out in the LSD profiles, as the number of significant oscillation modes detected in the LPVs of the LSD profiles is less than for the single \ion{Fe}{II} line. Also the amplitudes and the S/N levels of these modes are reduced significantly. To improve our analysis a method to identify the modes in blended lines would have to be developed. Also, observations in a wider spectral range could help solve the problem of blended lines. The lines that \citet{Zima2006b} used for their analysis of the $\delta$\,Sct star FG\,Vir, lie outside the observed spectral range of the McDonald data. Observing a wider spectral range with a stable, high-resolution spectrograph would allow to follow up on the additional spectral feature detected in the LSD profile of 4\,CVn (see Fig.~\ref{F_LSD}), and could possibly lead to identification of the origin of the latter.

Accurate mode identification of many observed frequencies is a crucial input for successful modeling of stellar oscillations. Analyses of LPVs are important as they can constrain the azimuthal order $m$ of the pulsation modes. However, current shortcomings in the methodology prevent unambiguous and accurate mode identification for 4\,CVn. Including a complete description of rotation and the use of blended lines into mode-identification techniques will make this possible in the future.

\begin{acknowledgements}
      This research has received funding from the Fonds zur F\"orderung der wissenschaftlichen Forschung (FWF), Austria P17441-N02, from the European
Community's Seventh Framework Programme FP7-SPACE-2011-1, project number 312844 (SPACEINN), as well as from the Belgian Science Policy Office (Belspo, C90309: CoRoT Data Exploitation). It is partly based on observations made with the Mercator Telescope, operated on the island of La Palma by the Flemish Community, at the Spanish Observatorio del Roque de los Muchachos of the Instituto de Astrof\'isica de Canarias and obtained with the HERMES spectrograph, which is supported by the Fund for Scientific Research of Flanders (FWO), Belgium , the Research Council of KU Leuven, Belgium, the Fonds National Recherches Scientific (FNRS), Belgium, the Royal Observatory of Belgium, the Observatoire de Genève, Switzerland and the Th\"uringer Landessternwarte Tautenburg, Germany. SB is supported by the Foundation for Fundamental Research on Matter (FOM), which is part of the Netherlands Organisation for Scientific Research (NWO).
\end{acknowledgements}

\bibliographystyle{aa} 
\bibliography{references}

\newpage

\appendix

\section{Amplitude and phase profiles}
\label{A_amp.phiProfiles}

\begin{figure}[h]
\includegraphics[width=88mm]{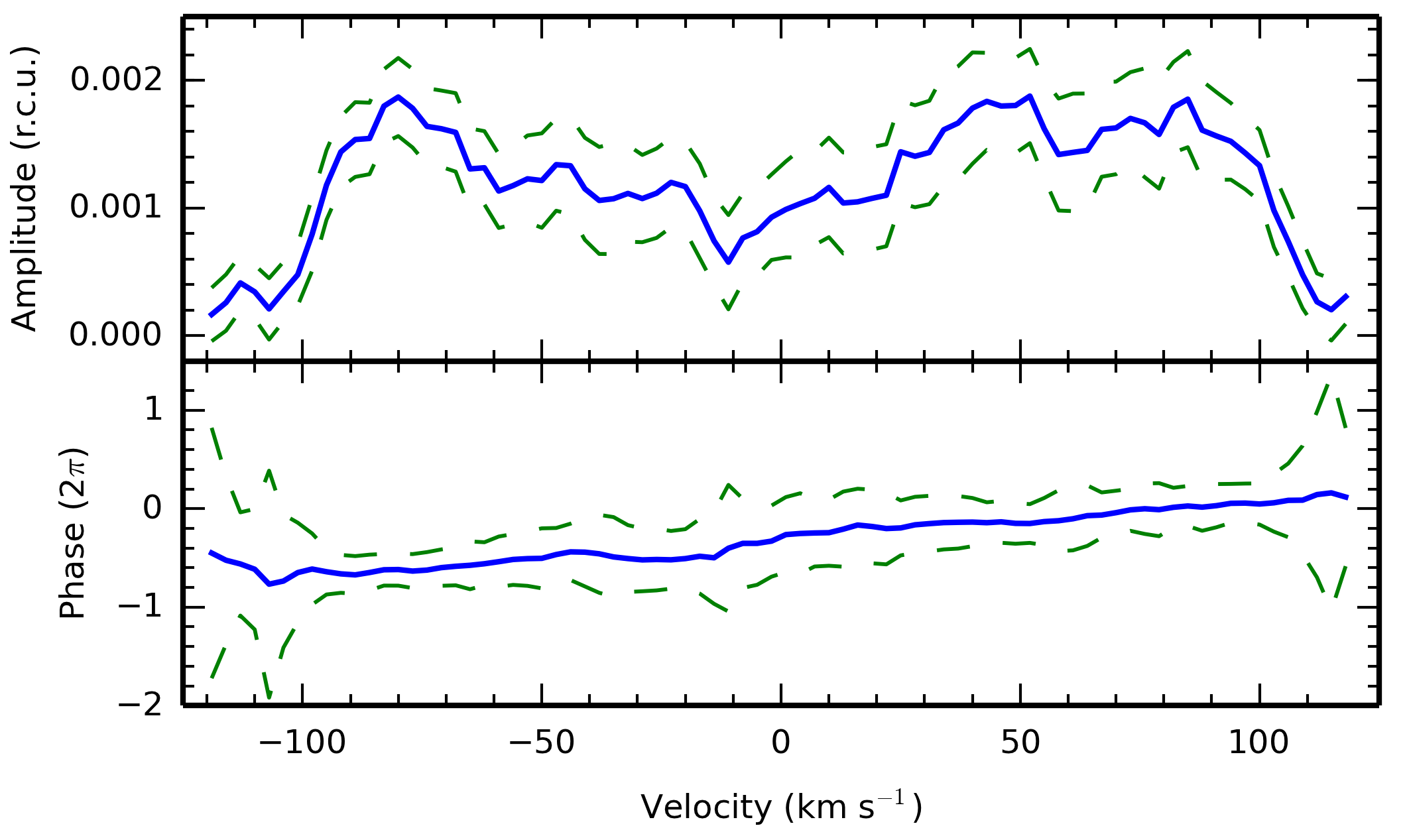}
\caption{Amplitude (upper panel) and phase (lower panel) across the line for frequency $f_8=6.6801~\mathrm{d}^{-1}$. Observations are shown as the blue, solid line and errors of the observations are shown as the green, dashed line.}
\label{F_zapF8}
\end{figure}

\begin{figure}[h]
\includegraphics[width=88mm]{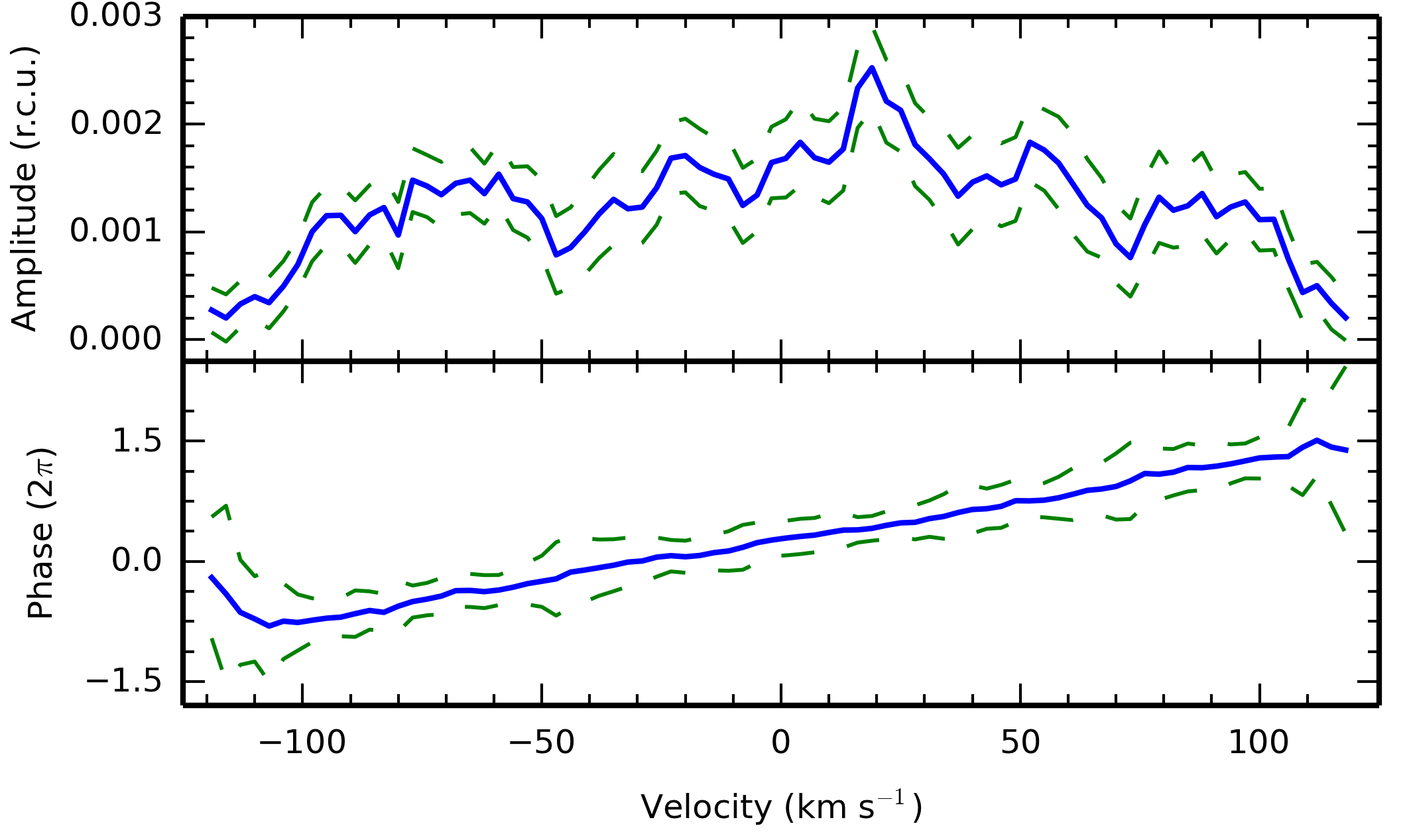}
\caption{Same as Fig.~\ref{F_zapF8} but for frequency $f_9=4.0743~\mathrm{d}^{-1}$.}
\label{F_zapF9}
\end{figure}

\begin{figure}[h]
\includegraphics[width=88mm]{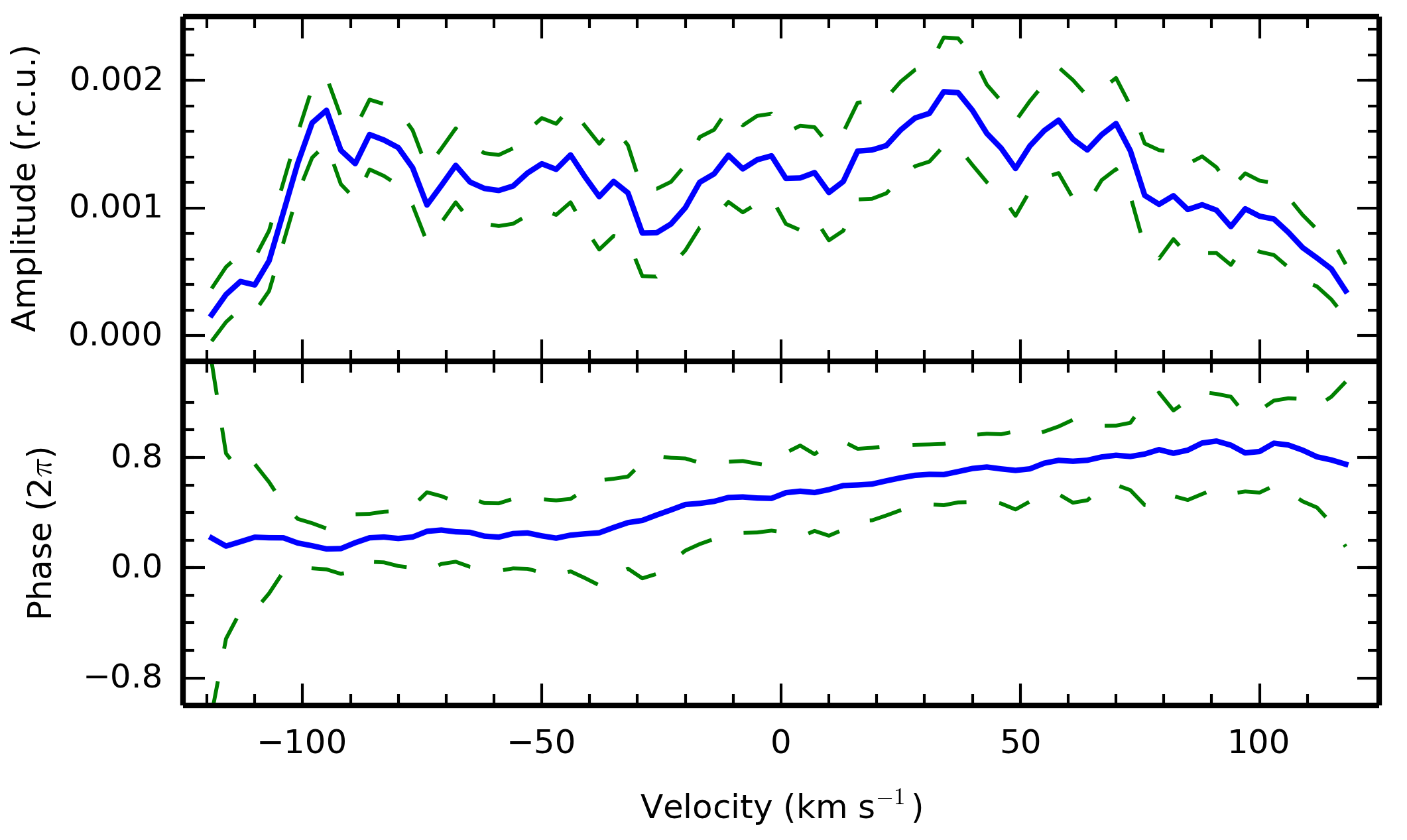}
\caption{Same as Fig.~\ref{F_zapF8} but for frequency $f_{10}=6.1171~\mathrm{d}^{-1}$.}
\label{F_zapF10}
\end{figure}

\begin{figure}[h]
\includegraphics[width=88mm]{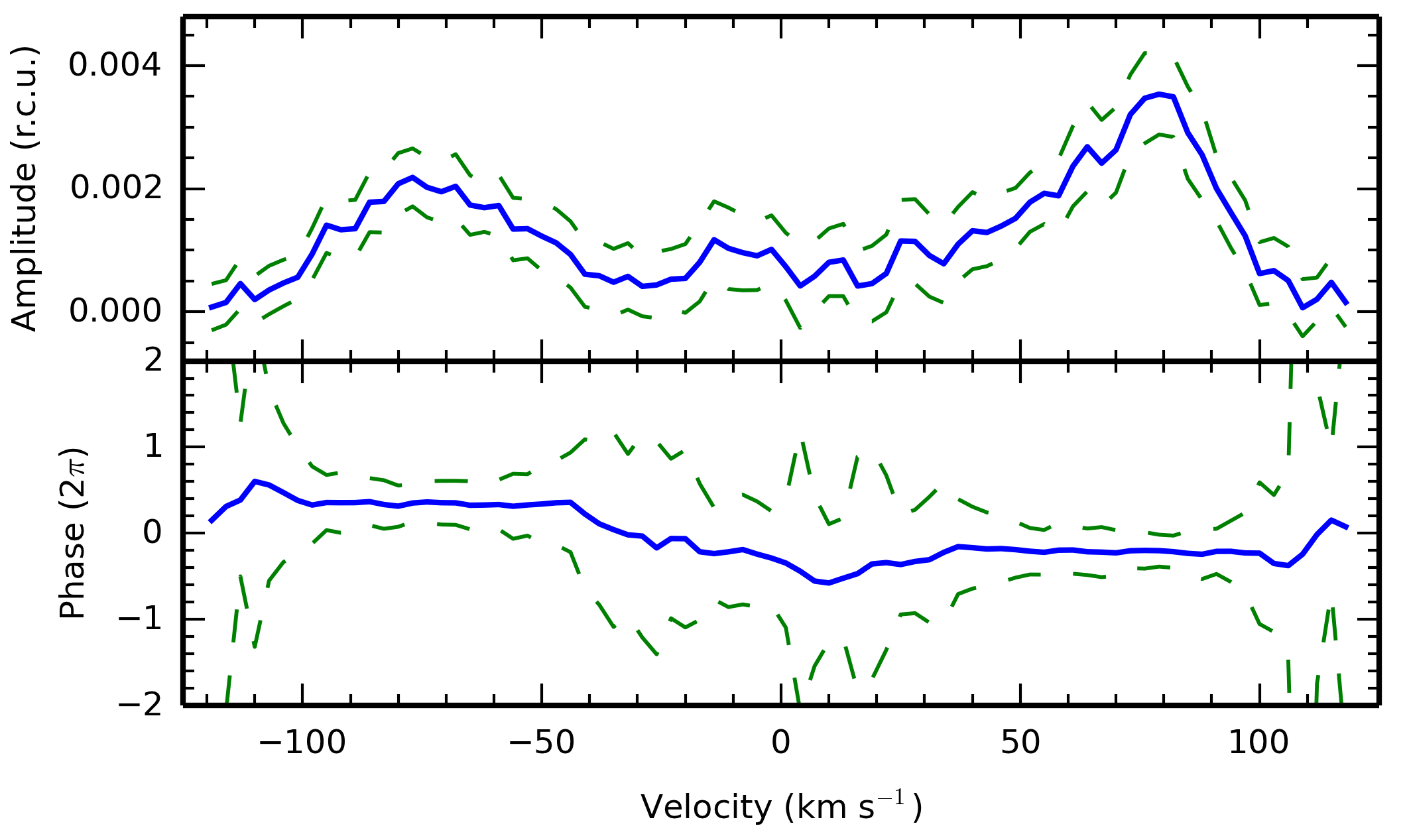}
\caption{Same as Fig.~\ref{F_zapF8} but for frequency $f_{11}=6.975~\mathrm{d}^{-1}$.}
\label{F_zapF11}
\end{figure}

\begin{figure}[h]
\includegraphics[width=88mm]{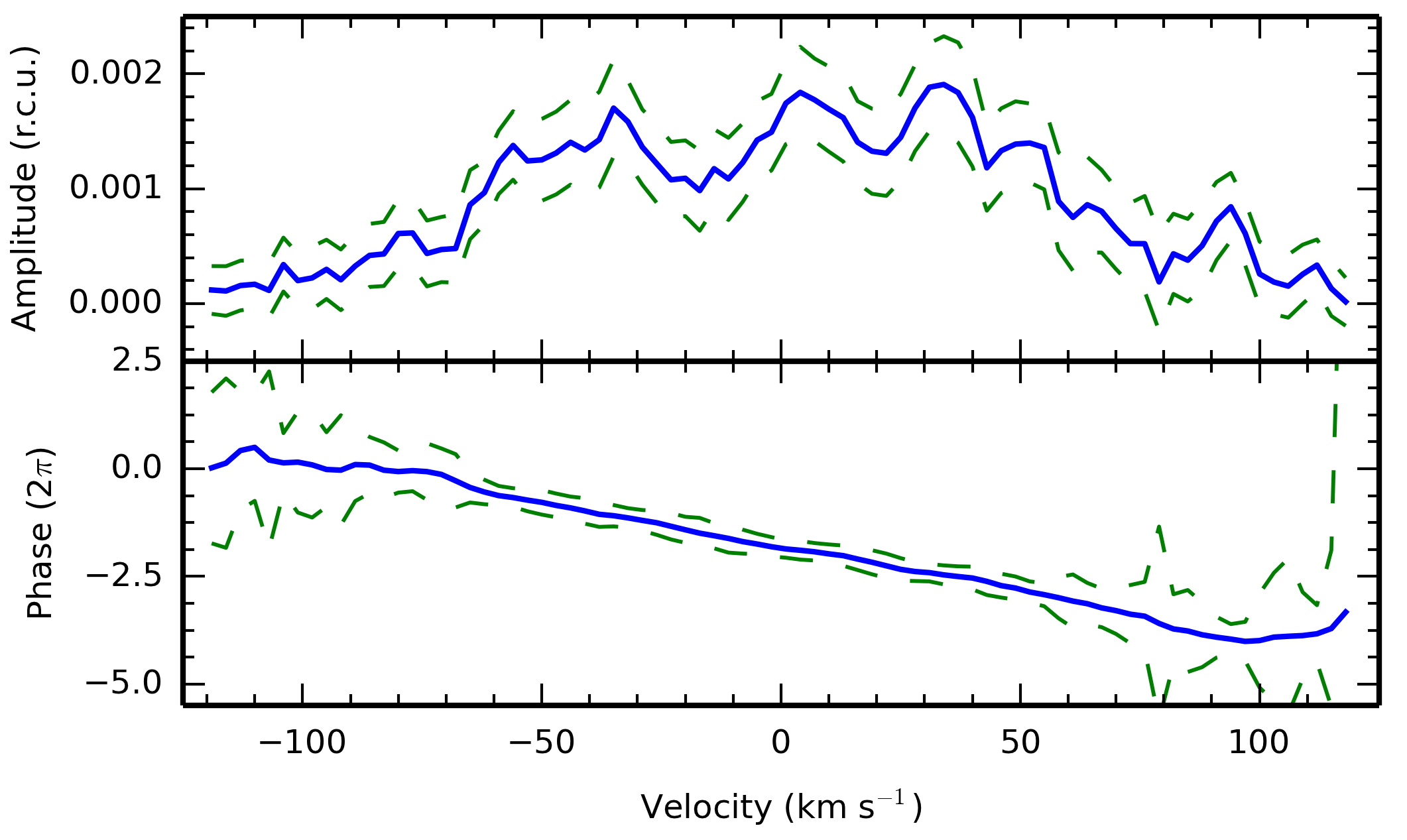}
\caption{Same as Fig.~\ref{F_zapF8} but for frequency $f_{12}=10.1702~\mathrm{d}^{-1}$.}
\label{F_zapF12}
\end{figure}

\begin{figure}[h]
\includegraphics[width=88mm]{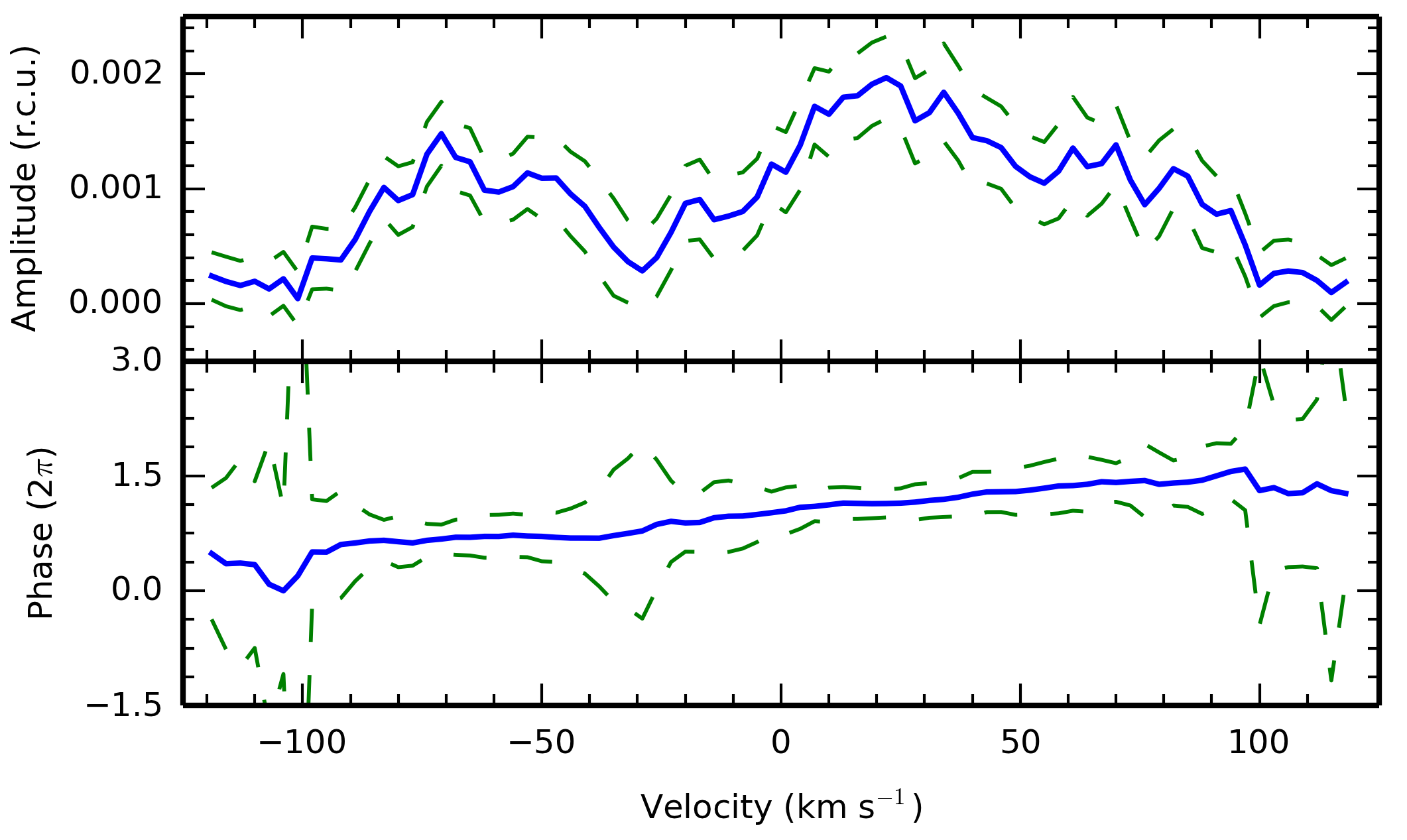}
\caption{Same as Fig.~\ref{F_zapF8} but for frequency $f_{13}=6.1910~\mathrm{d}^{-1}$.}
\label{F_zapF13}
\end{figure}

\begin{figure}[h]
\includegraphics[width=88mm]{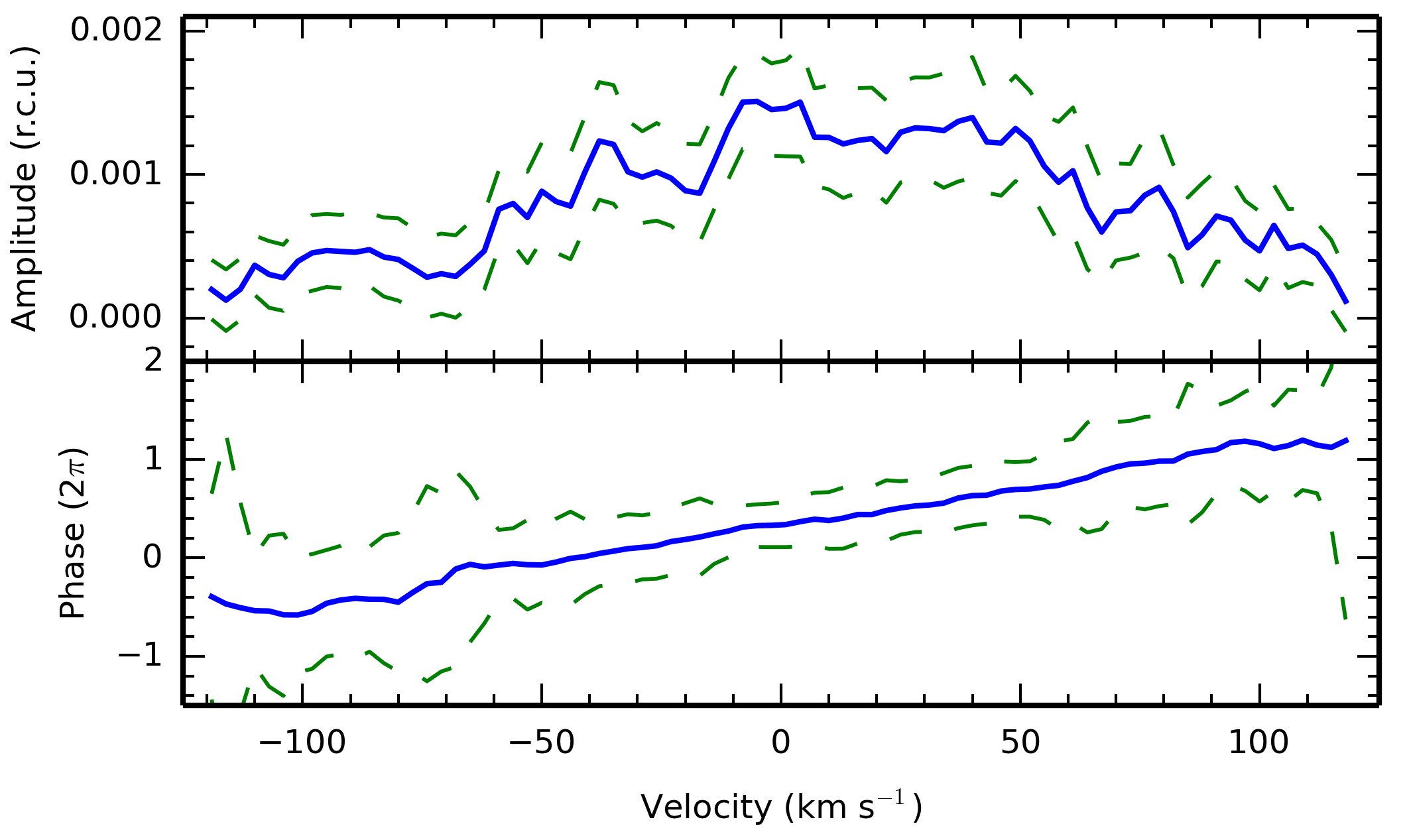}
\caption{Same as Fig.~\ref{F_zapF8} but for frequency $f_{14}=12.4244~\mathrm{d}^{-1}$.}
\label{F_zapF14}
\end{figure}

\clearpage

\begin{figure}[h]
\includegraphics[width=88mm]{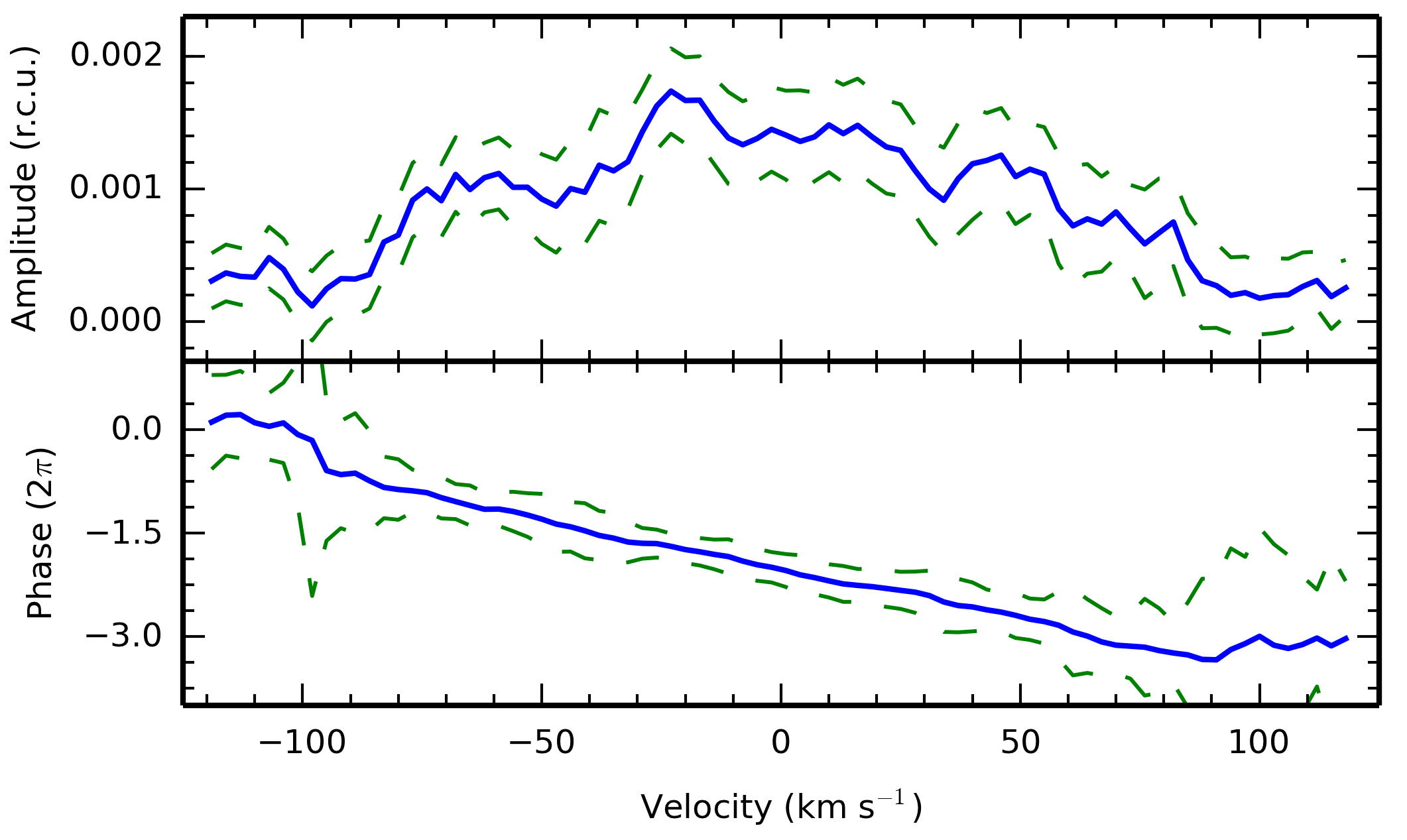}
\caption{Same as Fig.~\ref{F_zapF8} but for frequency $f_{15}=9.4113~\mathrm{d}^{-1}$.}
\label{F_zapF15}
\end{figure}

\begin{figure}[h]
\includegraphics[width=88mm]{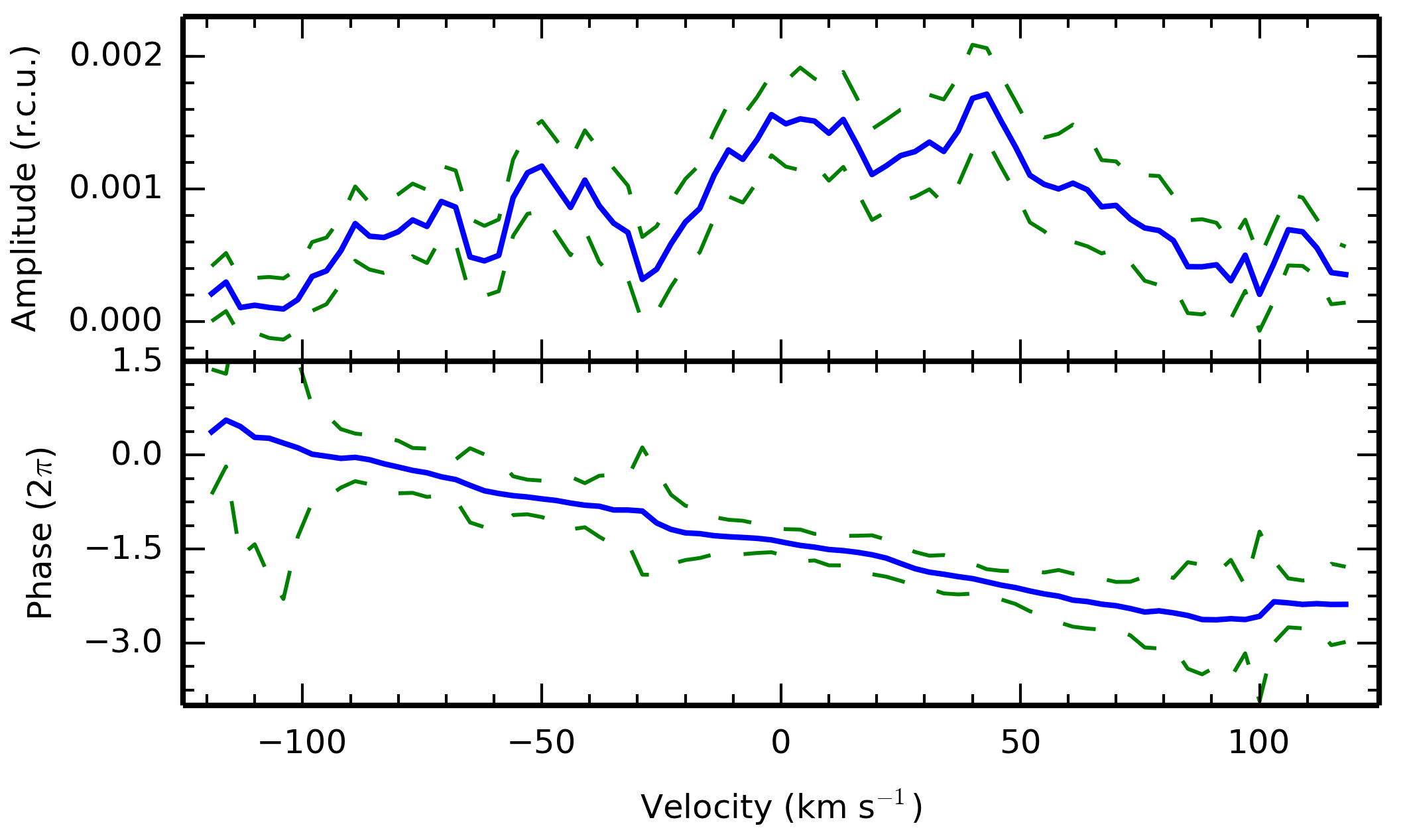}
\caption{Same as Fig.~\ref{F_zapF8} but for frequency $f_{16}=9.7684~\mathrm{d}^{-1}$.}
\label{F_zapF16}
\end{figure}

\begin{figure}[h]
\includegraphics[width=88mm]{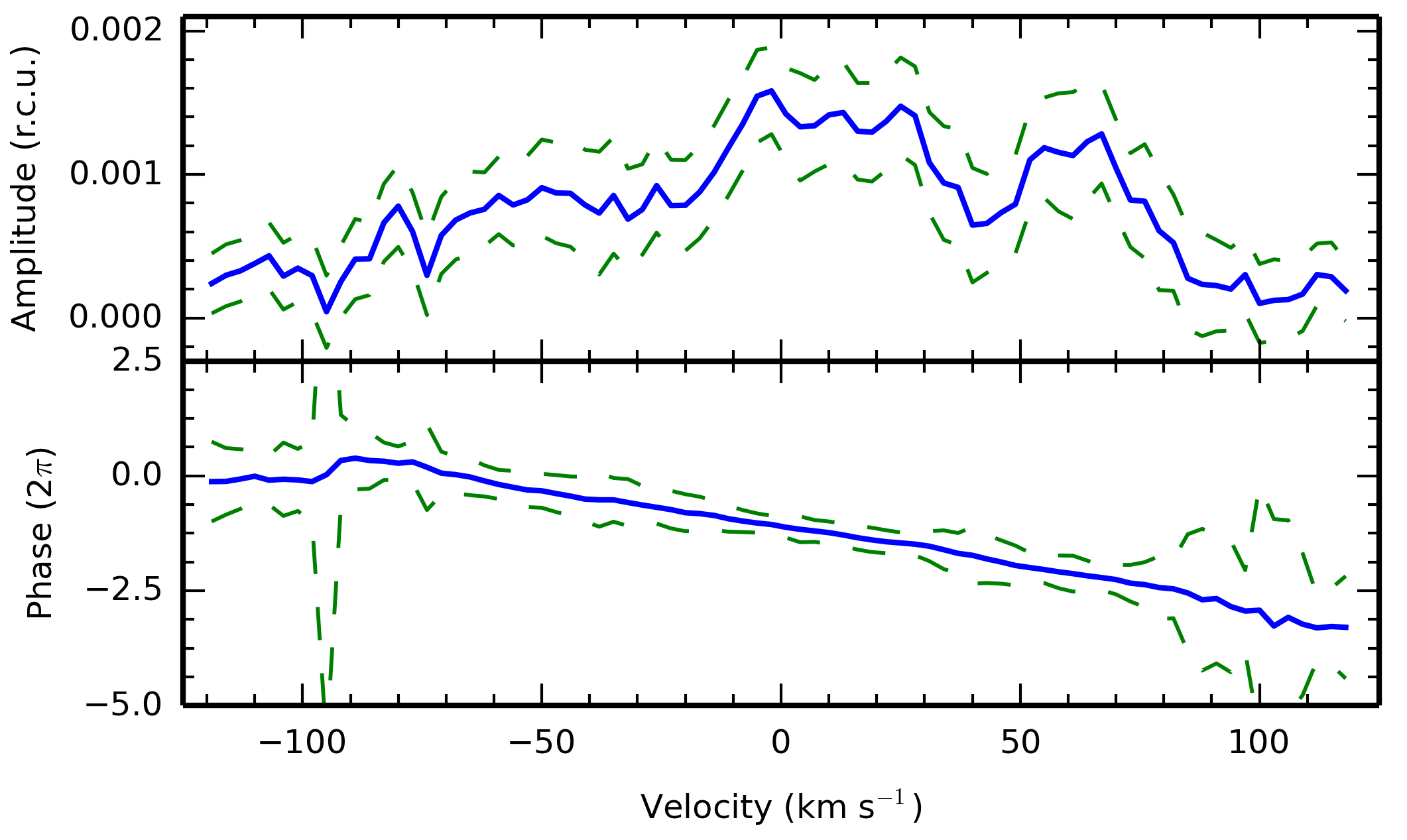}
\caption{Same as Fig.~\ref{F_zapF8} but for frequency $f_{17}=10.0372~\mathrm{d}^{-1}$.}
\label{F_zapF17}
\end{figure}

\begin{figure}[h]
\includegraphics[width=88mm]{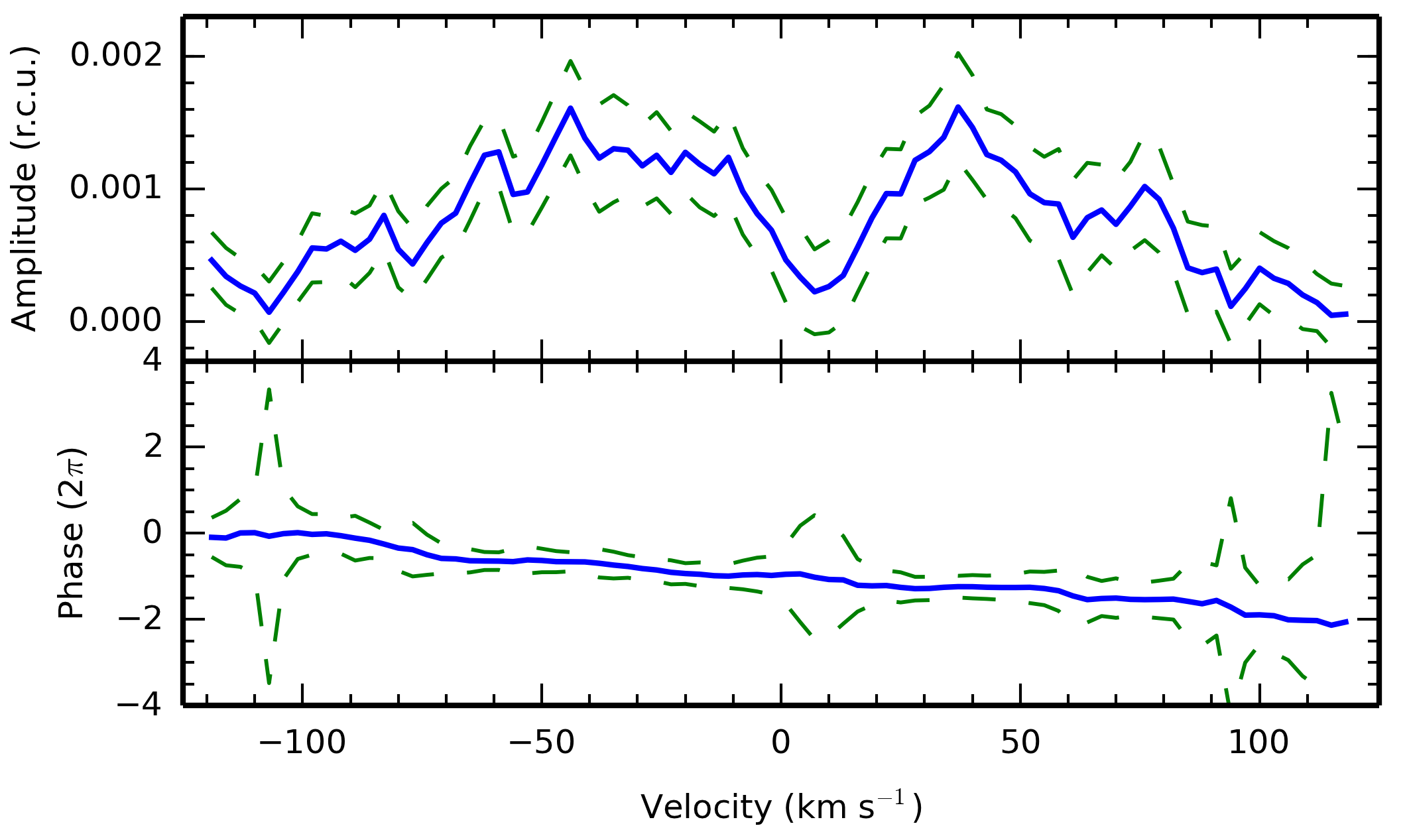}
\caption{Same as Fig.~\ref{F_zapF8} but for frequency $f_{18}=6.4030~\mathrm{d}^{-1}$.}
\label{F_zapF18}
\end{figure}

\begin{figure}[h]
\includegraphics[width=88mm]{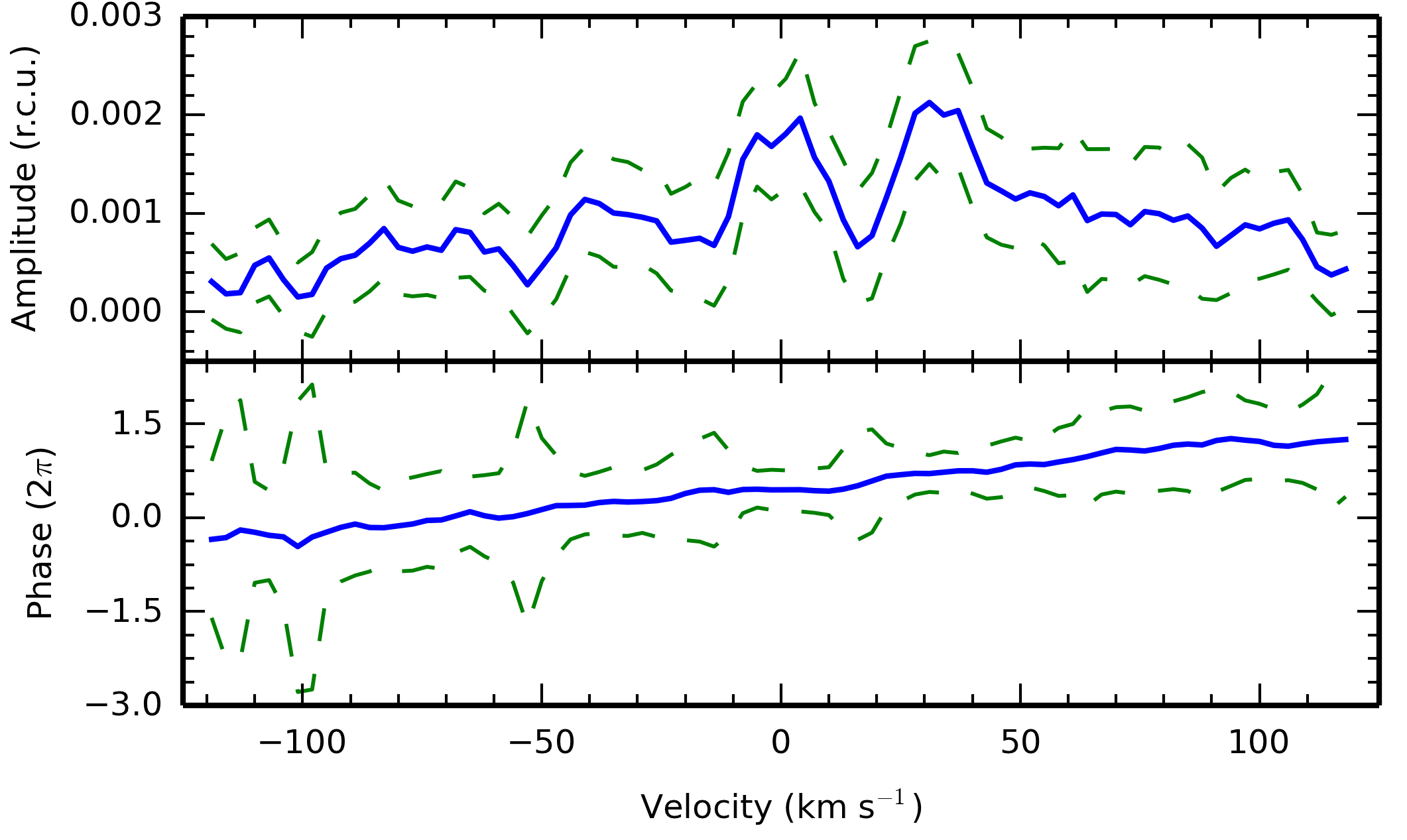}
\caption{Same as Fig.~\ref{F_zapF8} but for frequency $f_{19}=13.388~\mathrm{d}^{-1}$.}
\label{F_zapF19}
\end{figure}

\begin{figure}[h]
\includegraphics[width=88mm]{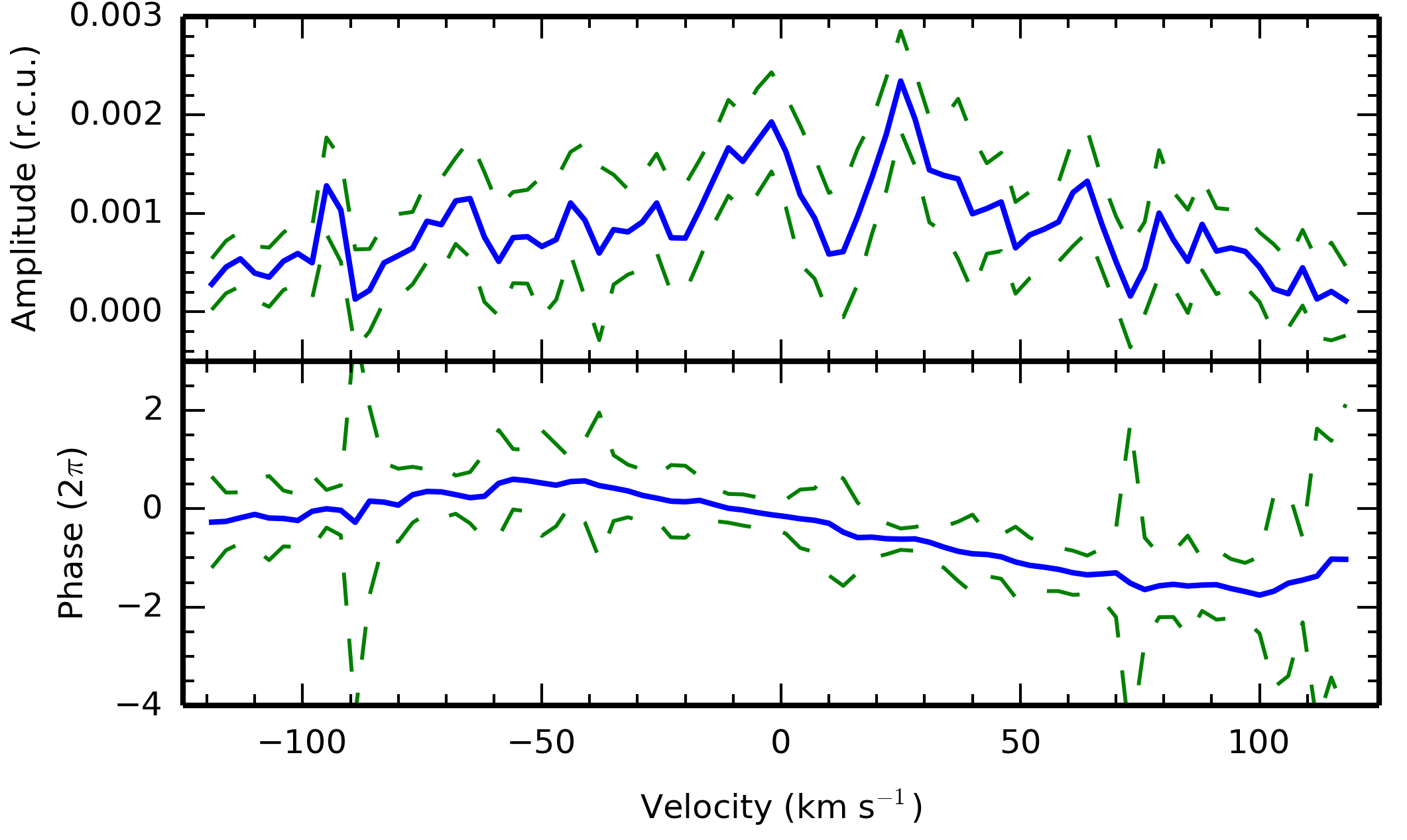}
\caption{Same as Fig.~\ref{F_zapF8} but for frequency $f_{20}=10.016~\mathrm{d}^{-1}$.}
\label{F_zapF20}
\end{figure}

\clearpage

\section{Comparison of LSD profiles and single-line profiles}
\label{A_LSDcomp}

\begin{figure}[h]
\includegraphics[width=88mm]{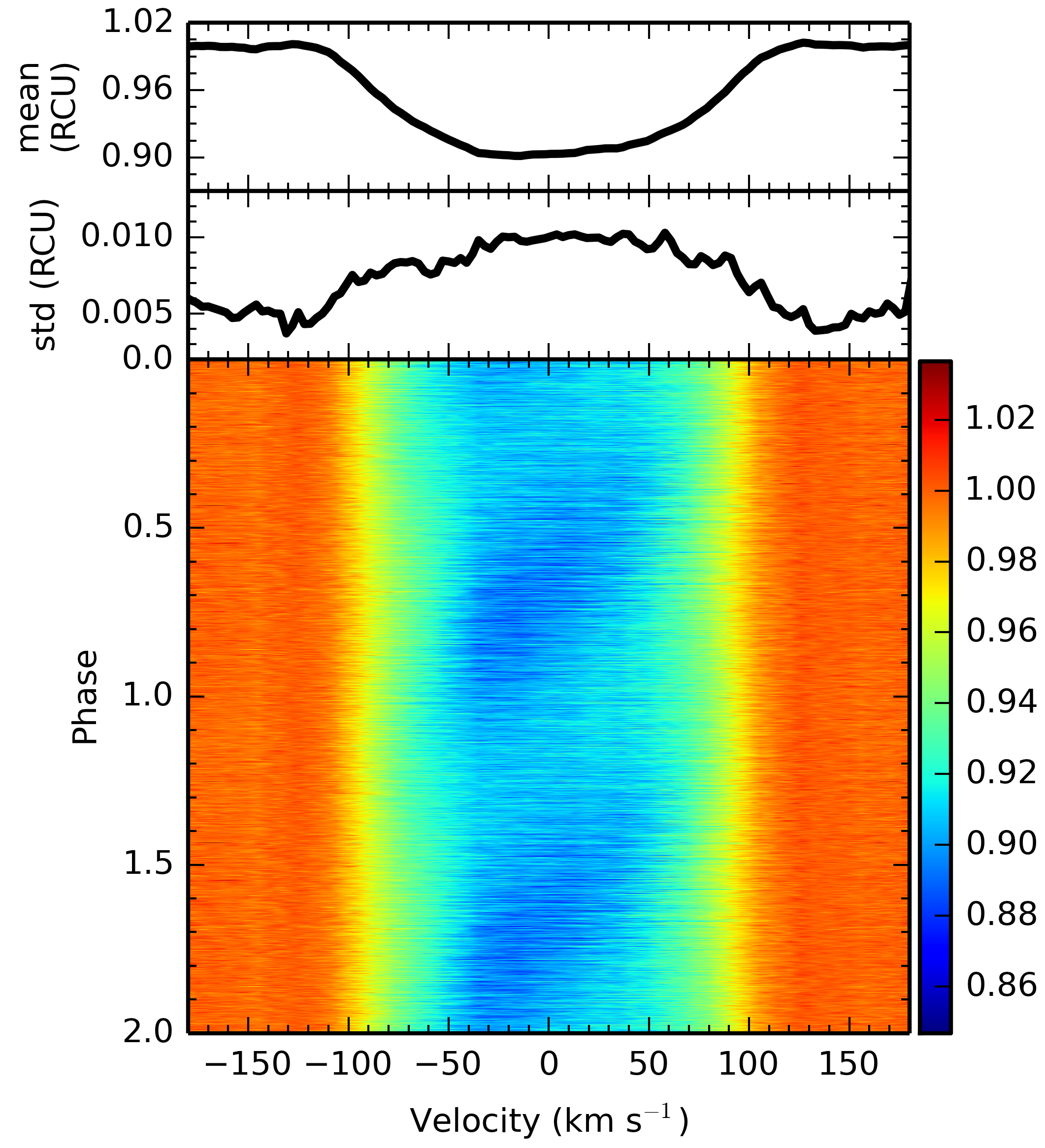}
\caption{Upper panel: Mean profile of the \ion{Fe}{II} line at $4508.288\mathrm{~\AA}$ in the season 2010. Center panel: Standard deviation of the \ion{Fe}{II} line. Lower panel: Color image of the LPVs of the \ion{Fe}{II} line, phase folded on the frequency $f_1=7.3764~\mathrm{d}^{-1}$. The amplitude is color coded and can be read off from the color bar.}
\label{F_LPVimage_single}
\end{figure}

\begin{figure}[h]
\includegraphics[width=88mm]{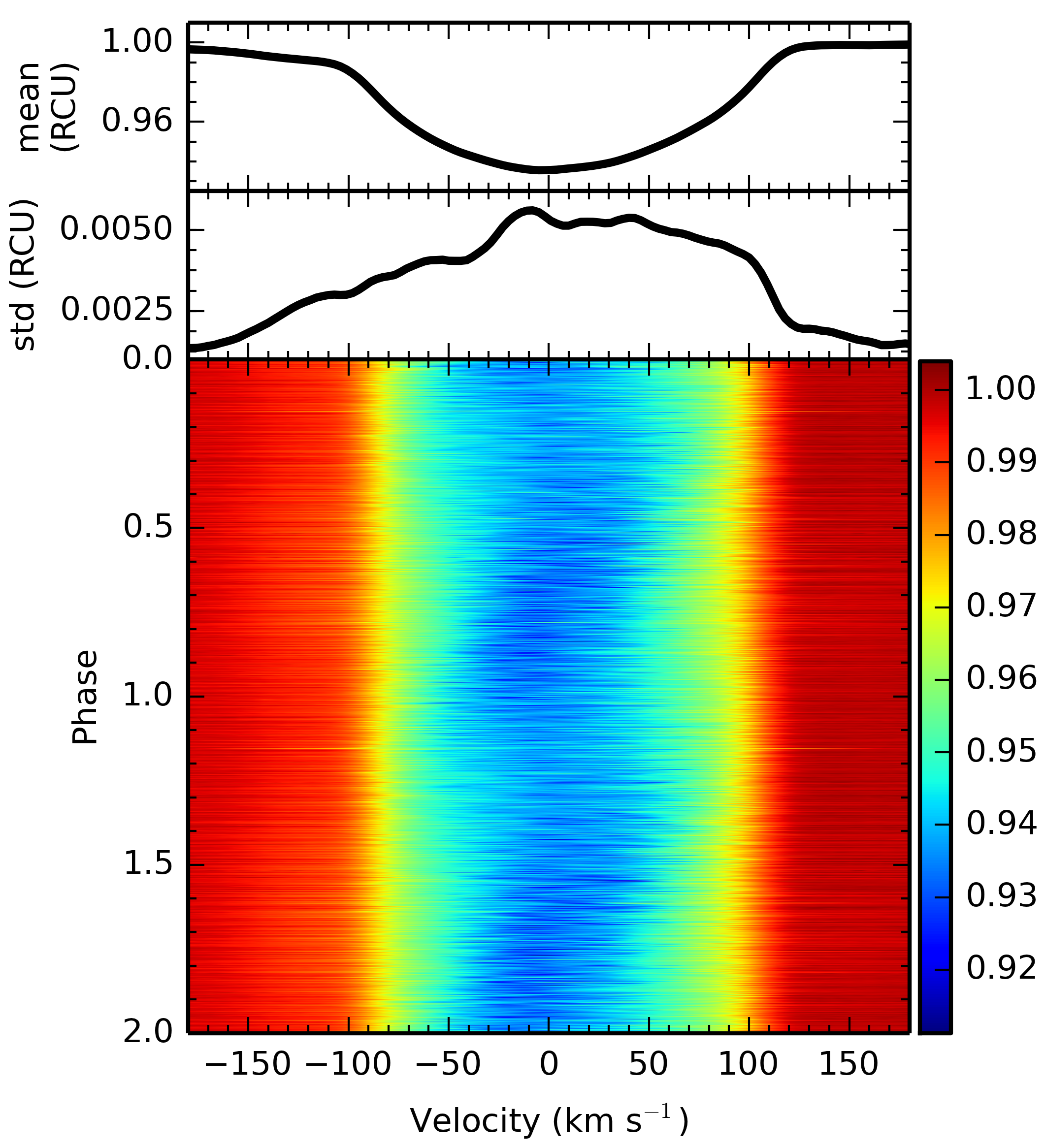}
\caption{Same as Fig.~\ref{F_LPVimage_single} for the standard LSD profiles calculated from all lines in the line mask.}
\label{F_LPVimage_LSDnormal}
\end{figure}

\begin{figure}[h]
\includegraphics[width=88mm]{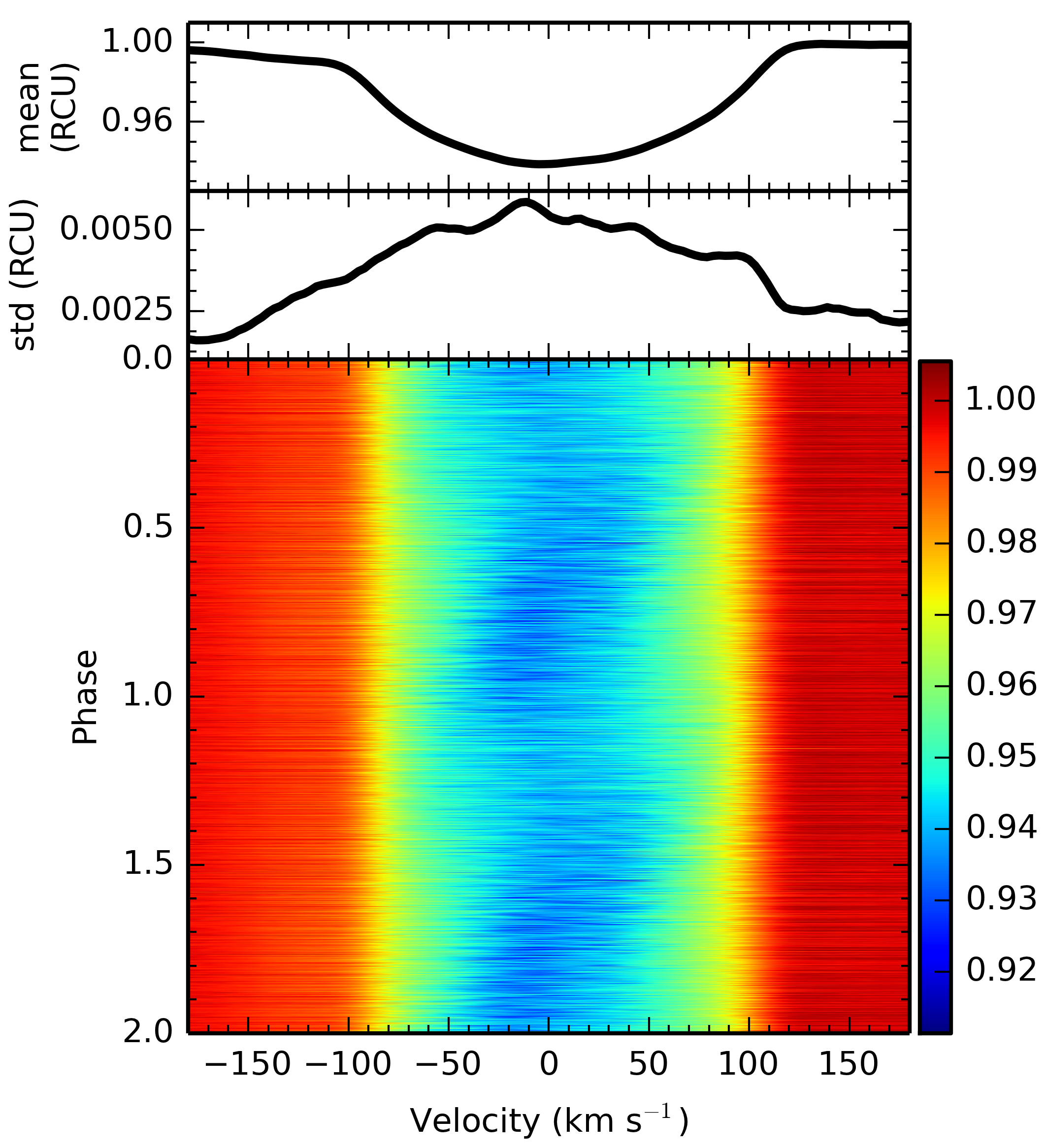}
\caption{Same as Fig.~\ref{F_LPVimage_single} for the LSD profiles calculated from all Fe lines in the line mask only.}
\label{F_LPVimage_LSDFe}
\end{figure}

\clearpage

\section{Results of the mode identification of $f_1=7.3764~\mathrm{d}^{-1}$ and $f_2=5.8496~\mathrm{d}^{-1}$}
\label{A_MIresults}

\begin{table*}
\caption{Results of the mode identification using set 1 of stellar parameters ($M=2.0~\mathrm{M}_\odot$, $R=3.72~\mathrm{R}_\odot$, $\log g=3.6$, and $T_\mathrm{eff}=7050$~K).}
\label{T_resultsMIset1}
\centering
\begin{tabular}{c | r r r r r || c | r r r r r }
\hline
\hline
\multicolumn{6}{c ||}{$f_1=7.3764~\mathrm{d}^{-1}$} & \multicolumn{6}{c}{$f_2=5.8496~\mathrm{d}^{-1}$} \\
\hline
$(\ell_1,m_1)$ & \multicolumn{1}{c}{$i$} & \multicolumn{1}{c}{$v_{eq}\sin i$} & \multicolumn{1}{c}{$v_{eq}$} & \multicolumn{1}{c}{$v_{eq}/v_{crit}$} & \multicolumn{1}{c ||}{$\chi^2$} & $(\ell_2,m_2)$ & \multicolumn{1}{c}{$i$} & \multicolumn{1}{c}{$v_{eq}\sin i$} & \multicolumn{1}{c}{$v_{eq}$} & \multicolumn{1}{c}{$v_{eq}/v_{crit}$} & \multicolumn{1}{c}{$\chi^2$} \\
 & \multicolumn{1}{c}{(deg)} & \multicolumn{1}{c}{(km\,s$^{-1}$)} & \multicolumn{1}{c}{(km\,s$^{-1}$)} & & & & \multicolumn{1}{c}{(deg)} & \multicolumn{1}{c}{(km\,s$^{-1}$)} & \multicolumn{1}{c}{(km\,s$^{-1}$)} & & \\
\hline
$(3,-2)$ & 66.3 & 106.5 & 116.3 & 0.36 & 28.6 & $(3,3)$ & 61.0 & 106.9 & 122.2 & 0.38 & 29.0 \\
$(3,-3)$ & 88.9 & 106.7 & 106.7 & 0.33 & 29.7 & $(3,2)$ & 36.2 & 106.8 & 180.8 & 0.56 & 30.8 \\
$(3,-1)$ & 41.5 & 106.6 & 160.9 & 0.50 & 34.6 & $(3,1)$ & 51.5 & 106.6 & 136.2 & 0.43 & 35.3 \\
$(2,-2)$ & 51.5 & 106.6 & 136.2 & 0.43 & 38.3 & $(2,2)$ & 44.1 & 106.8 & 153.5 & 0.48 & 39.6 \\
$(2,-1)$ & 38.3 & 106.5 & 171.8 & 0.54 & 47.7 & $(2,1)$ & 35.1 & 106.7 & 185.6 & 0.58 & 65.6 \\
$(1,-1)$ & 43.6 & 106.6 & 154.6 & 0.48 & 68.7 & $(1,1)$ & 24.6 & 106.7 & 256.3 & 0.80 & 68.3 \\
\hline
\end{tabular}
\tablefoot{The results are ordered by increasing $\chi^2$. The critical velocity $v_{crit}=320.3~\mathrm{km\,s}^{-1}$.}
\end{table*}

\begin{table*}
\caption{Results of the mode identification using set 2 of stellar parameters ($M=1.5~\mathrm{M}_\odot$, $R=3.75~\mathrm{R}_\odot$, $\log g=3.45$, and $T_\mathrm{eff}=6950$~K).}
\label{T_resultsMIset2}
\centering
\begin{tabular}{c | r r r r r || c | r r r r r }
\hline
\hline
\multicolumn{6}{c ||}{$f_1=7.3764~\mathrm{d}^{-1}$} & \multicolumn{6}{c}{$f_2=5.8496~\mathrm{d}^{-1}$} \\
\hline
$(\ell_1,m_1)$ & \multicolumn{1}{c}{$i$} & \multicolumn{1}{c}{$v_{eq}\sin i$} & \multicolumn{1}{c}{$v_{eq}$} & \multicolumn{1}{c}{$v_{eq}/v_{crit}$} & \multicolumn{1}{c ||}{$\chi^2$} & $(\ell_2,m_2)$ & \multicolumn{1}{c}{$i$} & \multicolumn{1}{c}{$v_{eq}\sin i$} & \multicolumn{1}{c}{$v_{eq}$} & \multicolumn{1}{c}{$v_{eq}/v_{crit}$} & \multicolumn{1}{c}{$\chi^2$} \\
 & \multicolumn{1}{c}{(deg)} & \multicolumn{1}{c}{(km\,s$^{-1}$)} & \multicolumn{1}{c}{(km\,s$^{-1}$)} & & & & \multicolumn{1}{c}{(deg)} & \multicolumn{1}{c}{(km\,s$^{-1}$)} & \multicolumn{1}{c}{(km\,s$^{-1}$)} & & \\
\hline
$(3,-2)$ & 61.5 & 106.7 & 121.4 & 0.44 & 28.7 & $(3,3)$ & 38.3 & 106.9 & 172.5 & 0.62 & 30.4 \\
$(3,-3)$ & 90.0 & 106.8 & 106.8 & 0.39 & 30.5 & $(3,2)$ & 34.1 & 107.0 & 190.9 & 0.69 & 30.8 \\
$(3,-1)$ & 43.0 & 106.7 & 156.5 & 0.57 & 36.0 & $(2,2)$ & 36.7 & 106.9 & 178.9 & 0.65 & 36.8 \\
$(2,-2)$ & 58.3 & 106.8 & 125.5 & 0.45 & 38.5 & $(3,1)$ & 48.9 & 106.6 & 141.5 & 0.51 & 39.1 \\
$(2,-1)$ & 43.0 & 106.6 & 156.3 & 0.57 & 47.7 & $(2,1)$ & 34.1 & 106.8 & 190.5 & 0.69 & 64.0 \\
$(1,-1)$ & 43.6 & 106.7 & 154.7 & 0.56 & 69.1 & $(1,1)$ & 23.0 & 106.7 & 273.1 & 0.99 & 66.9 \\
\hline
\end{tabular}
\tablefoot{The results are ordered by increasing $\chi^2$. The critical velocity $v_{crit}=276.3~\mathrm{km\,s}^{-1}$.}
\end{table*}

\begin{table*}
\caption{Results of the mode identification using set 3 of stellar parameters ($M=2.4~\mathrm{M}_\odot$, $R=5.6~\mathrm{R}_\odot$, $\log g=3.32$, and $T_\mathrm{eff}=6800$~K).}
\label{T_resultsMIset3}
\centering
\begin{tabular}{c | r r r r r || c | r r r r r }
\hline
\hline
\multicolumn{6}{c ||}{$f_1=7.3764~\mathrm{d}^{-1}$} & \multicolumn{6}{c}{$f_2=5.8496~\mathrm{d}^{-1}$} \\
\hline
$(\ell_1,m_1)$ & \multicolumn{1}{c}{$i$} & \multicolumn{1}{c}{$v_{eq}\sin i$} & \multicolumn{1}{c}{$v_{eq}$} & \multicolumn{1}{c}{$v_{eq}/v_{crit}$} & \multicolumn{1}{c ||}{$\chi^2$} & $(\ell_2,m_2)$ & \multicolumn{1}{c}{$i$} & \multicolumn{1}{c}{$v_{eq}\sin i$} & \multicolumn{1}{c}{$v_{eq}$} & \multicolumn{1}{c}{$v_{eq}/v_{crit}$} & \multicolumn{1}{c}{$\chi^2$} \\
 & \multicolumn{1}{c}{(deg)} & \multicolumn{1}{c}{(km\,s$^{-1}$)} & \multicolumn{1}{c}{(km\,s$^{-1}$)} & & & & \multicolumn{1}{c}{(deg)} & \multicolumn{1}{c}{(km\,s$^{-1}$)} & \multicolumn{1}{c}{(km\,s$^{-1}$)} & & \\
\hline
$(3,-2)$ & 69.4 & 106.8 & 114.1 & 0.40 & 29.7 & $(2,2)$ & 31.5 & 107.0 & 204.8 & 0.72 & 31.6 \\
$(3,-3)$ & 87.9 & 107.0 & 107.1 & 0.37 & 31.6 & $(3,2)$ & 79.4 & 106.5 & 108.3 & 0.38 & 34.0 \\
$(3,-1)$ & 42.0 & 106.8 & 159.6 & 0.56 & 37.4 & $(3,3)$ & 23.0 & 107.2 & 274.4 & 0.96 & 43.8 \\
$(2,-2)$ & 57.8 & 106.9 & 126.3 & 0.44 & 38.7 & $(3,1)$ & 40.9 & 107.0 & 163.4 & 0.57 & 44.3 \\
$(2,-1)$ & 44.1 & 106.8 & 153.5 & 0.54 & 48.5 & $(2,1)$ & 30.9 & 107.0 & 208.4 & 0.73 & 57.4 \\
$(1,-1)$ & 43.6 & 106.9 & 155.0 & 0.54 & 69.1 & $(1,1)$ & 23.0 & 107.0 & 273.8 & 0.96 & 61.7 \\
\hline
\end{tabular}
\tablefoot{The results are ordered by increasing $\chi^2$. The critical velocity $v_{crit}=286.0~\mathrm{km\,s}^{-1}$.}
\end{table*}

\clearpage

\begin{figure}[p]
\includegraphics[width=88mm]{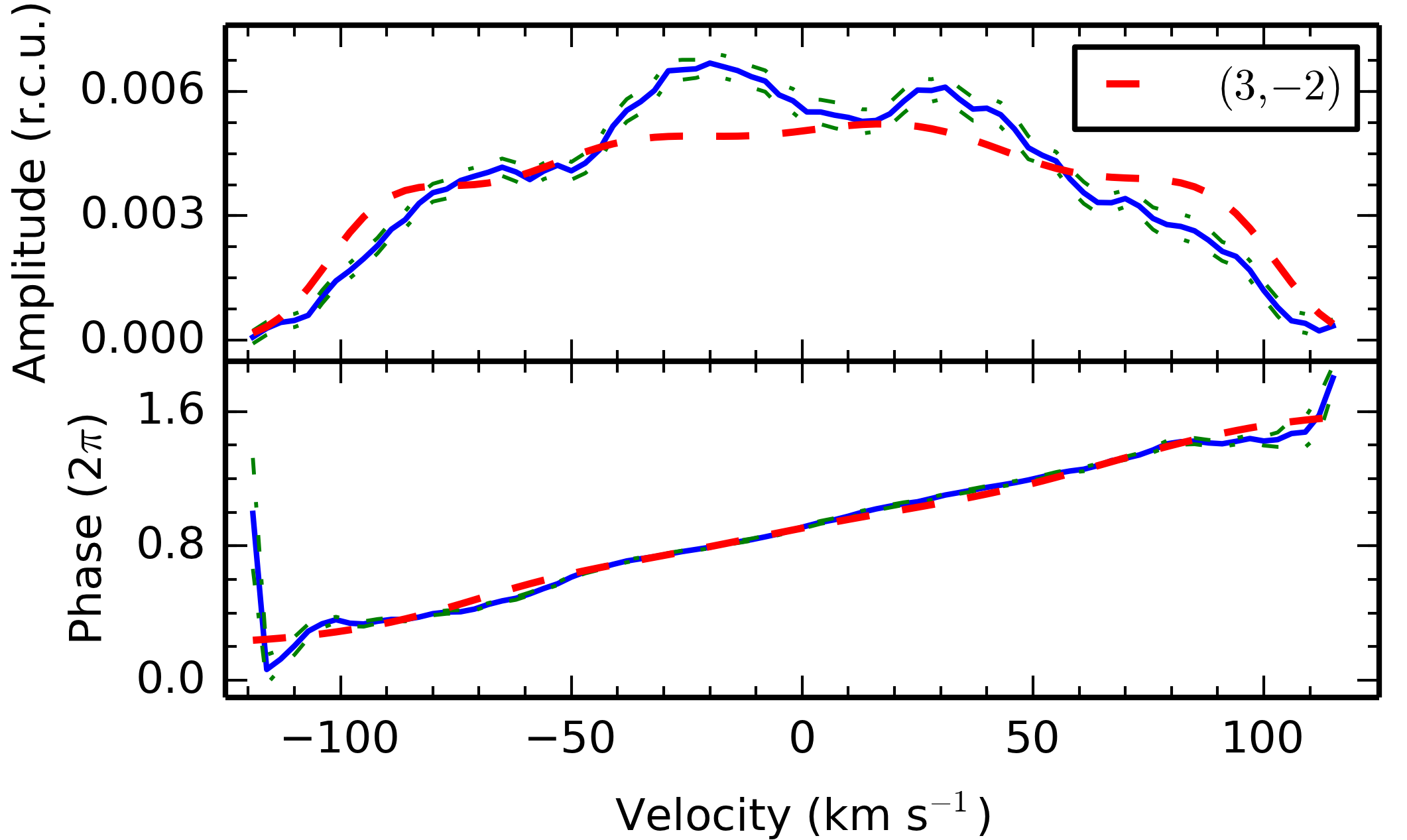}
\caption{Fit of the amplitude (upper panel) and phase (lower panel) across the line of the frequency $f_1$ with a $(\ell,m)=(3,-2)$ mode. The observed profiles are calculated from the phase-folded data set. Observations are shown as the blue, solid line, the errors of the observations are shown as the green, dash-dotted line, and the fit is shown as the red, dashed line. The fit has a $\chi^2$ of 28.6.}
\label{F_phiF1zapFit:l3m-2}
\vspace{12pt}
\includegraphics[width=88mm]{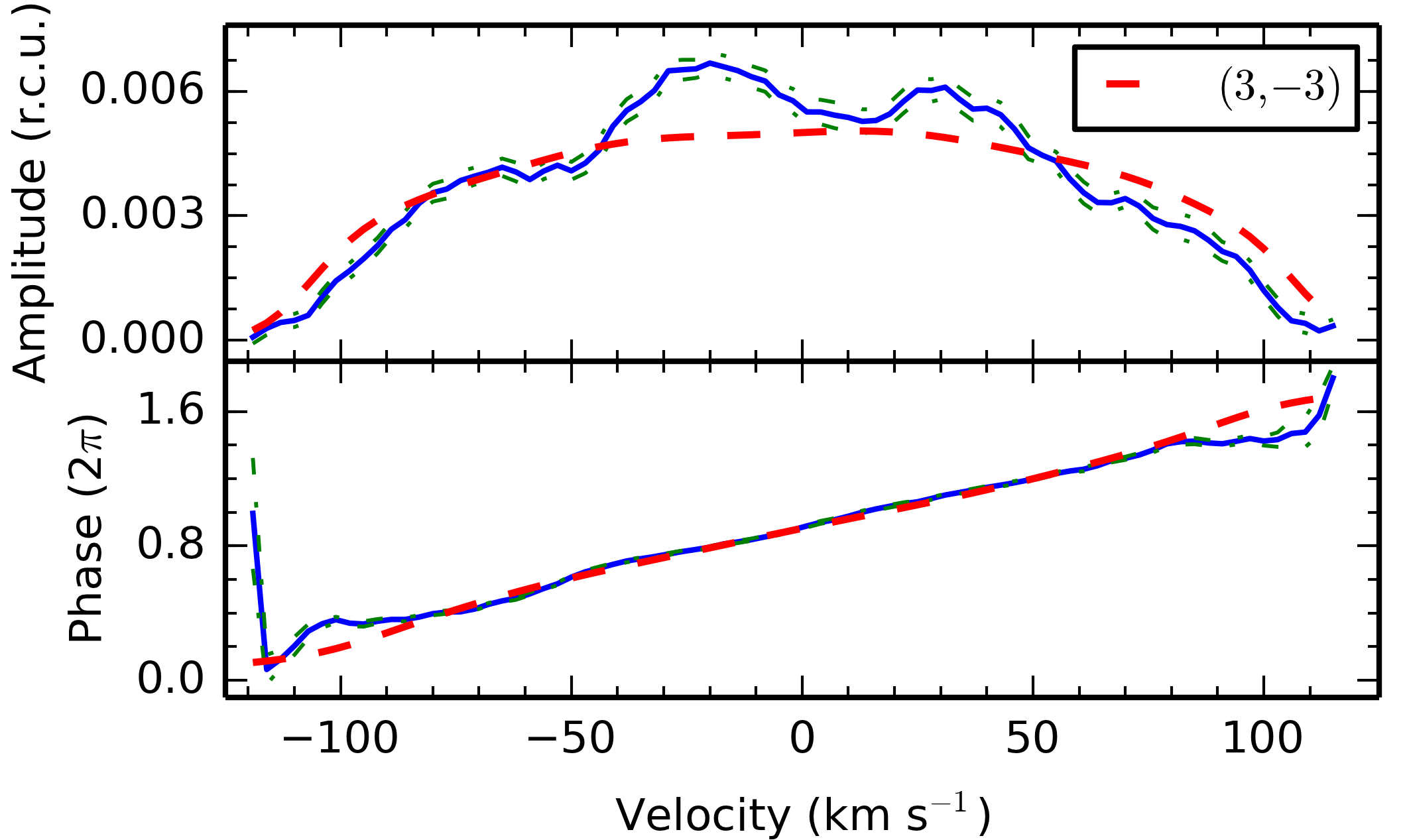}
\caption{Same as Fig.~\ref{F_phiF1zapFit:l3m-2}, but with a $(\ell,m)=(3,-3)$ mode. The fit has a $\chi^2$ of 29.7.}
\label{F_phiF1zapFit:l3m-3}
\vspace{12pt}
\includegraphics[width=88mm]{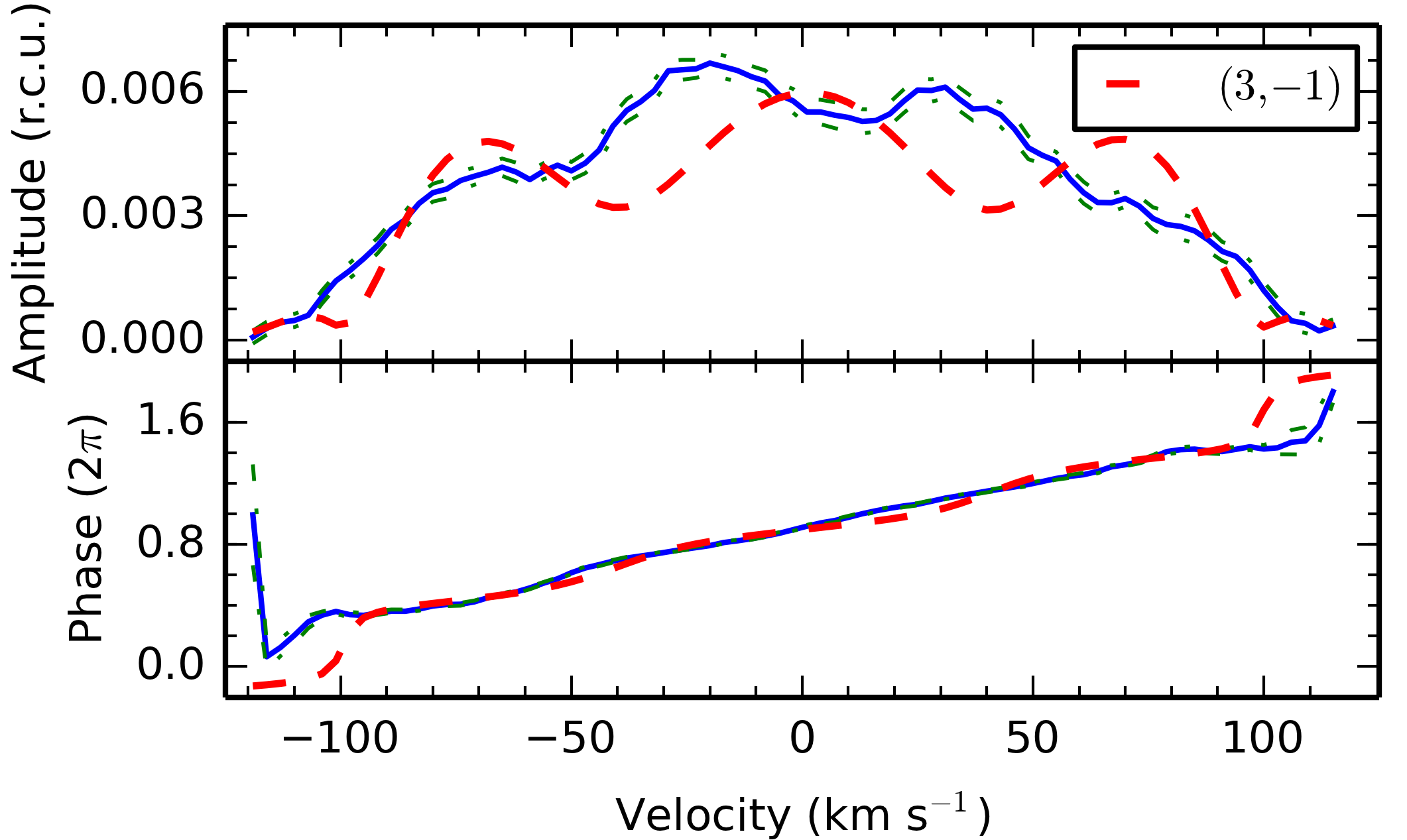}
\caption{Same as Fig.~\ref{F_phiF1zapFit:l3m-2}, but with a $(\ell,m)=(3,-1)$ mode. The fit has a $\chi^2$ of 34.6.}
\label{F_phiF1zapFit:l3m-1}
\end{figure}

\begin{figure}[p]
\includegraphics[width=88mm]{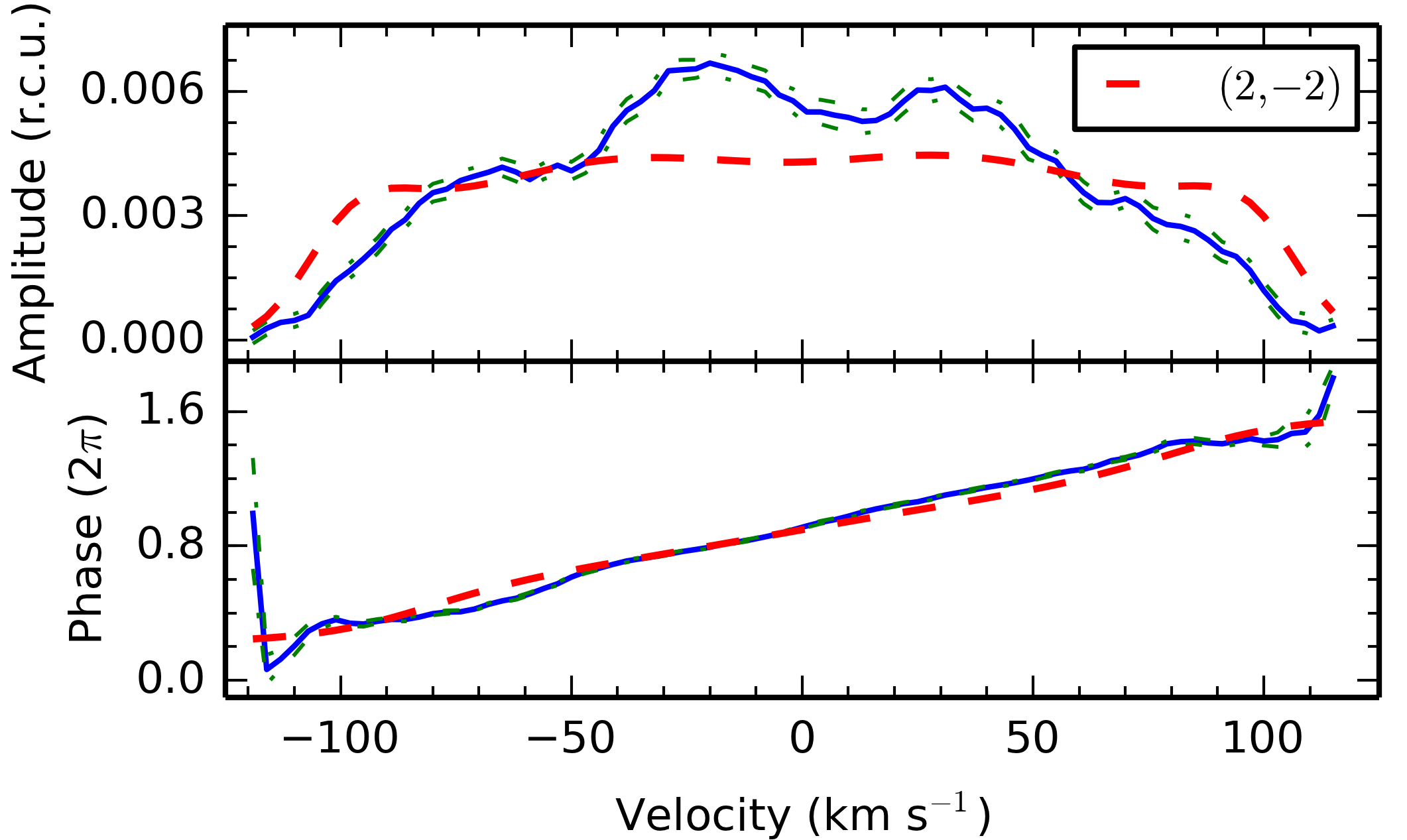}
\caption{Same as Fig.~\ref{F_phiF1zapFit:l3m-2}, but with a $(\ell,m)=(2,-2)$ mode. The fit has a $\chi^2$ of 38.3.}
\label{F_phiF1zapFit:l2m-2}
\vspace{12pt}
\includegraphics[width=88mm]{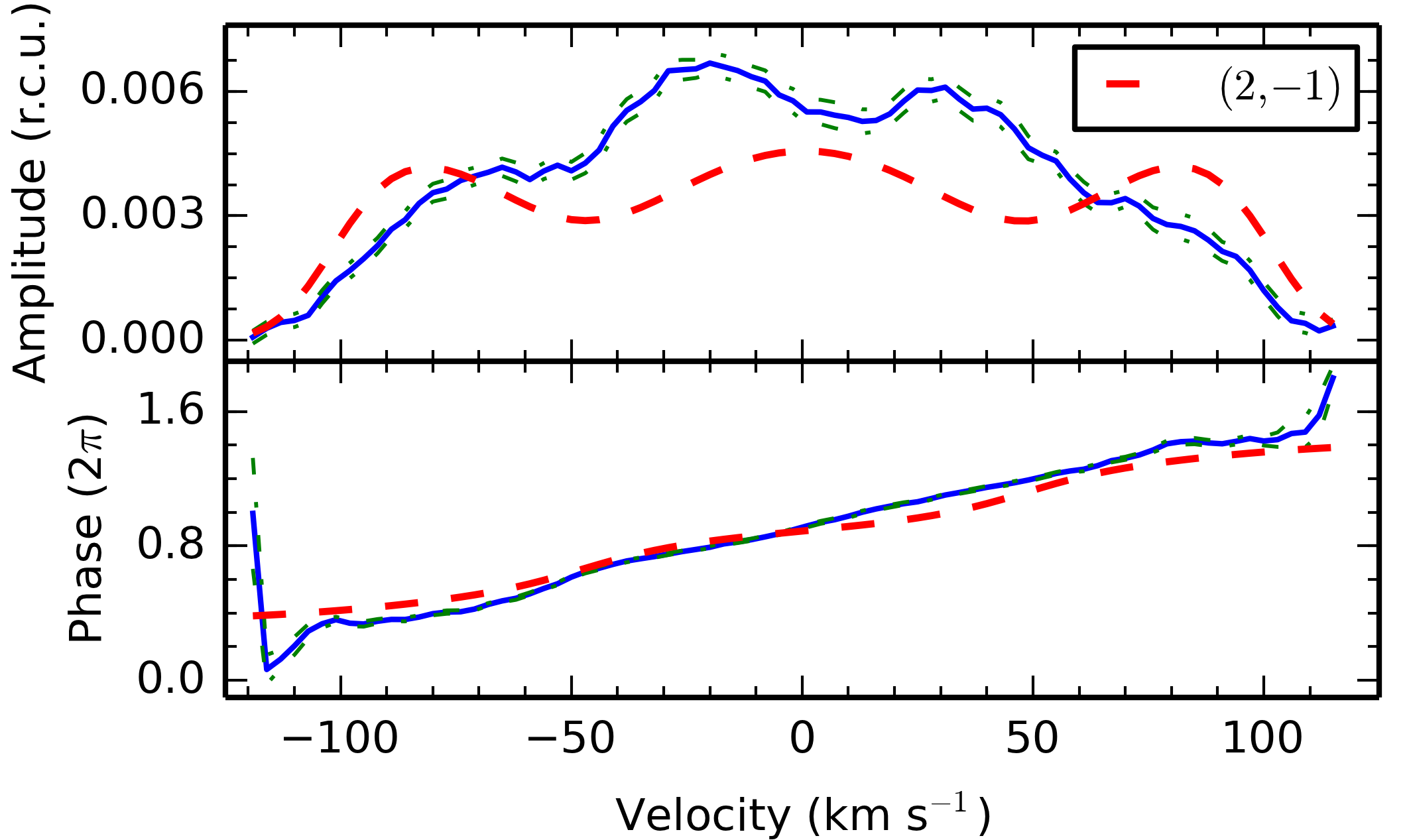}
\caption{Same as Fig.~\ref{F_phiF1zapFit:l3m-2}, but with a $(\ell,m)=(2,-1)$ mode. The fit has a $\chi^2$ of 47.7.}
\label{F_phiF1zapFit:l2m-1}
\vspace{12pt}
\includegraphics[width=88mm]{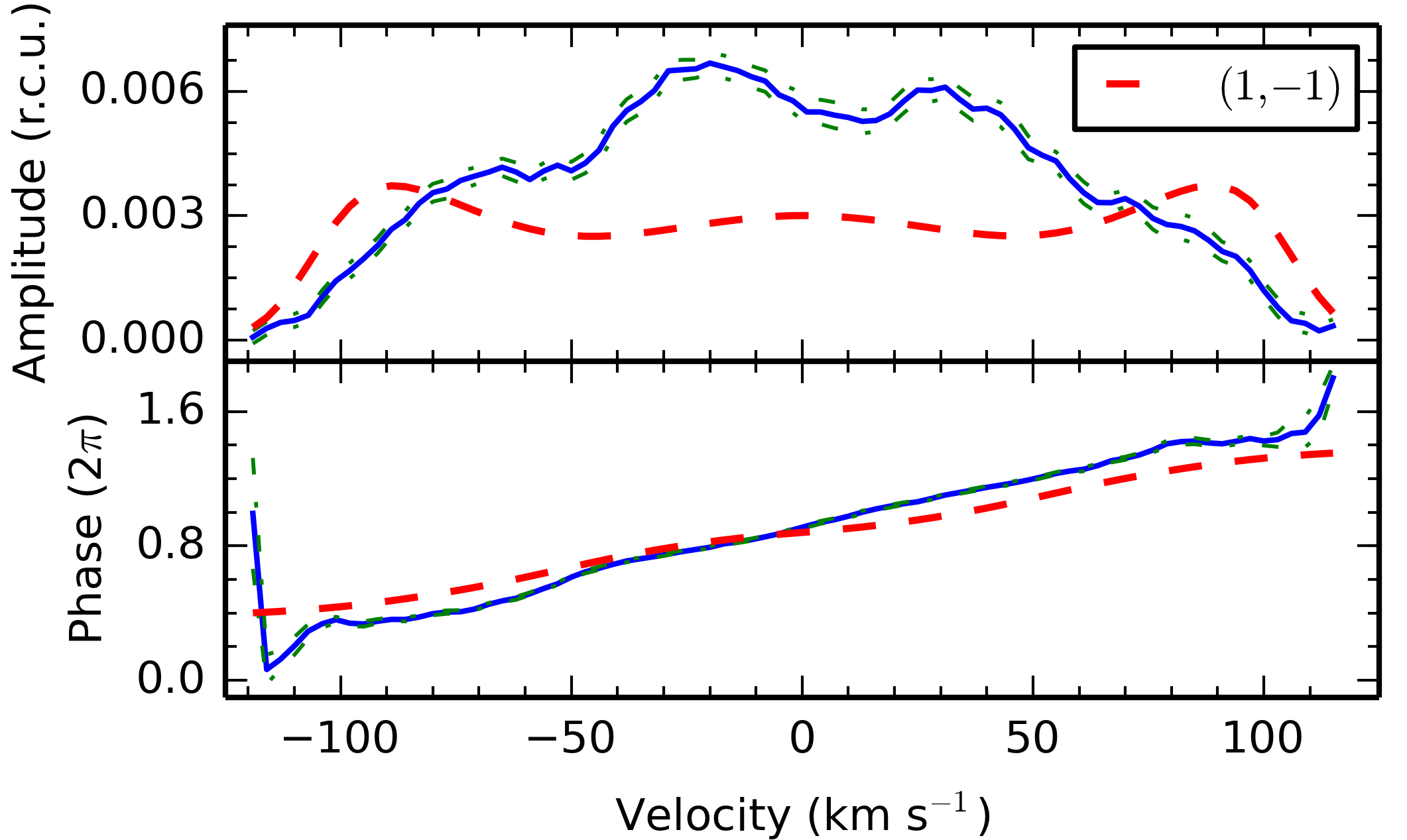}
\caption{Same as Fig.~\ref{F_phiF1zapFit:l3m-2}, but with a $(\ell,m)=(1,-1)$ mode. The fit has a $\chi^2$ of 68.7.}
\label{F_phiF1zapFit:l1m-1}
\end{figure}

\begin{figure}[p]
\includegraphics[width=88mm]{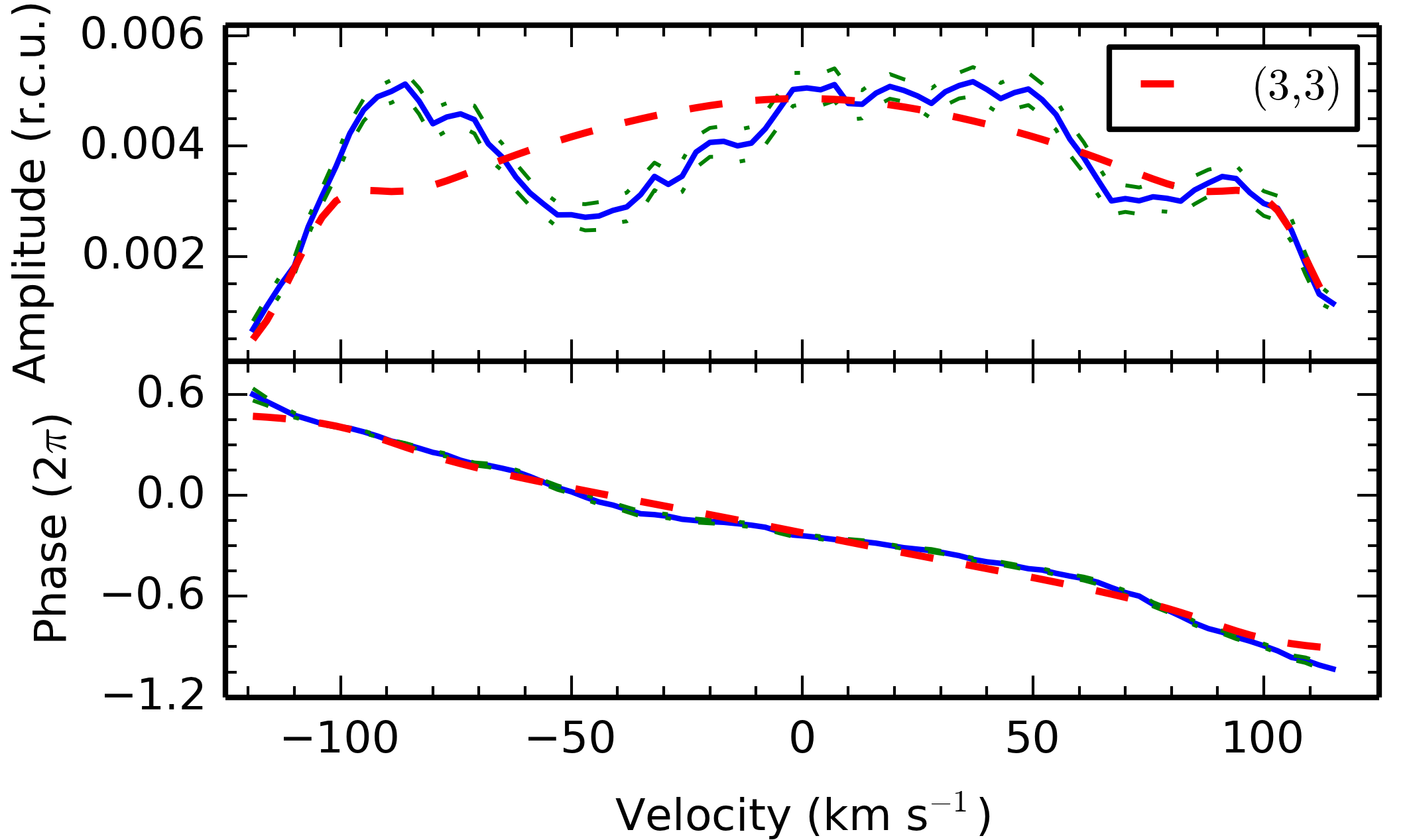}
\caption{Fit of the amplitude (upper panel) and phase (lower panel) across the line of the frequency $f_2$ with a $(\ell,m)=(3,3)$ mode. The observed profiles are calculated from the phase-folded data set. Observations are shown as the blue, solid line, the errors of the observations are shown as the green, dash-dotted line, and the fit is shown as the red, dashed line. The fit has a $\chi^2$ of 29.}
\label{F_phiF2zapFit:l3m3}
\vspace{12pt}
\includegraphics[width=88mm]{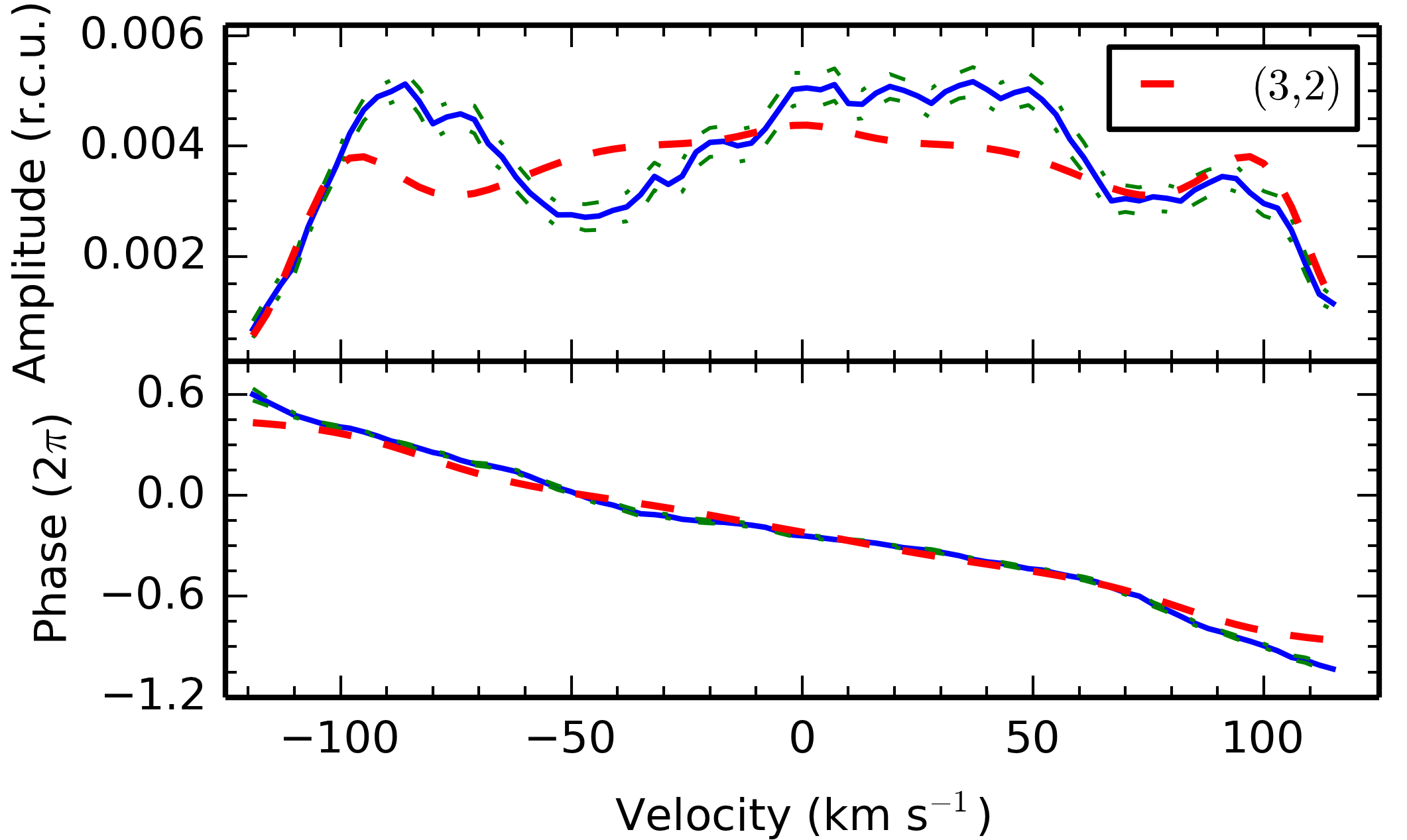}
\caption{Same as Fig.~\ref{F_phiF2zapFit:l3m3}, but with a $(\ell,m)=(3,2)$ mode. The fit has a $\chi^2$ of 30.8.}
\label{F_phiF2zapFit:l3m2}
\vspace{12pt}
\includegraphics[width=88mm]{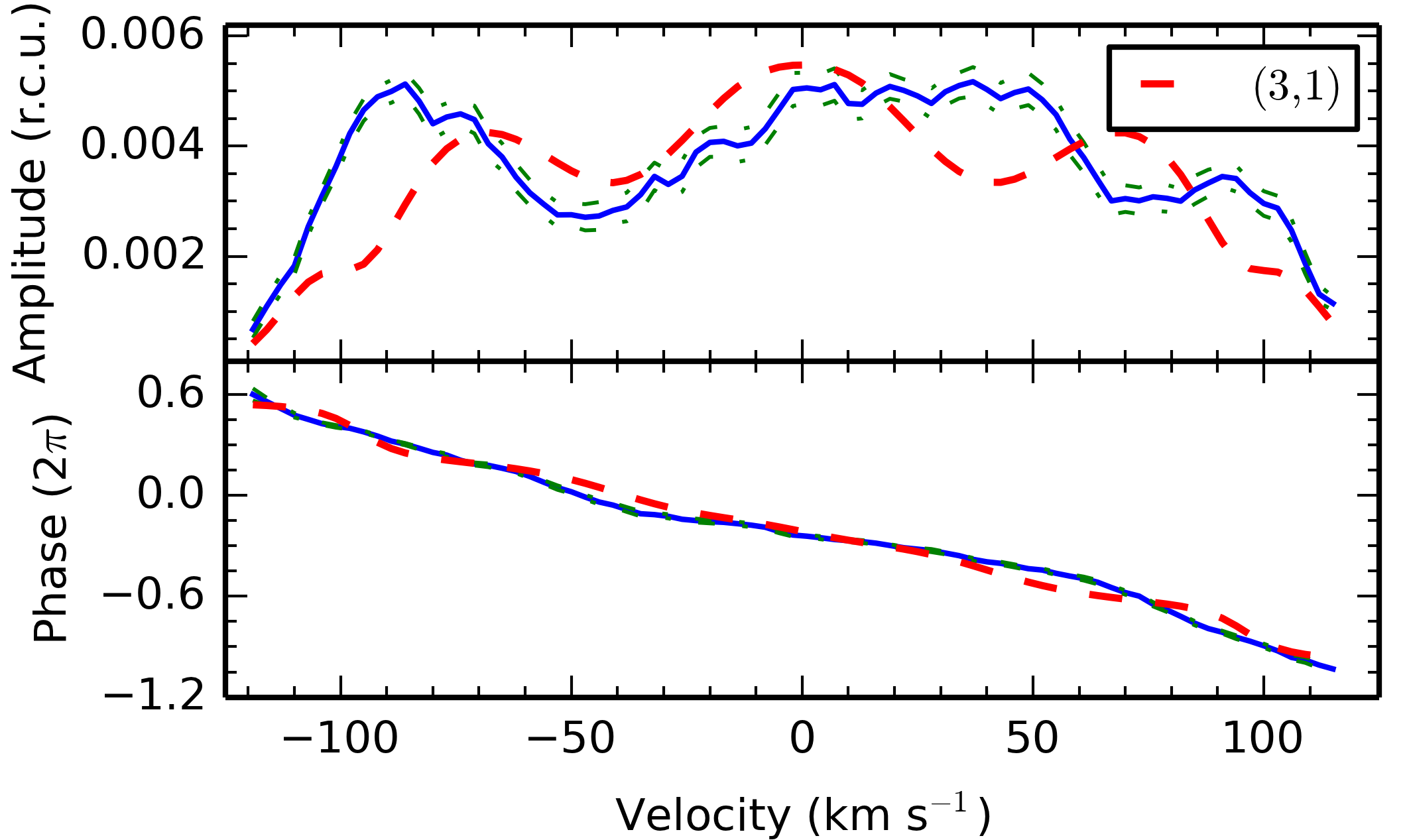}
\caption{Same as Fig.~\ref{F_phiF2zapFit:l3m3}, but with a $(\ell,m)=(3,1)$ mode. The fit has a $\chi^2$ of 35.3.}
\label{F_phiF2zapFit:l3m1}
\end{figure}

\begin{figure}[p]
\includegraphics[width=88mm]{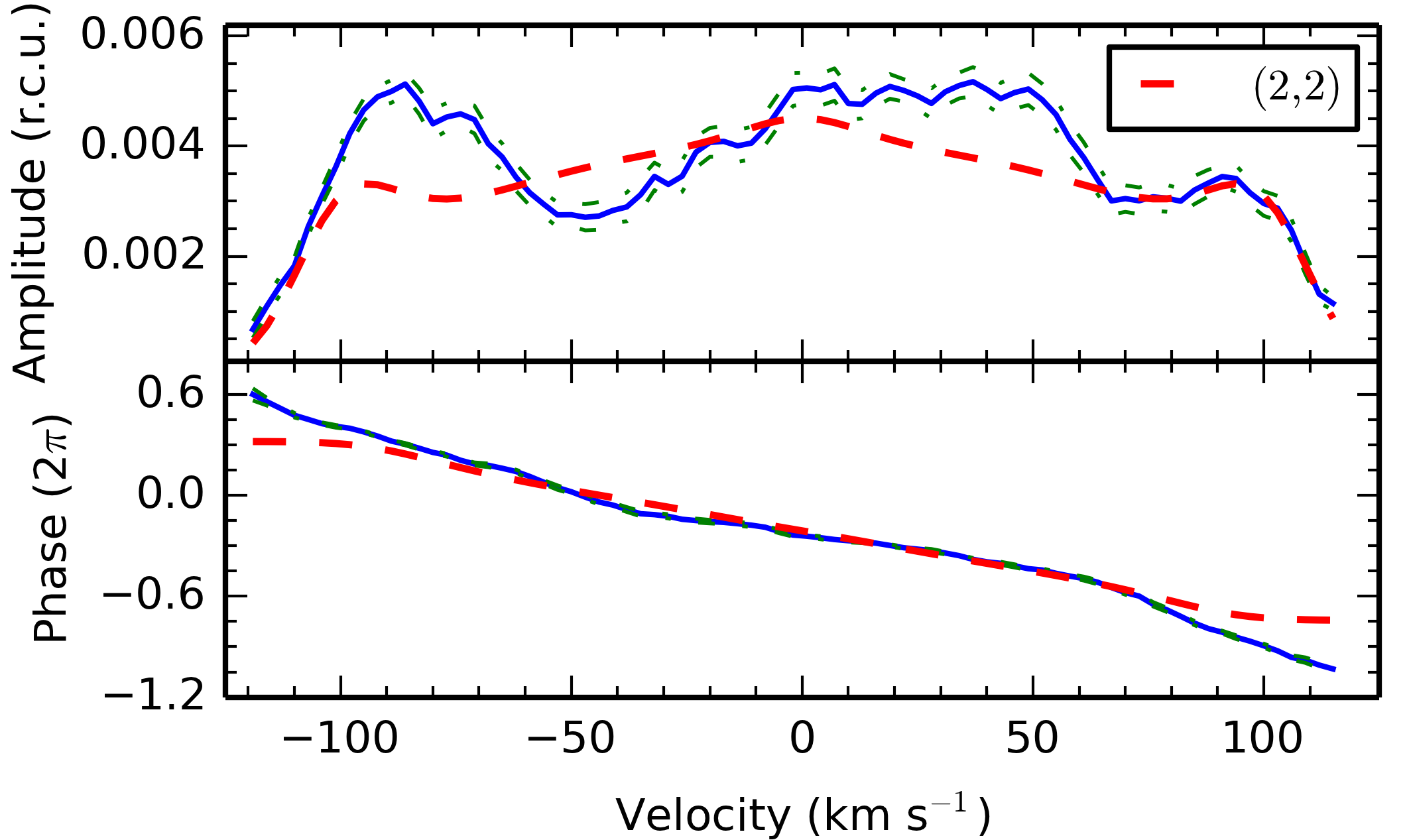}
\caption{Same as Fig.~\ref{F_phiF2zapFit:l3m3}, but with a $(\ell,m)=(2,2)$ mode. The fit has a $\chi^2$ of 39.6.}
\label{F_phiF2zapFit:l2m2}
\vspace{12pt}
\includegraphics[width=88mm]{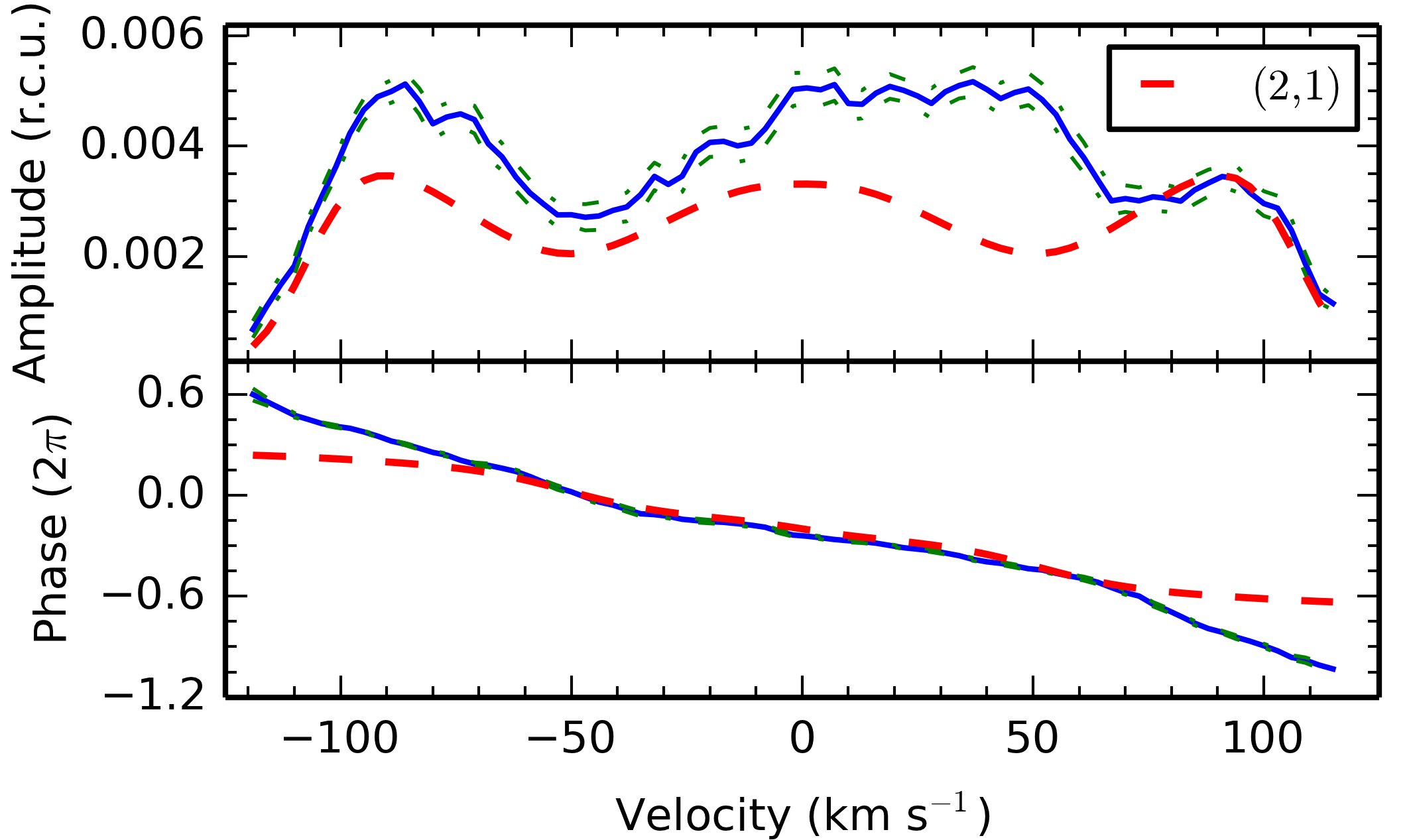}
\caption{Same as Fig.~\ref{F_phiF2zapFit:l3m3}, but with a $(\ell,m)=(2,1)$ mode. The fit has a $\chi^2$ of 65.6.}
\label{F_phiF2zapFit:l2m1}
\vspace{12pt}
\includegraphics[width=88mm]{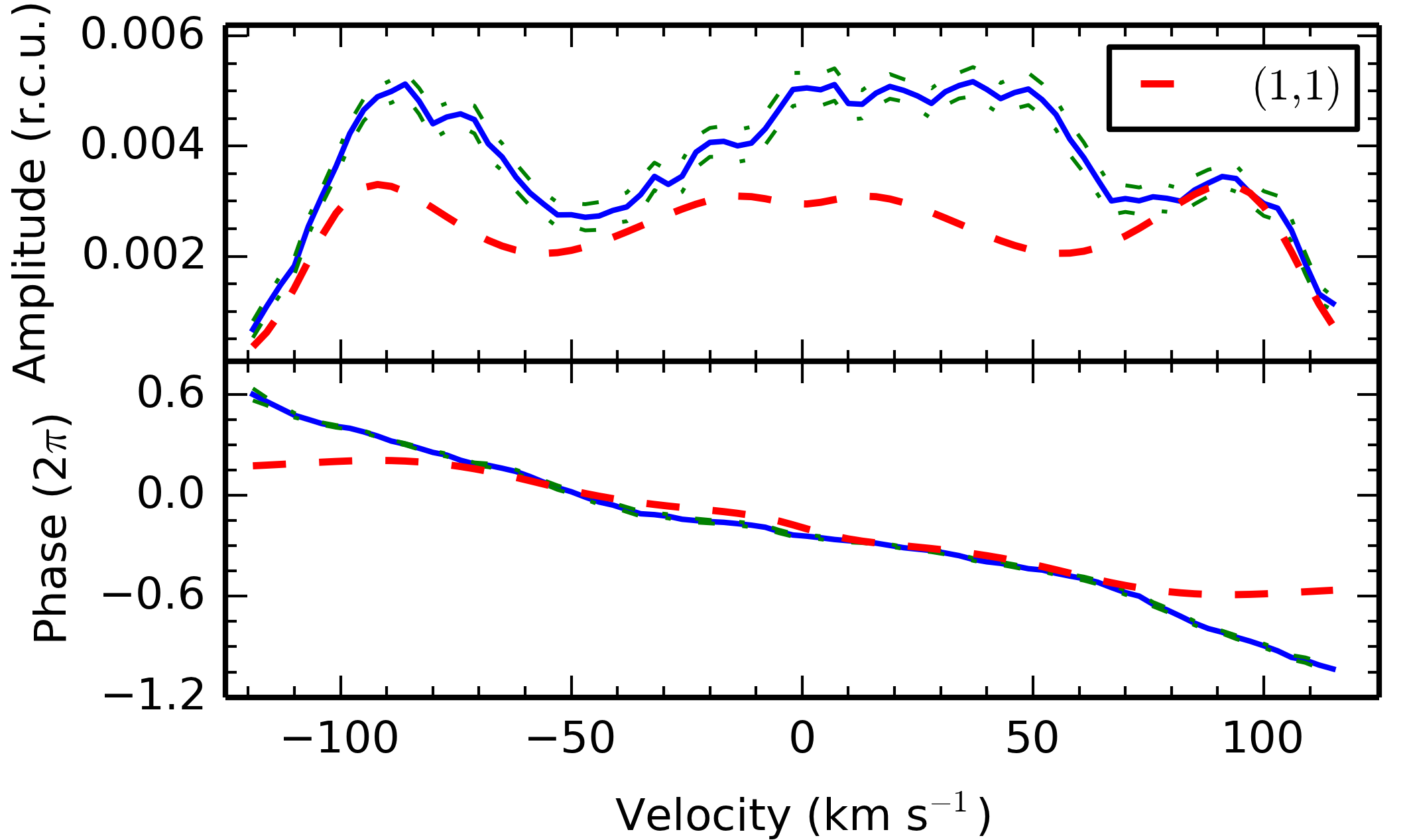}
\caption{Same as Fig.~\ref{F_phiF2zapFit:l3m3}, but with a $(\ell,m)=(1,1)$ mode. The fit has a $\chi^2$ of 68.3.}
\label{F_phiF2zapFit:l1m1}
\end{figure}

\end{document}